\documentclass[preprint,review,12pt]{elsarticle}

 \usepackage{graphics}
\usepackage{caption}
\usepackage{graphicx}
\usepackage{amsmath}
\usepackage{amssymb}
\usepackage{floatrow}
\usepackage{float}
\usepackage{graphicx,subcaption}
\usepackage{subeqnarray}
\usepackage{multirow,bigdelim}
\usepackage{booktabs}
\usepackage{lscape}
\usepackage{array}
\newcolumntype{C}[1]{>{\centering\let\newline\\\arraybackslash\hspace{0pt}}m{#1}}

\DeclareFloatFont{tiny}{\tiny}
 \usepackage{epsfig}

\usepackage{amssymb}

 \usepackage{amsthm}

\usepackage{lineno}

 \biboptions{compress}

\makeatletter
\newcommand{\thickhline}{%
    \noalign {\ifnum 0=`}\fi \hrule height 1pt
    \futurelet \reserved@a \@xhline
}
\newcolumntype{"}{@{\hskip\tabcolsep\vrule width 1pt\hskip\tabcolsep}}
\makeatother

\begin{document}

\begin{frontmatter}


\title{ Semi-implicit direct forcing immersed boundary method for incompressible viscous thermal flow problems:  a Schur complement approach}

\author{Yuri Feldman}

\address{Department of Mechanical Engineering, Ben-Gurion University of the Negev,\\
P.O. Box 653, Beer-Sheva 84105, Israel}

\begin{abstract}
An extended immersed boundary method utilizing a semi-implicit direct forcing approach for the simulation of confined incompressible viscous thermal flow problems is presented. The method utilizes a Schur complement approach to enforce the kinematic constraints of no-slip and the corresponding thermal boundary conditions for immersed surfaces. The developed methodology can be straightforwardly adapted to any existing incompressible time marching solver based on a segregated pressure-velocity coupling. The method accurately meets the thermal and the no-slip boundary conditions on the surfaces of immersed bodies for the entire range of Rayleigh numbers $10^3\leqslant Ra\leqslant10^6$. Strategies for further increasing the computational efficiency of the developed approach are discussed. The method has been extensively verified by applying it for the simulation of a number of representative fully 3D confined natural convection steady and periodic flows. Complex dynamic phenomena typical of this kind of flow including  vortical structures and convection cells and instability characteristics, were simulated and visualized and the results were found to compare favorably with results known from literature.
\end{abstract}


\end{frontmatter}


\section{Introduction}
\label{Intro}
Accurate simulation of viscous flows in the presence of  bodies of complex geometry is critical, both in pure science and in engineering. The traditional, and probably still the most popular, methodology for resolving the characteristics of viscous flows in the vicinity of immersed bodies utilizes a body-conformal grid precisely matching the discretized body surface with the adjacent flow domain (up to the  discretization error). However, the ever-increasing complexity of the flow configurations studied seriously challenges the efficiency and precision of numerical methods based on the body-conformal approach, as the grid orthogonality and skewness may have a significant effect on the accuracy and the stability of the methods. The immersed boundary (IB) method, in which the body surface is determined by the location of a set of discrete Lagrangian points that do not necessarily coincide with the underlying Eulerian grid, comprises an attractive alternative to the body-conformal grid approach. The method, initially developed  by Peskin \cite{peskin1972JCP} for the simulation of the blood flow in the mitral valve in the heart, has become very popular over the last three decades and comprises the basis for this rapidly developing field of computational science.

Accurately imposing a body-force field in the vicinity of an immersed surface to enforce the no-slip kinematic constraints is critical for any IB method. Among a number of existing methods, the present study focuses on the direct forcing approach, which enforces the desired value of velocity directly on the boundary without involving any dynamical process. The direct forcing approach, initially formulated by Mohd-Yusof \cite{Mohd-Yusof1997}, has gained popularity over the years due to its simple implementation and robustness. In the direct forcing approach, the Lagrangian body-force field can be calculated either explicitly or implicitly. The fully explicit formulation is typically  implemented on the basis of the segregated pressure-velocity coupling utilized for the  solution of incompressible Navier Stokes (NS) equations. In this case, the kinematic no-slip constraints are applied to the intermediate non-solenoidal velocity, which is further corrected to meet the divergence free constraint (see e.g. \cite{Mohd-Yusof1997, faldun2000JCP,le2007computmeth}).  The explicit approach has been adopted in various engineering applications, as detailed in the comprehensive review of Mittal and Iaccarino \cite{mittal2005AnnRev}. Recently the approach has also been extended to thermal flow problems \cite{yoon2010heattransf,ren2012compufluid,ren2013IJHMT,gulb2015IJHMT}. Explicit calculation of the Lagrangian forces can be  straightforwardly implemented without requiring any modification of the existing time marching solvers, which explains its high popularity. Despite its evident attractiveness, the explicit treatment has two major drawbacks. First, the existence of local mass leakage through boundaries of the immersed body and second, the pointwise character of the calculation of Lagrangian forces and heat fluxes, which does not reflect their mutual interaction in a given time step. While the first limitation can be successfully remedied by simply decreasing the value of time step, the second one can seriously deteriorate the precision of the performed analysis, as it violates the intrinsic elliptic character of NS equations. In particular, the issue is of critical importance in confined natural convection unsteady flows characterized by low and moderate Rayleigh numbers, which are the focus of the present study.

The limitations of the explicit direct forcing IB method motivated further research aimed at improving the existing explicit formulations. Su et al. \cite{su2007compufluid} proposed a new forcing procedure based on a solution of a banded linear system, coupling together all the Lagrangian markers. A similar approach was then implemented by Ren et al. \cite{ren2012compufluid,ren2013IJHMT},  who implicitly evaluated  all the Lagrangian forces and heat sources by assembling them into a single system of equations. The studies of Wang et al. \cite{wang2008MTPHFLOW} and Kempe et al. \cite{kempe2012JCP,kempe2015JCP}, who introduced multi-direct forcing (MDF) schemes based on iterative enhancement of Euler-Lagrange coupling, are also worth mentioning. Although the developed approaches succeeded in improving the accuracy of the imposition of no-slip and thermal constraints, non-negligible inaccuracies attributed to the explicit treatment of the diffusion term in the NS and energy equations still remain. Moreover, correction of velocity and temperature fields by the MDF schemes locally deteriorates the momentum and the thermal balances close to the immersed boundaries and can result in non-negligible inaccuracies in the estimation of local Nusselt numbers and drag and lift coefficients.

The fully implicit implementation of the Lagrangian forces and heat fluxes offers an alternative to the fully explicit direct forcing approach. The Lagrangian forces and heat fluxes implicitly embedded into the corresponding energy and momentum equations play the role of distributed Lagrange multipliers, which reflect the impact of the immersed body on the surrounding flow. Together with additional equations connecting between the corresponding Eulerian and Lagrangian temperature and velocity fields, they form the whole closed system of the coupled equations. Solution of the system provides high fidelity results for a broad range of Reynolds and Rayleigh numbers. Pioneering work in this field is due to Golowinski et al. \cite{glowinski1998ComputMethApplMechEngrg}, who applied  the distributed Lagrange multiplier method (DLM) for the simulation of 2D flow around a moving disc. The method  was then extended to the simulation of particulate flows \cite{Glowinski2001JCP,Yu2002JNNFM,Yu2004JFM} and to the simulation of fluid/flexible-body interactions \cite{Yu2005JCP}. Another approach was proposed by Taira and Colonius \cite{taira2007JCP}, who utilized the distributed Lagrange multiplier to simultaneously  satisfy the divergence-free and no-slip kinematic constraints by the solution of the modified Poisson equation in the framework of the projection method. The developed approach was then applied in various engineering fields, including active and passive flow control \cite{taira2007JCP,choi2015jfm}, optimization of performance of  a hot air balloon \cite{samanta2010aiaa} and the dynamic interactions between rigid-body systems and incompressible viscous flows \cite{wang2015JCP}.

The latest theoretical developments of the fully coupled Lagrange multiplier approach are: due to the study of Kallemov et al. \cite{kallemov2016CAMcOS}, who presented  a novel IB formulation for modeling flows around fixed or moving rigid bodies that is suitable for a broad range of Reynolds numbers, including steady Stokes flow; the formulation developed by Stein et al. \cite{Stein2016JCP}, who established the immersed boundary smooth extension (IBSE) method and  demonstrated the superiority of its  convergence for a wide spectrum of equations with both  Dirichlet and Neumann boundary conditions; the work of Feldman and Gulberg \cite{feldman2016JCP}, comprising an extended fully pressure-velocity-Lagrange multipliers coupled formulation of the IB method, capable of performing accurate linear stability analysis of incompressible viscous and thermal 2D and axi-symmetric flows; and the recent study of Liska and Colonius \cite{Liska2016JCP} who proposed novel time integration schemes for the efficient solution of the discrete momentum equations coupled with the discrete divergence free and no-slip constraints. It is also worth mentioning the recent works of Bao et al. \cite{Bao2017JCP}, who succeeded in substantially improving the volume conservation for fluid-structure interaction applications in periodic domains, as well as Stein et al. \cite{Stein2017JCP}, who further extended the IBSE method to allow for the imposition of a divergence free  constraint.

Although the fully implicit treatment of the Lagranginan forces and heat fluxes succeeds in accurately satisfying the kinematic constraints on the surface of an immersed body, it often includes non-trivial intermediate stages requiring substantial modification of time stepping codes which were originally developed without IB capability. In our present work we pursue a different concept, where the IB approach is not seen as a stand alone solver, but, rather, it comprises a methodology of enforcing boundary conditions. One of the main purposes of the present study is to demonstrate the ways of plugging the IB approach in any of many available open source CFD packages or in-house developed solvers in a modular fashion. To achieve this we apply a semi-implicit approach, which is applicable for the whole family of pressure-velocity segregated solvers based, for example, on SIMPLE or fractional step algorithms. The idea is to implicitly couple the Lagrangian forces with an intermediate non-solenoidal velocity field, further updated by the following correction step. The concept has already been successfully implemented for the investigation of isothermal viscous flows by Park et al. \cite{Park2016JCP}, who combined a block $LU$ decomposition technique with Taylor series expansion to construct the implicit IB forcing in a recurrence form,  and by Le et al. \cite{lee2008ComputMethApplMechEngrg}, who used the FISHPACK \cite{fishpack} library for the solution of modified Helmholtz and Poisson equations. The present study demonstrates how the developed methodology can be embedded into the generic incompressible NS solver based on the SIMPLE algorithm \cite{vitoshkin2013} in order to endow it with the IB capability. Although in this case the kinematic constraints are satisfied by the machine zero precision only for the intermediate non-solenoidal velocity field, it is shown that an inaccuracy accumulated in the corrected velocity field in the vicinity of an immersed surface is less than a discretization error. A number of representative fully 3D confined natural convection flows are simulated and the obtained results are favorably compared with the data available in the literature.

\section{The theoretical background}
\subsection{Predictor-corrector approach with incorporated IB functionality}
\label{ThBackgr}
We start with a brief description of the standard predictor-corrector approach based on the SIMPLE algorithm of Patankar and Spalding \cite{patankar197IJHMT} with  incorporated IB capability. The flow is governed by the non-dimensional incompressible NS and energy equations, in which the buoyancy effects are introduced by applying the Boussinesq approximation:
\begin{equation}
\nabla\cdotp \textbf{\textit{u}}= 0,
\label{eq:continuity}
\end{equation}
\begin{equation}
\frac{\partial{\textbf{\textit{u}}}}{\partial{\textit{t}}}+(\textbf{\textit{u}}\cdotp\nabla){\textbf{\textit{u}}}=
-\nabla{p}+\sqrt{\frac{Pr}{Ra}}\nabla^2 \textbf{\textit{u}}+\theta {\vec{e}}_{z} + \textbf{\textit{f}},
\label{eq:momentun}
\end{equation}
\begin{equation}
\frac{\partial{\theta}}{\partial{\textit{t}}}+(\textbf{\textit{u}}\cdotp\nabla){\theta}=
\frac{1}{\sqrt{PrRa}}\nabla^2\theta+ \textit{q},
\label{eq:energy}
\end{equation}
where \textbf{\textit{u}}=\textit{(u,v,w)}, \textit{p}, \textit{t}, and $\theta$ are the non-dimensional velocity, pressure, time and temperature, respectively, and ${\vec{e}}_{z}$ is a unit vector in the vertical (\textit{y}) direction. In accordance with the Boussinesq approximation the density field is given by $\rho$=$\rho_{0}$(1-$\beta(T-T_{c})$), where $\rho_{0}$ is the density value corresponding to the thermal equilibrium. The equations are normalized  by utilizing \textit{L}, \textit{U}=$\sqrt {g\beta L \Delta T}$, \textit{t}=${L{/}U}$, and \textit{P}=$\rho {U}^2$ for length, velocity, time, and pressure scales, respectively.  The Rayleigh, $Ra$, and Prandtl, $Pr$, numbers are $Ra$=${\frac{g \beta}{\nu \alpha} \Delta T {L}^3}$ and $Pr$=$\nu$${/}$$\alpha$, respectively, where $\nu$ is the kinematic viscosity, $\alpha$ is the thermal diffusivity and $\Delta T$ is the temperature difference between the maximal and the minimal temperatures of the problem.

\begin{figure}
\centering
\caption{A schematic description of the basic principles of the IB method on the staggered Eulerian grid.}
\includegraphics[width=0.7\textwidth,clip=]{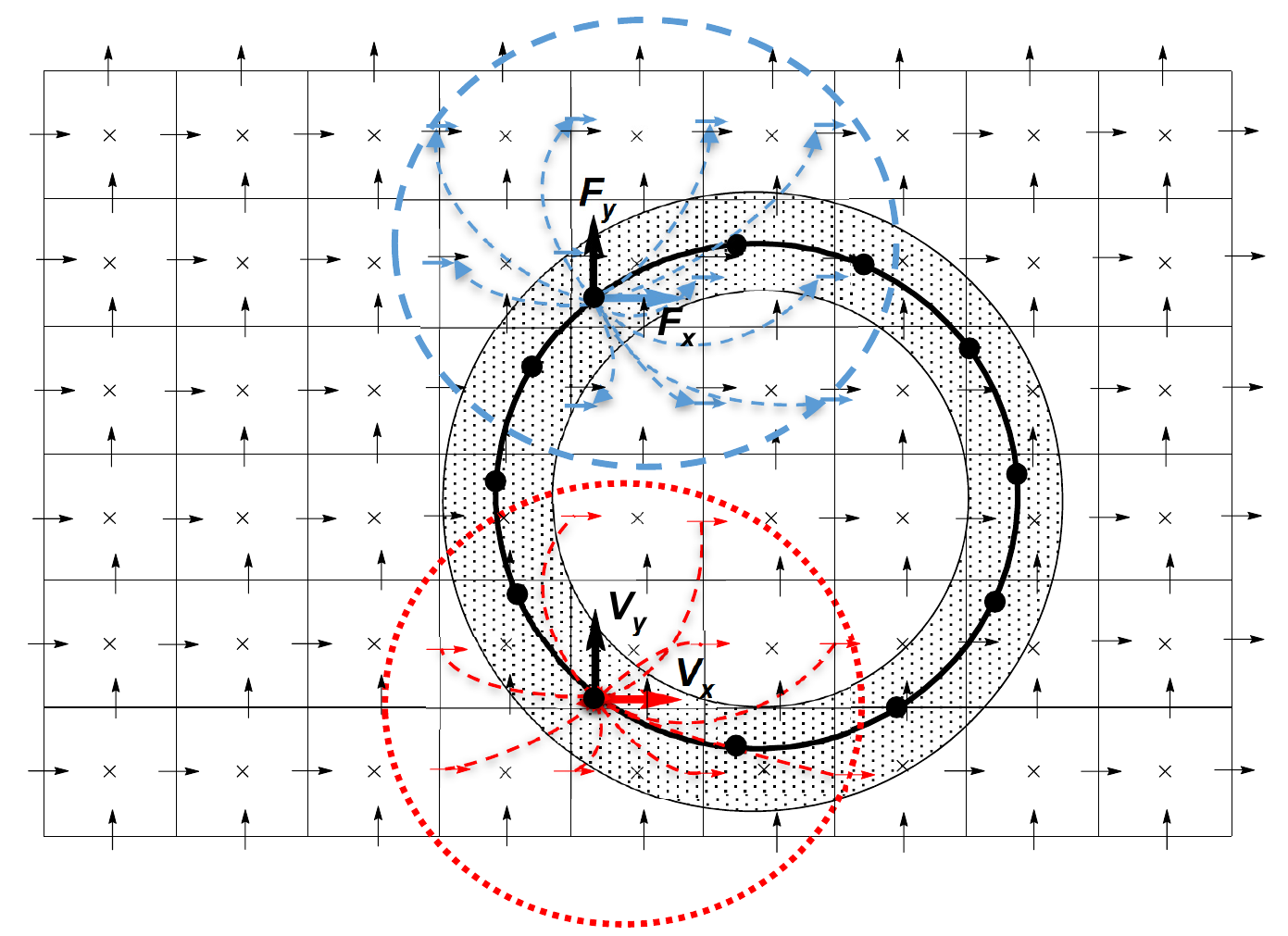}
\label{fig:PhysModel}
\end{figure}

The approach is applied to the simulation of unsteady natural convection incompressible flow in the presence of an immersed body of irregular shape. The surface of the immersed body is determined by a set of discrete Lagrangian points, whose location does not necessarily coincide with the underlying Eulerian staggered grid, as shown in Fig. \ref{fig:PhysModel}.

To provide the best accuracy, the method utilizes a uniform grid in the vicinity of the immersed body surface. In this region, the distance between the neighboring points of the immersed body surface, $\Delta l$, and the width of a grid cell, $\Delta \textbf{\textit{X}}$, should be approximately the same. Away from the body, non-uniform discretization can be used. The impact of the immersed body on the nearby flow is reflected by the force, $\textit{\textbf{f}}$, and the power, $q$, densities, each related to the corresponding virtual volume located within the shadowed region confined by internal and external concentric circles and confining every Lagrangian point (see Fig. \ref{fig:PhysModel}). The densities enter as sources into the momentum, Eq. (\ref{eq:momentun}), and energy, Eq. (\ref{eq:energy}), equations. The values of the sources are not known a priori and are part of the solution along with the velocity, pressure and temperature fields. To enforce the non-slip and thermal (temperature or heat flux) boundary conditions on the surface of the immersed body, the nearby velocity and temperature fields in the vicinity of every Lagrangian point (dotted circle) are interpolated to the location of the Lagrangian points.

Next, two adjoint operators are defined to exchange information between Lagrangian points and the Eulerian grid: namely, an interpolation operator $\textit{\textbf{I}}$ interpolating the values of  Eulerian velocities $\textbf{\textit{u}}(\textit{\textbf{x}}_\textit{i})$ and temperatures  $\theta(\textit{\textbf{x}}_\textit{i})$ to the nearby Lagrangian points $\textbf{\textit{X}}^{k}$; and a regularization operator $\textit{\textbf{R}}$ smearing the values of the Lagrangian force $\textbf{\textit{F}}^{\textit{k}}(\textbf{\textit{X}}^{k})$ and the power $\textit{Q}^{k}(\textbf{\textit{X}}^{k})$ densities to the nearby Eulerian grid:
\begin{subeqnarray}
&\textbf{\textit{R}}(\textbf{\textit{F}}^{\textit{k}}(\textit{\textbf{X}}^{\textit{k}}),\textit{Q}^{\textit{k}}(\textbf{\textit{X}}^{\textit{k}}))=\int_{S}(\textbf{\textit{F}}^{\textit{k}}(\textit{\textbf{X}}^{k}),\textit{Q}^{k}(\textbf{\textit{X}}^{k}))
\cdot\delta(\textit{\textbf{x}}_\textit{i}-\textit{\textbf{X}}^{k})dV_S^k,\\
&\textbf{\textit{I}}(\textbf{\textit{u}}(\textit{\textbf{x}}_\textit{i}),\textit{$\theta$}(\textit{\textbf{x}}_\textit{i}))=\int_{\Omega}(\textbf{\textit{u}}(\textbf{\textit{x}}_i),\theta(\textit{\textbf{x}}_i))\cdot\delta(\textbf{\textit{X}}^{k}-\textit{\textbf{x}}_\textit{i})
dV_{\Omega_i},\quad
\label{RegInt}
\end{subeqnarray}
where $S$ corresponds to all the cells belonging to the immersed body surface, $\Omega$ corresponds to a group of flow domain cells located in the close vicinity of the immersed body surface, $d{V}_S^k$ corresponds to the virtual volume surrounding each Lagrangian point $k$, and $dV_{\Omega_i}$ is the volume of the corresponding cell of the Eulerian flow domain, whose velocity and temperature values are explicitly involved in enforcing the boundary conditions at point $k$ of the immersed body. Note that both interpolation and regularization operators use convolutions with the same discrete Dirac delta function $\delta$ of the form
\begin{equation}
  \delta(r)=\begin{cases}
    \frac{1}{6\Delta r} \Bigg [ 5-3\frac{|r|}{\Delta r}-\sqrt{-3 \Big(1-\frac{|r|}{\Delta r}\Big)^2+1} \Bigg ]& \text{for $0.5\Delta r\leq |r|\leq 1.5\Delta r$},\\
    \frac{1}{3\Delta r} \Bigg [ 1+\sqrt{-3 \Big(\frac{|r|}{\Delta r}\Big)^2+1} \Bigg ]& \text{for $|r|\leq 0.5\Delta r$},\\
    0 & \text{otherwise},\\
  \end{cases}
  \label{eq:DeltaFunc}
\end{equation}
introduced by Roma et al. in \cite{roma1999JCP}. Here $\Delta r$ is the cell width in the $r$ direction. The chosen delta function, successfully utilized in a large number of previous studies (see e.g. \cite{kempe2012JCP, taira2007JCP, feldman2016JCP, uhlmann2005JCP}), has been specifically designed for performing calculations on the staggered grids and became popular due to its compact kernel (only three cells in each direction of the computational domain are involved). Following the SIMPLE method \cite{patankar197IJHMT}, the NS and energy equations, Eqs. (\ref{eq:continuity}- \ref{eq:energy}), are discretized as:
\begin{equation}
\frac{3\theta^{n+1}}{2\Delta t}  - \frac{1}{\sqrt{\textit{Pr}\textit{Ra}}}\textbf{\textit{L}}(\theta^{n+1})-\textbf{\textit{R}}(\textit{Q}^{\textit{k}}(\textit{\textbf{X}}^{\textit{k}})) =  \frac{4\theta^n-\theta^{n-1}}{2\Delta t} -\textit{\textbf{N}}(\theta^n,\textbf{\textit{u}}^n)+ \theta^{n}{\vec{e}}_{z},
\label{eq:DiscrEnergy}
\end{equation}
\begin{equation}
\frac{3\textbf{\textit{u}}^*}{2\Delta t}-\sqrt{\frac{Pr}{Ra}}\textbf{\textit{L}}(\textbf{\textit{u}}^*)-\textbf{\textit{R}}(\textit{\textbf{F}}^{\textit{k}}(\textit{\textbf{X}}^{\textit{k}})) =\frac{4\textbf{\textit{u}}^n-\textbf{\textit{u}}^{n-1}}{2\Delta t}-\textit{\textbf{N}}(\textbf{\textit{u}}^n)+\theta^{n+1} {\vec{e}}_{z}-\nabla p^n,
\label{eq:DiscrMomentPredict}
\end{equation}
\begin{equation}
\triangle(\delta p)=\frac{3}{2\Delta t} \nabla \cdot \textbf{\textit{u}}^*,
\label{eq:DiscPoisson}
\end{equation}
\begin{equation}
\textit{\textbf{u}}^{n+1}=\textit{\textbf{u}}^*-\frac{2\Delta t}{3}\nabla(\delta p),\quad \quad p^{n+1}=p^n+\delta p,
\label{eq:Correction}
\end{equation}
where the second order backward finite difference scheme was utilized for discretizing the temporal derivatives, while a standard staggered mesh second-order conservative finite-volume formulation \cite{patankar1980} was used for discretizing all the spatial derivatives. Operators $\textbf{\textit{N}}$ and  $\textbf{\textit{L}}$ read for non-linear convective and linear Laplacian terms of the NS equations, respectively. For the confined natural convection flows which are the focus of the present study Eqs. (\ref{eq:DiscrEnergy} - \ref{eq:Correction}) are solved with the Dirichlet (or Neumann) boundary conditions for the temperature field and non-slip boundary conditions for all velocity components. Neumann boundary conditions (zero value of derivative in the normal to the boundary direction) with a single reference Dirichlet point introduced anywhere inside the computational domain are used for the solution of the Poisson equation, Eq. (\ref{eq:DiscPoisson}), formulated for the pressure correction field. Following the formalism of the SIMPLE method the pressure field in the predictor, Eq. (\ref{eq:DiscrMomentPredict}), is taken from the previous time step and the solution is first obtained for the non-solenoidal velocity field $\textit{\textbf{u}}^*$. Then, the pressure correction field is calculated by the solution of Eq. (\ref{eq:DiscPoisson}). Finally, Eqs. (\ref{eq:Correction}) are used to correct the pressure and the velocity values at the next time step. Until now, nothing was said about the terms $\textbf{\textit{R}}(\textit{Q}^{\textit{k}}(\textit{\textbf{X}}^{\textit{k}}))$ and $\textbf{\textit{R}}(\textit{\textbf{F}}^{\textit{k}}(\textit{\textbf{X}}^{\textit{k}})) $ implicitly entering into the left-hand sides of the energy, Eq. (\ref{eq:DiscrEnergy}), and the momentum, Eq. (\ref{eq:DiscrMomentPredict}), equations. These terms are  related to the regularized power, $q$, and force, $\textit{\textbf{f}}$,  densities imposing temperature (or heat flux) and kinematic no-slip constraints, respectively, on the surface of an immersed body. Implicit treatment of the above terms implies introducing additional relationships to achieve closure of the overall system of Eqs. (\ref{eq:DiscrEnergy} - \ref{eq:Correction}), as detailed in the next section.
\subsection{Formalism of the implicit direct forcing IB method}
Closure of the system of Eqs. (\ref{eq:DiscrEnergy} - \ref{eq:Correction}) is achieved by introducing additional relationships comprising thermal and kinematic no-slip constraints for the energy and momentum equations, respectively. The full system of equations extended with IB capability and allowing for calculation of temperature, $\theta^{n+1}$, and intermediate velocity, $\textit{\textbf{u}}^*$, followed by the correction of velocity, $\textbf{\textit{u}}$, and the pressure, $p$, fields can be compactly written as:
\begin{subeqnarray}
&
    \begin{bmatrix}
    \textbf{\textit{H}}_{\theta,\textbf{\textit{u}}}  & \textbf{\textit{R}}(\textit{Q}^{\textit{k}}(\textit{\textbf{X}}^{\textit{k}}),\textit{\textbf{F}}^{\textit{k}}(\textit{\textbf{X}}^{\textit{k}})) \\

    \textbf{\textit{I}}(\textit{$\theta$}(\textit{\textbf{x}}_\textit{i}),\textbf{\textit{u}}^*(\textit{\textbf{x}}_\textit{i}))  & 0

   \end{bmatrix}
   \begin{bmatrix}
    \theta^{n+1}, \textbf{\textit{u}}^* \\
    \textit{Q},\textit{\textbf{F}}
  \end{bmatrix}
  =
   \begin{bmatrix}
    RHS_{\theta, \textbf{\textit{u}}}^{n-1,n} \\
    \theta_b, \textit{\textbf{U}}_b \\
  \end{bmatrix},\quad \\
  &
  \triangle(\delta p)=\frac{3}{2\Delta t} \nabla \cdot \textbf{\textit{u}}^*,\\
  &
  \textit{\textbf{u}}^{n+1}=\textit{\textbf{u}}^*-\frac{2\Delta t}{3}\nabla(\delta p),\quad \quad p^{n+1}=p^n+\delta p,
\label{SchuCompact}
\end{subeqnarray}
where $\textbf{\textit{H}}_{\theta,\textbf{\textit{u}}}$ reads for the modified Helmholtz operator acting on the temperature or the velocity fields,  $\textbf{\textit{R}}(\textit{Q}^{\textit{k}}(\textit{\textbf{X}}^{\textit{k}}),\textit{\textbf{F}}^{\textit{k}}(\textit{\textbf{X}}^{\textit{k}}))$ corresponds to the entries obtained by regularization of the power and force densities determined at Lagrangian points of the immersed body to the nearby locations of Eulerian grid, and   $\textbf{\textit{I}}(\textit{$\theta$}(\textit{\textbf{x}}_\textit{i}),\textbf{\textit{u}}(\textit{\textbf{x}}_\textit{i}))$ corresponds to the entries obtained by interpolation of  Eulerian temperature and velocity fields to the Lagrangian points of the immersed body. Note that if discretized on a uniform staggered grid,  $\textbf{\textit{I}}(\textit{$\theta$}(\textit{\textbf{x}}_\textit{i}),\textbf{\textit{u}}(\textit{\textbf{x}}_\textit{i}))= \textbf{\textit{R}}^T(\textit{Q}^{\textit{k}}(\textit{\textbf{X}}^{\textit{k}}),\textit{\textbf{F}}^{\textit{k}}(\textit{\textbf{X}}^{\textit{k}}))$, although this property was not explicitly exploited while implementing the curently developed approach.
\section{The implementation details}
\subsection{The Schur complement approach}
Recalling that the current  study aims to develop a general methodology to acquire the previously developed pressure-velocity segregated solvers of incompressible NS equations with the IB capability, we next present further details regarding implementation of a solution of the system of Eqs. (\ref{SchuCompact}a). Utilizing the Schur complement approach, the system of Eqs. (\ref{SchuCompact}a) is analytically split into two smaller equivalent systems:
\begin{subeqnarray}
&[Q, \textbf{\textit{F}}]=[\textbf{\textit{I}}\textbf{\textit{H}}^{-1}\textbf{\textit{R}}]^{-1}[\textbf{\textit{I}}\textbf{\textit{H}}^{-1}\textit{RHS}_{\theta,\textbf{\textit{u}}}^{n-1,n}-\theta_b, \textbf{\textit{U}}_b],\\
&[\theta, \textbf{\textit{u}}^*]=\textbf{\textit{H}}^{-1}[\textit{RHS}_{\theta,\textbf{\textit{u}}}^{n-1,n}-\textbf{\textit{R}}[Q,\textbf{\textit{F}}]],
\label{SchurSplitted}
\end{subeqnarray}
solution of  which  is first performed for the distributed Lagrange multiplier terms \textit{Q} and \textbf{\textit{F}} (Eqs. (\ref{SchurSplitted}a)), and then for the corresponding Eulerian temperature, $\theta$, and intermediate velocity, $\textbf{\textit{u}}^*$, fields (Eqs. (\ref{SchurSplitted}b)). A closer  look at the structural characteristics of the matrices \textbf{\textit{H}}, \textbf{\textit{I}}, and \textbf{\textit{R}} gives rise to considerations regarding an efficient solution of Eqs. (\ref{SchurSplitted}a and \ref{SchurSplitted}b). Matrix \textbf{\textit{H}}, comprising  the modified Helmholtz operator built for the corresponding fields of temperature and velocity components, is the largest $m\times m$ matrix, where $m$ reads for the total number of unknowns on the Eulerian grid (typically $\textit{O}(10^6-10^7)$ for realistic 3D problems). Matrix \textbf{\textit{H}} is typically  a sparse matrix, and for the spatial and temporal discretiztions utilized in the present study has non-zero entries arranged along only 7 diagonals. For other spatial dicretizations which may be based on stencils of higher orders and also include additional non linear terms discretized in a semi-implicit manner, the matrix  \textbf{\textit{H}} will include more non-zero entries, but will still remain sparse, making it possible to perform simulations on dense grids with a reasonable memory consumption. In particular, the maximum amount of memory required to store all the four  \textbf{\textit{H}} matrices (one for the temperature and three for all the velocity components) with all auxiliary data did not exceed 6 Gb of the CPU RAM for the $300^3$ grid resolution. Matrices \textbf{\textit{I}} and \textbf{\textit{R}} containing entries related to the acquired IB capability are of  dimensions $n \times m$ and $m \times n$, respectively, where $n$ reads for the total number of Lagrangian points (typically $\textit{O}(10^3-10^4))$ for realistic 3D problems) determining the surfaces of all the immersed bodies involved in the simulation. The amount of non-zero entries in any $n-\textit{th}$ row or column of the matrices \textbf{\textit{I}} and \textbf{\textit{R}}, respectively, depends on a specific kernel of the discrete delta functions or on an order of interpolations utilized in the interpolation, \textbf{\textit{I}}, and the regularization, \textbf{\textit{R}}, operators, but on any account is not higher than $\textit{O}(10^2)$. As a result, the matrices \textbf{\textit{I}} and \textbf{\textit{R}} are\textit{ extremely}  sparse; they both are stored in compressed sparse row (CSR) format and their multiplication by any vector is further implemented by using standard routines from the Intel Math Kernel Library (MKL).

Recall next that the previously developed solver \cite{vitoshkin2013}, or any other solver for the solution of original (without IB functionality) NS equations $[\textbf{\textit{H}}_{\theta, \textbf{\textit{u}}}][\theta, \textbf{\textit{u}}]=[\textit{RHS}_{\theta, \textbf{\textit{u}}}]$, based on the pressure-velocity segregated approach, is available. The existing solver can be straightforwardly exploited in a black box manner for computing the product of the inverse modified Helmholtz operator, $\textit{\textbf{H}}^{-1}$, by any given vector of dimension $m \times 1$ comprising an essential part of the presently developed algorithm for the solution of the system of Eqs.(\ref{SchurSplitted}a  and  \ref{SchurSplitted}b), as detailed in Table \ref{Algorithm}.

\begin{table}[!htbp]
\fontsize{10}{10}
\selectfont
\centering
\caption{A detailed description of the major steps of the developed algorithm for solution of equations,  Eqs. (\ref{SchurSplitted}a and  \ref{SchurSplitted}b).}
\label{Algorithm}
\begin{tabular}{p{0.6cm}p{5.8cm}p{7cm}}
\thickhline
\multicolumn{3}{l}{The solution of $[Q,\textbf{\textit{F}}]=[\textbf{\textit{I}}\textbf{\textit{H}}^{-1}\textbf{\textit{R}}]^{-1}[\textbf{\textit{I}}\textbf{\textit{H}}^{-1}\textit{RHS}_{\theta,\textbf{\textit{u}}}^{n-1,n}-\theta_b,\textbf{\textit{U}}_b]$, (Eqs. (\ref{SchurSplitted}a))} \\
\thickhline
\\
1-a.  & Calculation of $\textbf{\textit{H}}^{-1}\textit{RHS}_{\theta,\textbf{\textit{u}}}^{n-1,n}$: &Employs the original solver \cite{vitoshkin2013}. The result is a $m \times 1$ vector. \\
\hline
2-a.  & Calculation of   $\textit{\textbf{I}}\textbf{\textit{H}}^{-1}\textit{RHS}_{\theta,\textbf{\textit{u}}}^{n-1,n}$:&Employs matrix-vector multiplication of the matrix \textbf{\textit{I}} stored in compressed sparse row (CSR) format by the vector  $\textbf{\textit{H}}^{-1}\textit{RHS}_{\theta,\textbf{\textit{u}}}^{n-1,n}$ obtained in 1-a. The result is a $n \times 1 $ vector.\\
\hline
3-a. & \parbox [t]{6.3cm} {Calculation of \\ $\textbf{\textit{I}}\textbf{\textit{H}}^{-1}\textit{RHS}_{\theta,\textbf{\textit{u}}}^{n-1,n}-\theta_b,\textbf{\textit{U}}_b$:} & Subtraction of  two $n \times 1$ vectors. The result is a $n \times 1 $ vector. \\
\hline
4-a.  & Calculation of $[\textbf{\textit{I}}\textbf{\textit{H}}^{-1}\textbf{\textit{R}}]$: & The procedure is repeated $n$ times for every  column $[\textbf{\textit{R}}]_n$ of the matrix \textbf{\textit{R}}. The original solver \cite{vitoshkin2013} is employed first to calculate $\textbf{\textit{H}}^{-1}[\textbf{\textit{R}}]_n$, followed by matrix-vector multiplication $\textbf{\textit{I}}\textbf{\textit{H}}^{-1}[\textbf{\textit{R}}]_n$. The final result is stored in the  column $[\textbf{\textit{I}}\textbf{\textit{H}}^{-1}\textbf{\textit{R}}]_n$ of the $n \times n$ matrix.  \\
\hline
5-a.  &\parbox [t]{6.3cm} {Calculation of \\ $[\textbf{\textit{I}}\textbf{\textit{H}}^{-1}\textbf{\textit{R}}]^{-1}  [\textbf{\textit{I}}\textbf{\textit{H}}^{-1}\textit{RHS}_{\theta,\textbf{\textit{u}}}^{n-1,n}-\theta_b,\textbf{\textit{U}}_b]$:} &Either by employing  $LU$ factorization of the small $n \times n$ matrix\footnote{We use an open source MUMPS package \cite{amestoy1998ComputMethApplMechEngrg,amestoy2001SIAMJMatr}.} or by any iterative (e.g. Krylov space or multigrid based) method\footnote{We use the bi-conjugate gradient (BICG) method.}.\\
\\
\thickhline
\multicolumn{3}{l}{The solution of $[\theta, \textbf{\textit{u}}^*]=\textbf{\textit{H}}^{-1}[\textit{RHS}_{\theta,\textbf{\textit{u}}}^{n-1,n}-\textbf{\textit{R}}[\textit{Q},\textbf{\textit{F}}]]$, (Eqs. (\ref{SchurSplitted}b))} \\
\thickhline
\\
1-b.  & Calculation of $\textbf{\textit{R}}[Q,\textbf{\textit{F}}]$: &Employs matrix-vector multiplication of the matrix \textbf{\textit{R}} stored in compressed sparse row (CSR) format by the vectors [\textit{Q},\textbf{\textit{F}}] calculated by the solution of Eqs. (\ref{SchurSplitted}a). \\
\hline
2-b. &\parbox [t]{6.3cm}  {Calculation of \\ $[\textit{RHS}_{\theta,\textbf{\textit{u}}}^{n-1,n}-\textbf{\textit{R}}[Q,\textbf{\textit{F}}]]$:} &Subtraction of  two $m \times 1$ vectors. The result is a $m \times 1 $ vector.\\
\hline
3-b. &\parbox [t]{6.3cm} {Calculation of \\ $\textbf{\textit{H}}^{-1}[\textit{RHS}_{\theta,\textbf{\textit{u}}}^{n-1,n}-\textbf{\textit{R}}[Q,\textbf{\textit{F}}]]$:} &Employs the original solver \cite{vitoshkin2013}. The result is a $m \times 1$ vector.\\
\thickhline
\end{tabular}
\end{table}
The present study focuses on the configurations characterized by the stationary surfaces of immersed bodies that make it possible to precompute the matrix $[\textbf{\textit{I}}\textbf{\textit{H}}^{-1}\textbf{\textit{R}}]$ (see step $4-a$ in Table \ref{Algorithm}), which  comprises the most time consuming step of the developed approach. The final solution of Eqs. (\ref{SchurSplitted}a) (see step $5-a$ in Table \ref{Algorithm}) can be either completed by a direct method, which will include $LU$ factorization of the matrix $[\textbf{\textit{I}}\textbf{\textit{H}}^{-1}\textbf{\textit{R}}]$ performed once at the beginning of the process followed by elimination and back substitution stages, or, alternatively, by utilizing any of the available iterative algorithms in each time step. Both options have been implemented in the present study. Namely, the direct $LU$ factorization of the matrix $[\textbf{\textit{I}}\textbf{\textit{H}}^{-1}\textbf{\textit{R}}]$ has been performed by utilizing an open source MUMPS package \cite{amestoy1998ComputMethApplMechEngrg,amestoy2001SIAMJMatr}, while the iterative solution of Eqs. (\ref{SchurSplitted}) was performed by the bi-conjugate gradient (BICG) method based on Krylov subspace iteration. Implementation of both direct and iterative algorithms for completing step $5-a$ of the developed approach increases its  flexibility, making it possible to switch between the algorithms depending on the hardware configuration, as well as  the  structure and the size of the matrix $[\textbf{\textit{I}}\textbf{\textit{H}}^{-1}\textbf{\textit{R}}]$\footnote{See the next section for more extensive discussion.}. It is remarkable that, similarly to the explicit direct forcing method \cite{Mohd-Yusof1997,kempe2012JCP}, the single time step integration implemented in the present implicit formulation also involves  dual employment of the original solution of the NS (without IB functionality), as follows from steps $1-a$ and $3-b$ in  Table \ref{Algorithm}.

\subsection{Optimization strategies}
With the aim of developing a generalized methodology capable of solution of highly resolved 3D flows, much effort was made to further optimize the developed algorithm in terms of memory consumption and CPU time. The study yielded a number of important optimization strategies, as detailed in the following.
\subsubsection{Optimization of memory consumption}
Optimization of the memory consumption is closely related to the way the $n \times n $ matrix $[\textbf{\textit{I}}\textbf{\textit{H}}^{-1}\textbf{\textit{R}}]$ was built. In general, this is an indefinite and  non-symmetric matrix, with a structure depending on a kernel of discrete Dirac delta functions utilized in the  interpolation and regularization operators, and on a spatial distribution of Lagrangian points. Furthemore, all the tests performed in the framework of the present study revealed that the matrix  $[\textbf{\textit{I}}\textbf{\textit{H}}^{-1}\textbf{\textit{R}}]$ is always full of non-zero entries with absolute values that are mostly very small ($O(10^{-16}$) and smaller). Storing all these non-zero entries is prohibitively expensive, so that it is necessary to set up a sparsing threshold. Recalling that the matrix $[\textbf{\textit{I}}\textbf{\textit{H}}^{-1}\textbf{\textit{R}}]$ is an operator determining the values of the Lagrangian forces and heat fluxes enforcing the kinematic no-slip and thermal constraints on the surfaces of immersed bodies, the threshold value satisfying the required precision can be adjusted manually for any specific flow configuration. In all the simulations performed in the present study the absolute threshold values were in the range of $(10^{-16} \div 10^{-25})$. For this range the thermal and kinematic no-slip constraints were met with up to 6 decimal digits, which should be less than the discretization error of the second order finite volume method utilized in the present study.

Another two factors which have a significant impact on the overall memory consumption are the value of the time step, $\Delta t$, and the Rayleigh number, $Ra$. The larger the time step the higher is amount of  entries of the matrix $[\textbf{\textit{I}}\textbf{\textit{H}}^{-1}\textbf{\textit{R}}]$ that are above the given threshold value and therefore have to be stored in the memory. It was found that for a time step of $( O(10^{-2}))$ the relative amount of non-zero entries of matrix $[\textbf{\textit{I}}\textbf{\textit{H}}^{-1}\textbf{\textit{R}}]$ was typically about $3\%$ of the total $n \times n$ matrix size. Decreasing the time step by a factor of 10 while keeping the same threshold value resulted in a steeper (by an order of magnitude or even more) decrease in the percentage of non-zero entries. Apart from the memory directly involved in the storage of non-zero entries of the matrix $[\textbf{\textit{I}}\textbf{\textit{H}}^{-1}\textbf{\textit{R}}]$, the time step and the $Ra$  values have a significant impact on the overall memory consumption when performing $LU$ factorization of the matrix. In fact,  the typical amount of memory required to store the factors of a single $[\textbf{\textit{I}}\textbf{\textit{H}}^{-1}\textbf{\textit{R}}]$ matrix\footnote{For 3D simulation the factors of four of such matrices need to be stored.} obtained on the $300^3$ grid  for the immersed body determined by approximately  $2\times 10^5$ points is about 40Gb for a time step of $( O(10^{-2}))$ and only about 4Gb for a time step of $(O(10^{-3}))$.  Detailed information regarding memory consumption versus the values of  $Ra$ number is given in the \textit{Results and discussion} section for all the  flow configurations considered in the study.
\subsubsection{Optimization of CPU time}
Optimization of CPU time was implemented on two levels. The first level is related to building the matrices $\textit{\textbf{I}}$ and $\textit{\textbf{R}}$ and precomputing the matrix $[\textbf{\textit{I}}\textbf{\textit{H}}^{-1}\textbf{\textit{R}}]$. Recalling that the originally developed solver \cite{vitoshkin2013} was accelerated by utilizing a multithreading parallelism based on the OpenMP approach, it was natural to use the same parallelism paradigm when extending the code with the IB capability. All the simulations were performed on a standard Linux server having 128 GB DDR3 shared memory and  2 Intel Xeon 12C processors, 24 threads each (48 threads in total). This also allowed us to exploit a built-in multithreading parallelism of the matrix-vector multiplication functions from the Intel MKL library. The typical  times required for precomputing the matrices $\textit{\textbf{I}}$ and $\textit{\textbf{R}}$  when running the simulation on 48 threads varied from a number of seconds to several minutes for the surfaces of immersed bodies determined by $O(10^3)$  and $O(10^5)$ Lagrangian points, respectively. Much more significant deviation in computational times was observed when precomputing the matrices  $[\textbf{\textit{I}}\textbf{\textit{H}}^{-1}\textbf{\textit{R}}]$ for all the velocity and temperature fields, which involved performing $n$ time integrations by the original solver for the three components of velocity and temperature. The typical  times required at this stage varied from a number of hours to a number of days  for $200^3$ and  $300^3$ grids, respectively.

The second level is related to the stage of solution of the system of  Eqs. (\ref{SchuCompact}-a). As has been already mentioned, the solution of the system of equations was performed by both direct and iterative solvers. In the case of the direct solver (i.e. MUMPS \cite{amestoy1998ComputMethApplMechEngrg, amestoy2001SIAMJMatr}) the $LU$ factorization should be performed once at the beginning of the process for the four $[\textbf{\textit{I}}\textbf{\textit{H}}^{-1}\textbf{\textit{R}}]$ matrices, corresponding to the three velocity components and temperature. Again, a built-in multithread parallelism of the solver was employed at this stage, and even for the most dense grids (consisting of $300^3$ finite volumes) the overall process was completed within less than a hour. The limitations of the direct solver show up at the intrinsically sequential forward elimination and back substitution stages which do not enjoy multithread parallelism. While on the coarse and moderate grids (up to $200^3$ finite volumes) the above drawback is not pronounced, on denser grids ($300^3$ finite volumes) it becomes critical and drastically deteriorates the overall performance of the computations involving a large number of time steps. In this case the remedy comes from utilizing an iterative solver which can be based either on a Krylov subspace iteration\footnote{For example, presently employed BICG algorithm.} or on multigrid algorithms. We found, however, that the solver based on the BICG algorithm suffers from poor convergence on non-uniform grids stretched close to the boundaries of the computational domain.

In summary, none of the checked direct and iterative approaches can be considered as the ultimate strategy for solution of the system of Eqs. (\ref{SchuCompact}a). In general, the optimized strategy can benefit from a combination of direct and iterative methods, and will always depend on the flow configuration considered. More details regarding the strategies applied to the analysis of the specific benchmark flows are provided in the following sections of the paper. It should also be noted that after the matrix $[\textbf{\textit{I}}\textbf{\textit{H}}^{-1}\textbf{\textit{R}}]$  has been built, the solution of Eqs. (\ref{SchuCompact}a) comprises an independent algebraic problem completely detached from the original solver and physical model and, therefore, can be separately solved by utilizing distributed memory parallelism, e.g. a hybrid or MPI based approach. Parallelized iterative solution of the system of Eqs. (\ref{SchuCompact}a) by utilizing both hybrid and MPI paradigms remained out of the scope of the present study and will be the focus of future work.
\section{Results and discussion}
The results obtained by the developed method have been extensively verified against available benchmark data provided for both steady and periodic confined natural convection flows.
\subsection{Natural convection from a hot sphere inside a cold cube }
\begin{figure}
\centering
\caption{Physical model of  a hot sphere inside a cold cube characterized by $R=0.2L$. The surface of the sphere is held at a constant hot temperature $\theta_h$, while all the cube walls are held at a constant temperature $\theta_c$. The position of the sphere along the cube centerline is determined by the non-dimensional parameter $\delta=(H-0.5L)/L$. }
\includegraphics[width=0.8\textwidth,clip=]{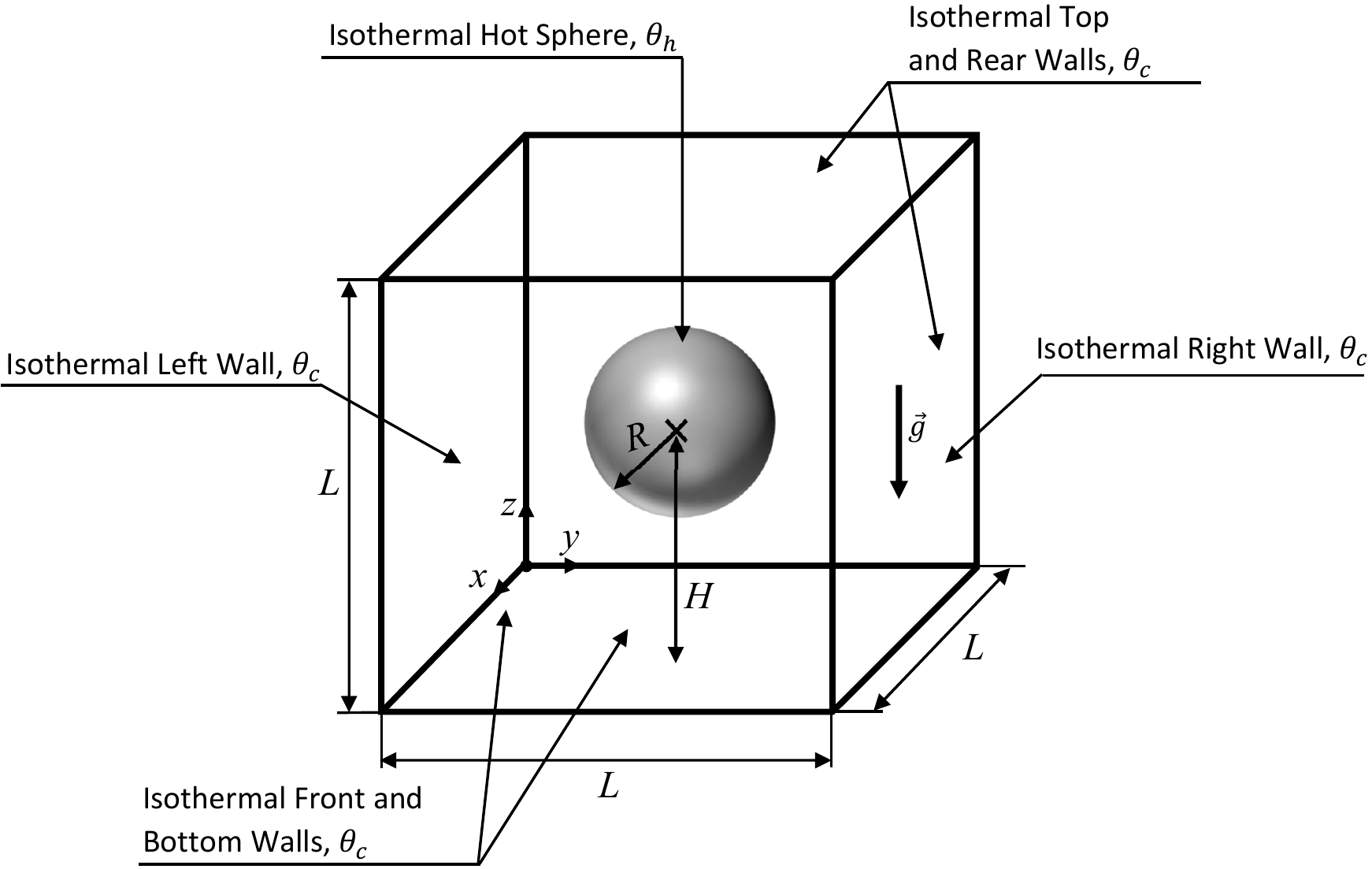}
\label{fig:SphereInsideCubePhysModel}
\end{figure}

\begin{figure}
\centering
\caption{Illustration of 100 Lagrangian points evenly distributed over the surface of a unit sphere by the non-iterative method of Leopardi \cite{Leopardi2006TRANS}.}
\includegraphics[width=0.35\textwidth,clip=]{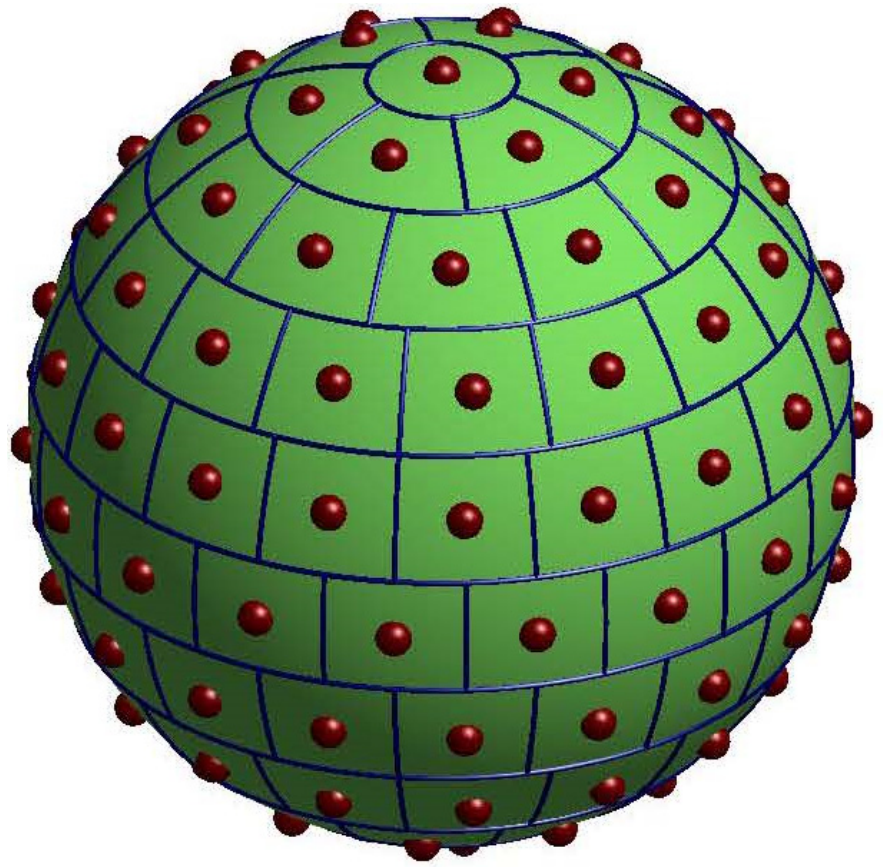}
\label{fig:PointsDistribution}
\end{figure}

Buoyancy convection from a hot sphere placed inside a cold cube is considered. The physical model of the problem is shown in Fig. \ref{fig:SphereInsideCubePhysModel}. The surface of the sphere is held at a constant hot temperature $\theta_h$, while the cold temperature $\theta_c$ is preserved at all the cube walls. The cube edge length $L$ is used for normalizing all the length scales of the problem. The hot sphere is of a constant radius $R=0.2L$. The center of the sphere is attached to the vertical centerline of the cube and its position is determined by a non-dimensional parameter $\delta=(H-0.5L)/L$, where $H$ is a non-dimensional distance between the center of the sphere and the bottom wall of the cube.
The best accuracy of the IB method is achieved by providing a uniform distribution of the Lagrangian points over the the surface of the sphere, implemented by the non-iterative method of Leopardi \cite{Leopardi2006TRANS}, so that each point is confined by a virtual surface of an equal area, as illustrated  in Fig. \ref{fig:PointsDistribution}.
Defining the Nusselt number, $Nu$, as the ratio of convective to conductive heat fluxes from the given surface, its average value for the surface of the sphere is obtained by:
\begin{equation}
\overline{Nu}=\sqrt{PrRa}\Delta x\overline{Q},
\label{HeatFlux}
\end{equation}
where the average heat flux $\overline{Q}$ is an arithmetic mean of all the non-dimensional heat fluxes $Q_k$ at each Lagrangian point $k$ of the immersed surface, intrinsically calculated by the the solution of the system of Eqs. (\ref{SchuCompact}).
The steady state flows were calculated for the values of $Ra=10^5$ and $Ra=10^6$ and $\delta=\pm0.25$. Convergence to steady state was assumed when the value of the $L_\infty$ norm calculated for the difference between the flow characteristics obtained at two consecutive time steps was less than $10^{-6}$ for all the flow fields. Details regarding  setup and characteristics of the numerical solution are given in Table \ref{SetupSphere}. Although direct algorithms are known for their high memory requirements for performing $LU$ factorization, the memory consumption for this specific configuration is quite modest, not exceeding 21Gb of RAM for the most memory consuming configuration. It is remarkable that the lower the $Ra$ value, the higher is the memory consumption, which approximately scales as $Ra^{-0.16}$ when the threshold value for the matrix  $[\textbf{\textit{I}}\textbf{\textit{H}}^{-1}\textbf{\textit{R}}]$ sparsing is equal to $10^{-22}$. Recalling that the fullness of matrix $[\textbf{\textit{I}}\textbf{\textit{H}}^{-1}\textbf{\textit{R}}]$ reflects the intensity of mutual interactions between the Lagrangian forces and heat fluxes, the above observation is not surprising, since the  intensity is higher for viscosity dominated flows. It is remarkable that for the moderate memory consumption typical of the present configuration the  time step duration is almost independent of the fullness of matrix $[\textbf{\textit{I}}\textbf{\textit{H}}^{-1}\textbf{\textit{R}}]$,  allowing for optimally exploiting advantages of the direct solver. Note also that if  the sphere is positioned far enough (more than two grid cells) from the cube boundaries the characteristics of the numerical solution do not depend on the spatial location of the sphere inside. For this reason the characteristics in Table \ref{SetupSphere} are presented for only a single value of $\delta=0$.
\begin{table}[]
\fontsize{10}{10}
\selectfont
\centering
\caption{Setup and characteristics of the numerical solution obtained on $200^3$ grid for the simulation of the natural convection from a hot sphere placed inside a cold cube characterized by the value of $R/L=0.2$, $\delta=0$.}
\label{SetupSphere}
\begin{tabular}{lllllll}
\hline
\multicolumn{1}{>{}m{1cm}}{$Ra$} & \multicolumn{1}{>{}m{2cm}}{Time step duration, [s]} & \multicolumn{1}{>{}m{2cm}}  {RAM, [Gb]}& \multicolumn{1}{>{}m{1cm}} {$\Delta t$} &   \multicolumn{1}{>{}m{3cm}} {Method of solution of Eqs. \ref{SchurSplitted}-a} &   \multicolumn{1}{>{}m{2cm}} {Number of time steps} &   \multicolumn{1}{>{}m{2cm}} {Sparsing threshold for $[\textbf{\textit{I}}\textbf{\textit{H}}^{-1}\textbf{\textit{R}}]$} \\
\hline
$10^3$ & 2.5 & 20.70&     \multicolumn{1}{c}{\multirow{4}{*}{$10^{-3}$}}&  \multicolumn{1}{c}{\multirow{4}{*}{Direct ($LU$)}} & $O(10^3)$ & \multicolumn{1}{c}{\multirow{4}{*}{$10^{-22}$}}\\
$10^4$ & 2.4 & 15.04&    \multicolumn{1}{c}{}                          &  \multicolumn{1}{c}{}                               & $O(10^3)$ & \multicolumn{1}{c}{}\\
$10^5$ & 2.3 & 9.734&     \multicolumn{1}{c}{}&  \multicolumn{1}{c}{} & $O(10^4)$ & \multicolumn{1}{c}{}\\
$10^6$ & 2.2 & 6.892&    \multicolumn{1}{c}{}                          &  \multicolumn{1}{c}{}                               & $O(10^4)$ & \multicolumn{1}{c}{}\\

\hline
\end{tabular}
\end{table}

Comparison between the obtained average $\overline{Nu}$ values and the corresponding values available in the literature is presented in Table \ref{NuAvSph}. An acceptable agreement is observed for both $Ra=10^5$ and $Ra=10^6$  and the entire range of $\delta$ values, successfully verifying the present calculations.
\begin{table}[]
\fontsize{11}{11}
\selectfont
\centering
\caption{Comparison between the present and the previously published  $\overline{Nu}$ values averaged over the surface of  hot a sphere placed within a cold cube.}
\label{NuAvSph}
\begin{tabular}{ccccccc}
\hline
{} & \multicolumn{3}{c}{$Ra=10^5$}   & \multicolumn{3}{c}{$Ra=10^6$}  \\
\cline{2-4}
\cline{5-7}
\multicolumn{1}{c}{$\delta$} & \multicolumn{1}{c} { Ref.\cite{yoon2010heattransf}} &   \multicolumn{1}{c} {Ref.\cite{gulb2015IJHMT}} & \multicolumn{1}{c}{Present} & \multicolumn{1}{c}  { Ref.\cite{yoon2010heattransf}} &   \multicolumn{1}{c} {Ref.\cite{gulb2015IJHMT}} & \multicolumn{1}{c}{Present}\\ \hline
-0.25    & 13.665  & 13.774   & 13.489    & 20.890   & 21.993     & 20.611   \\
-0.2     & 12.931  & 13.058   & 12.768    & 20.631   & 21.862     & 20.517   \\
-0.1     & 12.729  & 13.105   & 12.819    & 20.772   & 22.164     & 21.216   \\
0        & 12.658  & 13.415   & 13.160    & 20.701   & 23.344     & 21.589   \\
0.1      & 12.351  & 13.446   & 13.230    & 20.367   & 22.525     & 21.674   \\
0.2      & 12.254  & 13.635   & 13.462    & 19.664   & 22.208     & 21.487    \\
0.25     & 12.944  & 14.426   & 14.277    & 19.721   & 22.393     &21.757     \\ \hline

\end{tabular}
\end{table}
Natural convection flow is visualized by presenting vortical structures, determined by utilizing the $\lambda_2$ criterion proposed by Jeong and Hussain \cite{Jeong1995JFM}. According to the authors, the outermost outer surface of the vortex can be revealed by connecting the same negative $\lambda_2$ values close to zero. Following the recent study in \cite{seo2016}, the value of $\lambda_2=-0.1$ was chosen for the visualization of convection cells. Figure \ref{fig:HotSphereDelats} presents the visualization vortical structures obtained for the values of $Ra=10^5$ and $Ra=10^6$ and $\delta=0.25;0;-0.25$. As expected, the obtained fully 3D steady state flow is symmetric relative to the $X-Z$ and $Y-Z$ center planes, as well as relative to both main diagonal planes of the cube. For all the configurations a large circumferential convection cell is formed close to the top boundary of the cube. For $Ra=10^5$ the cell is characterized by a nearly toroidal shape, whereas for $Ra=10^6$ there is an entrainment of the bottom surface of the cell into the cell interior. The entrainment becomes more pronounced with decreasing the value of $\delta$. The value of $\delta$ is also directly correlated with the major radius of the toroidal convection cell. It can also be seen that for smaller values of $\delta$ the center of the convection cell acquires an elongated center-hollowed mushroom shape.
\clearpage
\begin{figure}[H]
\centering

      \begin{subfigure}{\textwidth}
        \includegraphics[width=0.3\textwidth,clip=]{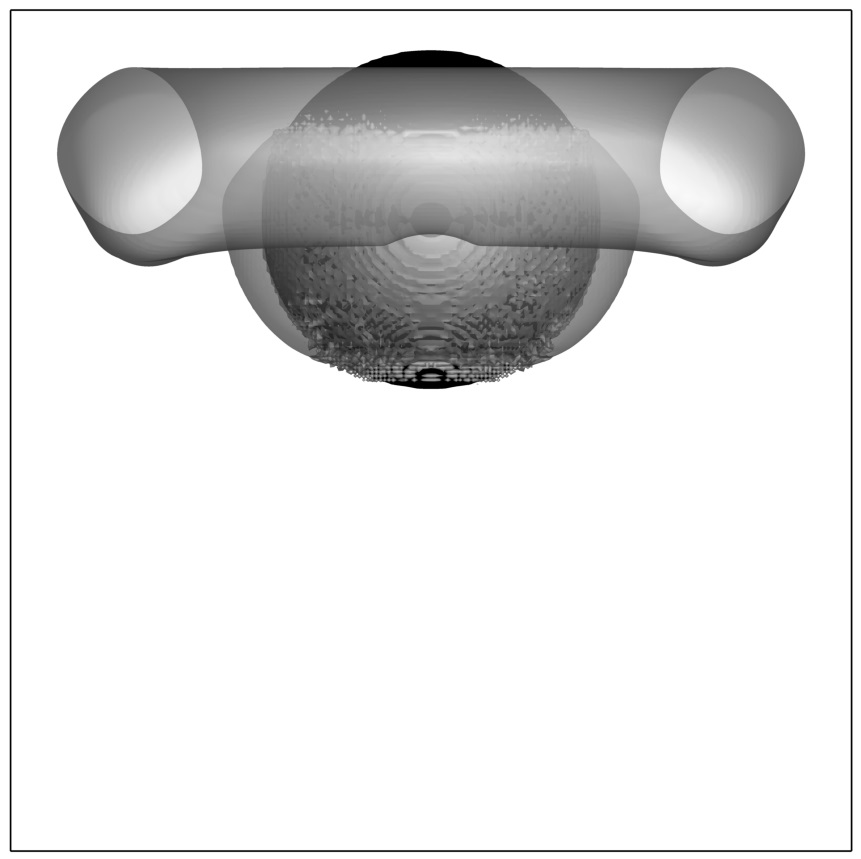}
        \includegraphics[width=0.3\textwidth,clip=]{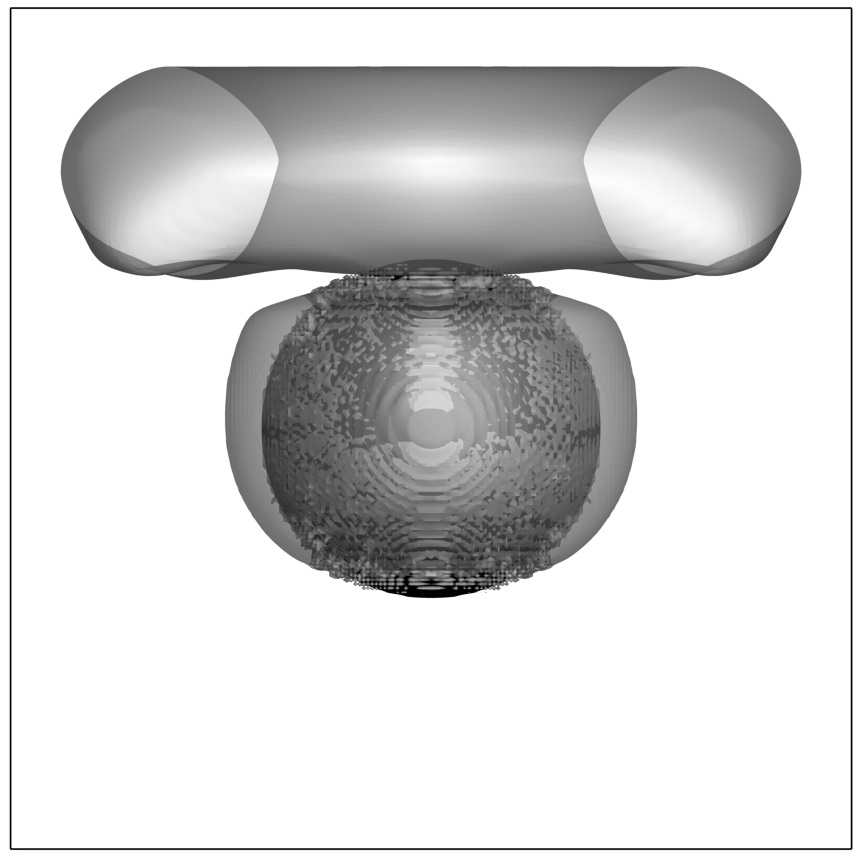}
        \includegraphics[width=0.3\textwidth,clip=]{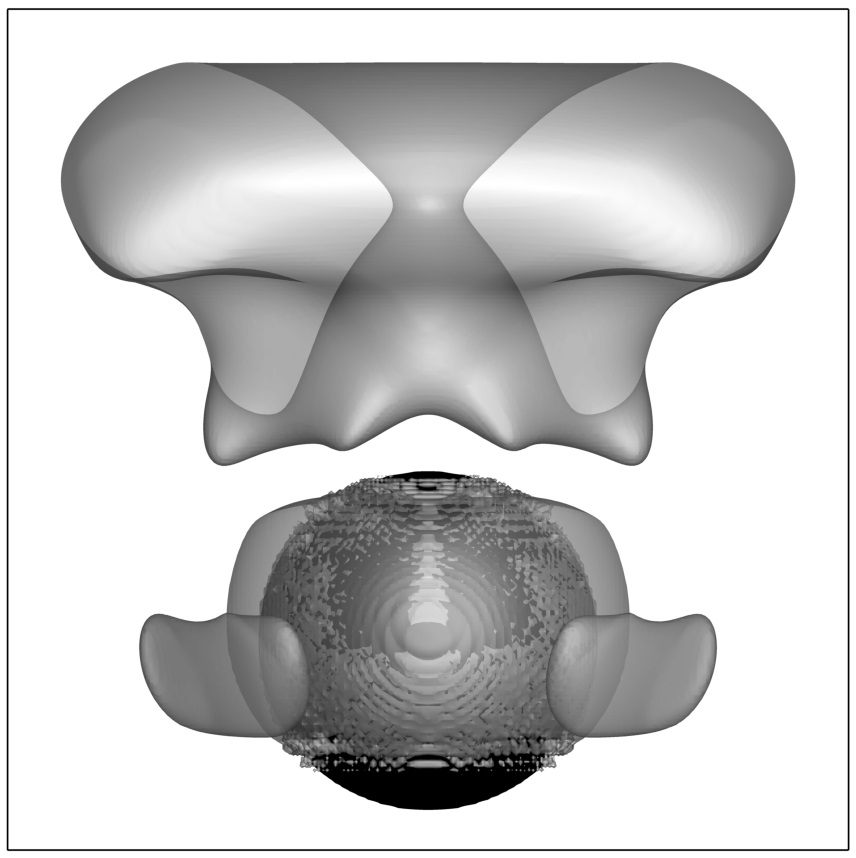}
         \caption{Isosurfaces of $\lambda_2=-0.1$, $Ra=10^5$,  $\delta=0.25;0;-0.25$ - front view  }
    \end{subfigure}
    \begin{subfigure}{\textwidth}
        \includegraphics[width=0.3\textwidth,clip=]{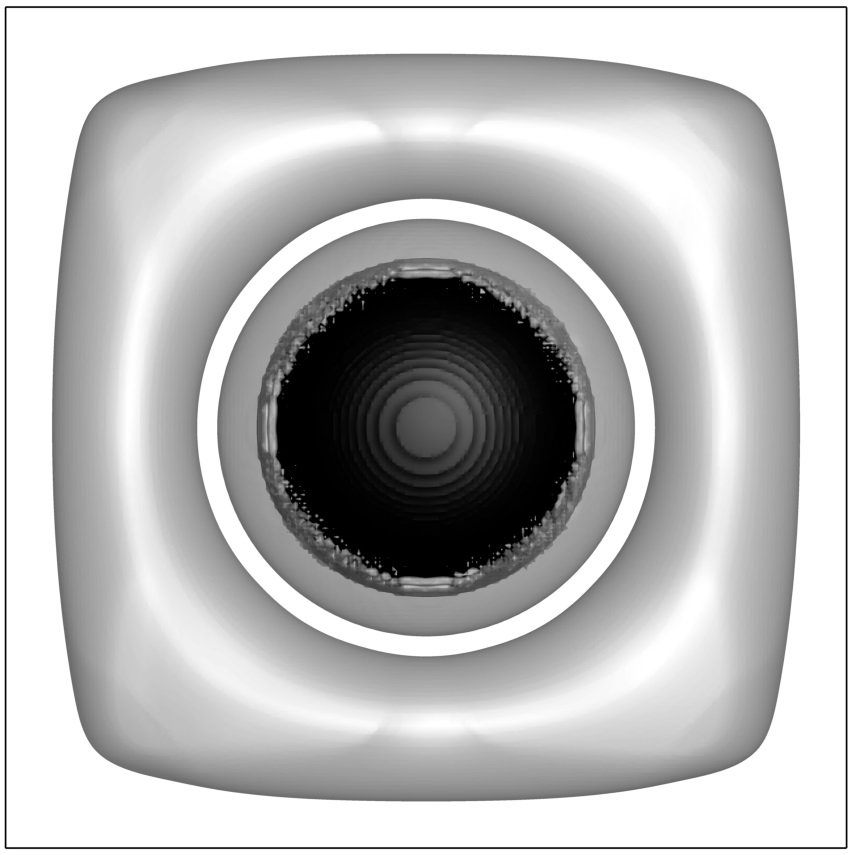}
        \includegraphics[width=0.3\textwidth,clip=]{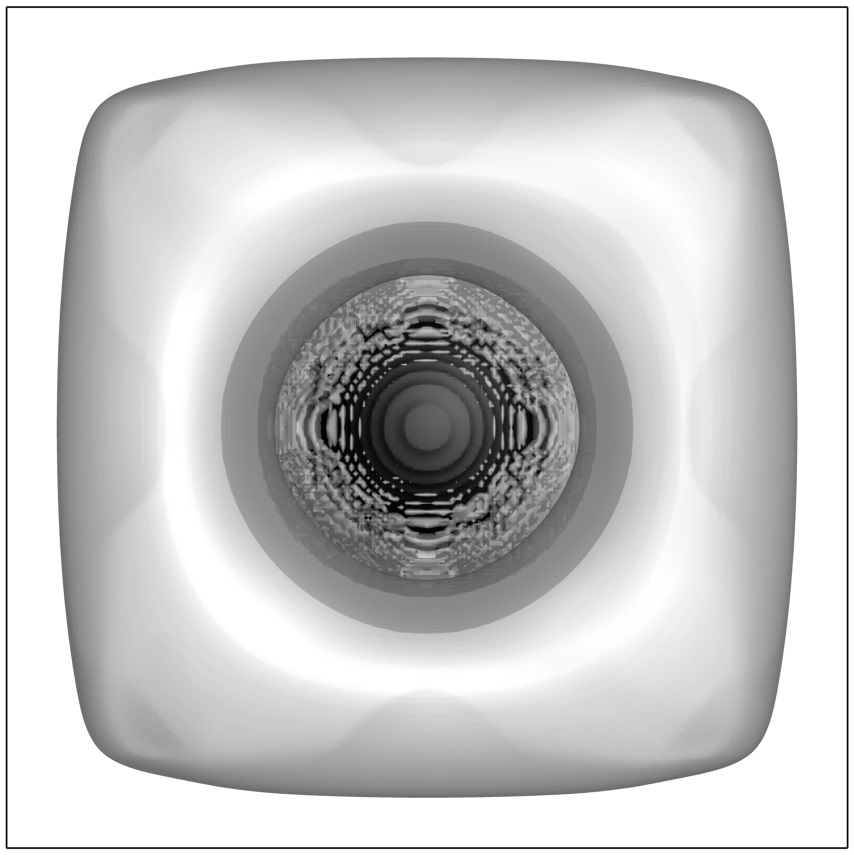}
        \includegraphics[width=0.3\textwidth,clip=]{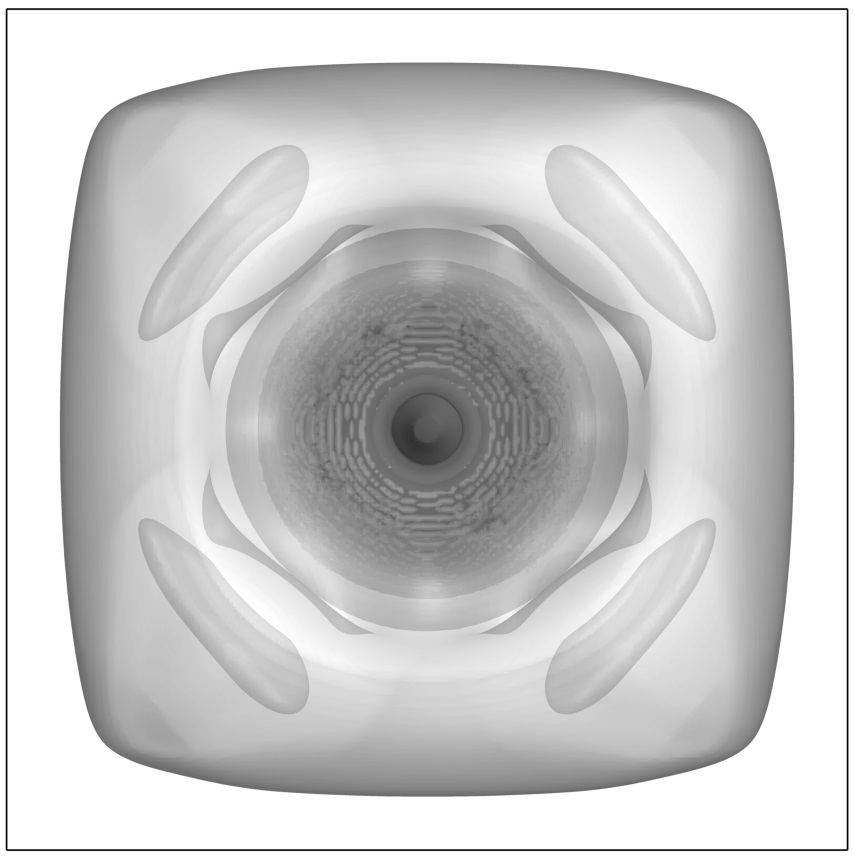}
        \caption{Isosurfaces of $\lambda_2=-0.1$, $Ra=10^5$,  $\delta=0.25;0;-0.25$ - top view}
    \end{subfigure}
   \begin{subfigure}{\textwidth}
        \includegraphics[width=0.3\textwidth,clip=]{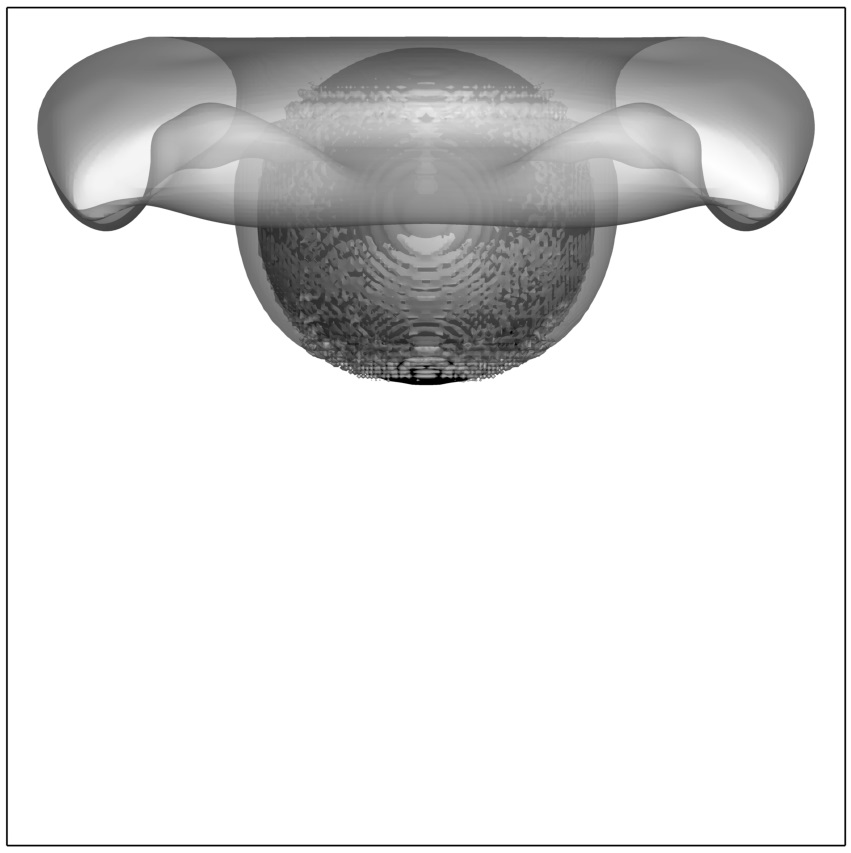}
        \includegraphics[width=0.3\textwidth,clip=]{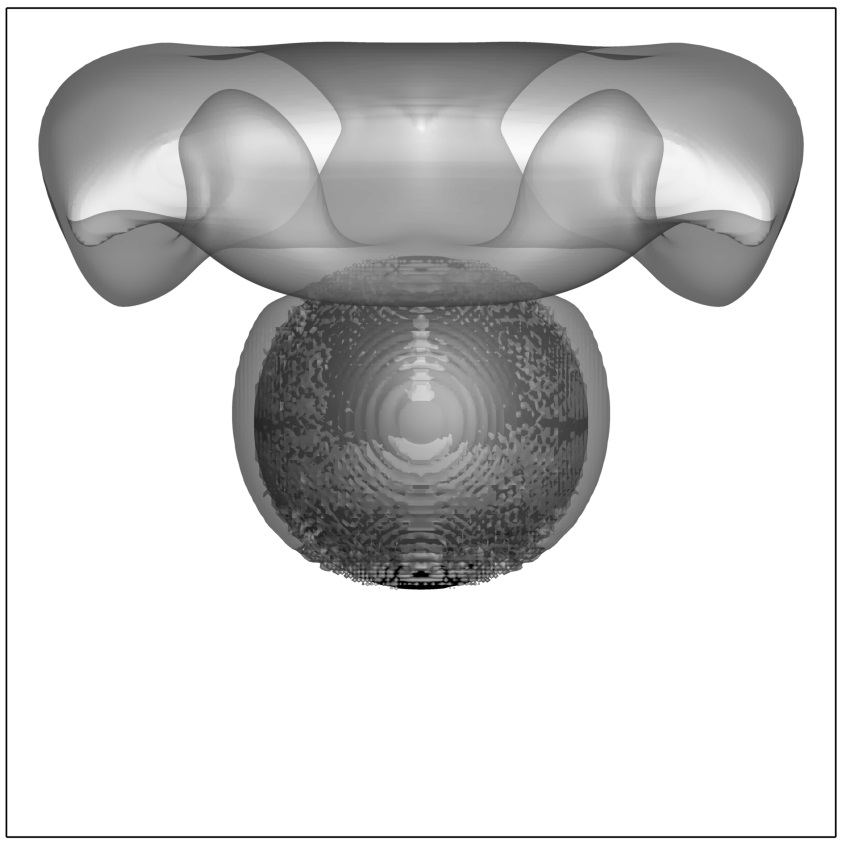}
        \includegraphics[width=0.3\textwidth,clip=]{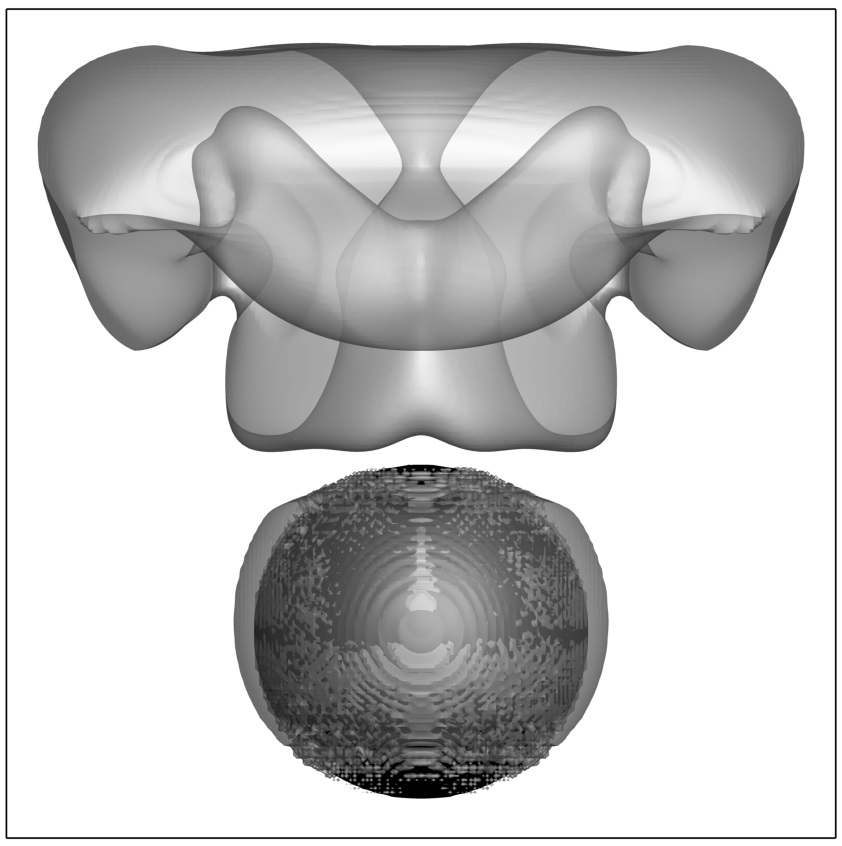}
         \caption{Isosurfaces of $\lambda_2=-0.1$, $Ra=10^6$,  $\delta=0.25;0;-0.25$ - front view}
    \end{subfigure}
    \end{figure}
\begin{figure}[H]
    \ContinuedFloat 
    \begin{subfigure}{\textwidth}
        \includegraphics[width=0.3\textwidth,clip=]{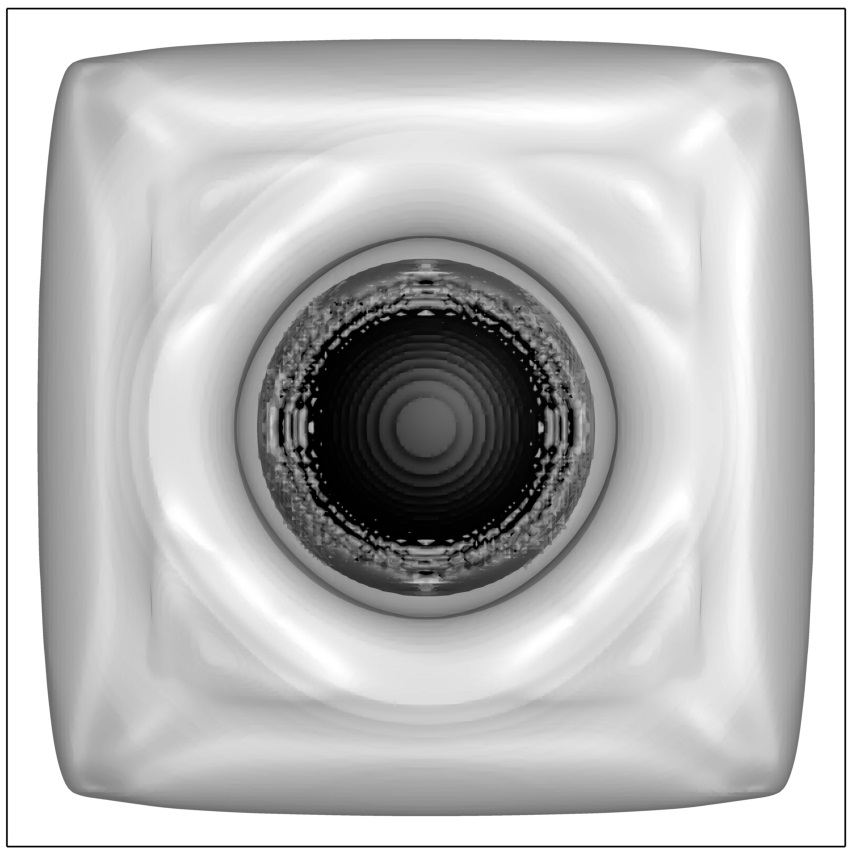}
        \includegraphics[width=0.3\textwidth,clip=]{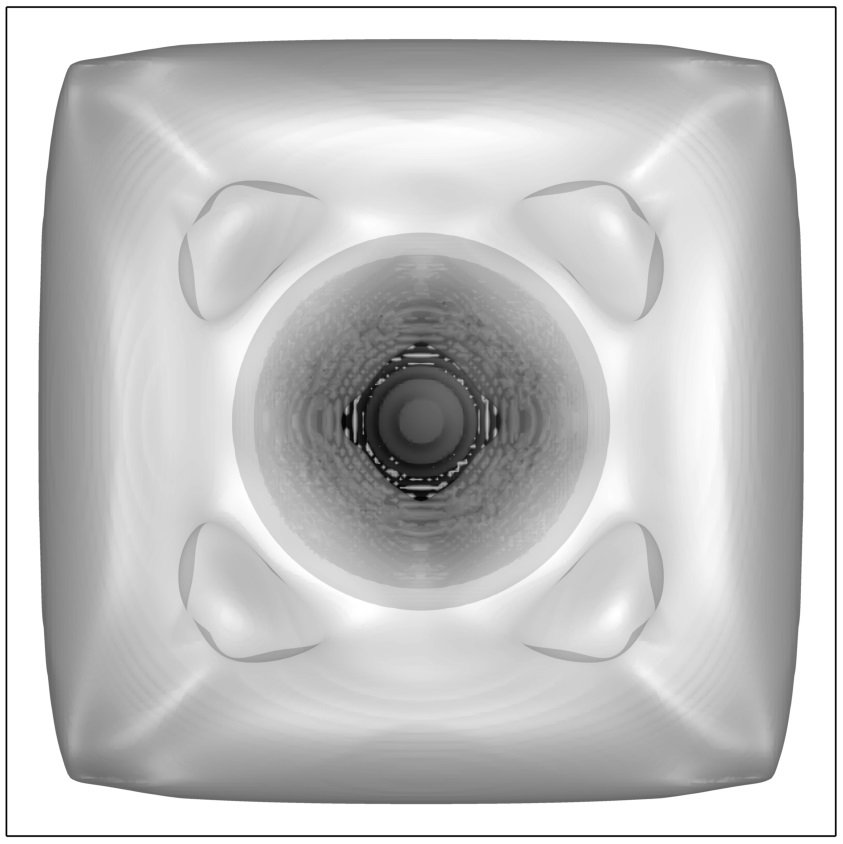}
        \includegraphics[width=0.3\textwidth,clip=]{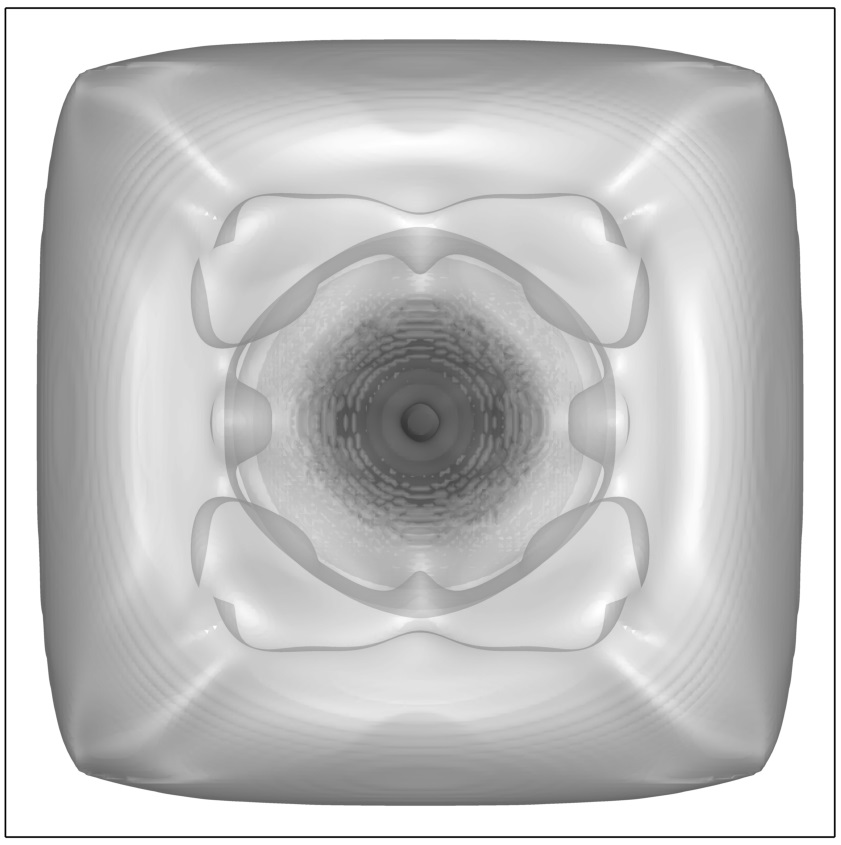}
        \caption{Isosurfaces of $\lambda_2=-0.1$, $Ra=10^6$,  $\delta=0.25;0;-0.25$ - top view}
    \end{subfigure}

    \caption{Visualization of convection cells for the natural convection flow from a hot sphere placed inside a cold cube.}
    \label{fig:HotSphereDelats}
\end{figure}
\subsection{Natural convection from a hot horizontal cylinder in a cold cube}
\begin{figure}
\centering
\caption{Physical model of  a hot cylinder inside a cold cube.}
\includegraphics[width=0.7\textwidth,clip=]{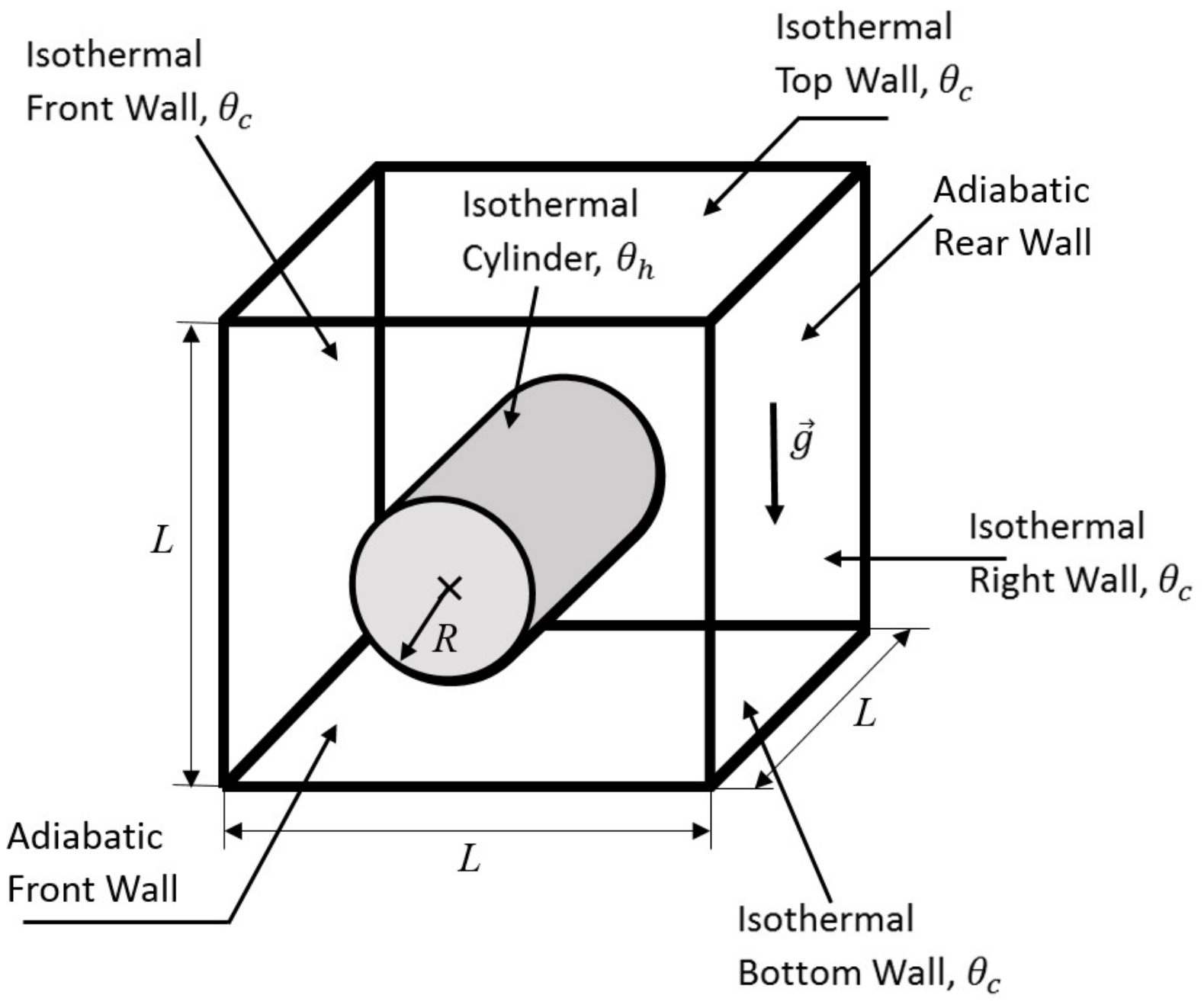}
\label{fig:PhysModelCylinder}
\end{figure}

The natural convection flow from a hot horizontal cylinder of radius $R$, whose axis coincides with the spanwise centerline of a cold cube of edge $L$ is considered. The cylinder extends across the whole width of the cube. The cylinder surface is held at a constant hot temperature, $\theta_h$, while its edges are attached to the adiabatic walls of the cube. All other cube walls are held at a constant cold temperature, $\theta_c$, and the gravity force acts downwards (see Fig. \ref{fig:PhysModelCylinder}). In the present study the configurations with radii varying in the range of $0.1\leq R/L\leq 0.4$ were investigated.
All the simulations were performed on $200^3$ uniform grids with a time step of $\Delta t=10^{-3}$. The convergence to steady state was assumed when the value of the $L_\infty$ norm calculated for the difference between the flow characteristics obtained at two consecutive time steps was less than $10^{-6}$ for all the flow fields. Special treatment should be applied when interpolating Eulerian temperatures on the Lagrangian points of the cylinder attached to both adiabatic walls of the cube, and when regularizing the volumetric Lagrangian heat fluxes to the adjacent Eulerian grid. This is because of the symmetric two-sided character of  the discrete Dirac delta function (see Eqs. \ref{eq:DeltaFunc}) utilized in the present study. Therefore, just a naive employment of the above delta function in interpolation (or regularization) operators acting from only one side of the cube wall will lead to an accumulation of computational errors. The remedy comes from the fact that the adiabatic boundary condition is equivalent to a mirror symmetry distribution of the temperature field with respect to the thermally insulated wall. For this reason one should double the impact of interpolated temperatures and regularized volumetric heat fluxes when employing the delta function from one side of the adiabatic boundary. Note also that no special treatment is required for the velocity fields at the same Lagrangian points, as they are simply skipped when building the corresponding $\textbf{\textit{R}}$ and $\textbf{\textit{I}}$ matrices. This is due to the fact that the non-slip kinematic constraints are automatically enforced by the imposed non-slip boundary conditions on the cube's walls.

It should be noted that our simulations resulted in steady state flows for the entire range of $R/L$ and $Ra$ values. This result differs from the data recently published in \cite{seo2016}, which reported  quasi-periodic and periodic flows for $R/L=0.3$ and $R/L=0.4$, respectively, at $Ra=10^6$. In order to exclude the dependence of the currently obtained solution on grid resolution and time step, additional simulations were performed for the above configurations for which the values of maximal grid size and time step were reduced by a factor of 2. Insignificant deviations (no more than 0.5\% for all the flow variables) were found between the newly obtained and the original steady state flows, which favorably verifies the independence of our results on grid size and time step values. The apparent reason for the discrepancy observed between the current and the the previously published \cite{seo2016} results can be attributed to the non-linear physics of the considered system, which can simultaneously exhibit a number of different states at the same value of $Ra$ number. Performing a global linear stability analysis which would formally prove the existence and character of the bifurcated flow reported in \cite{seo2016} remained out of the scope of the present study. In the following we will put forward a number of arguments supporting the possibility that the currently observed and the previously reported flows belong to different branches. With all these caveats an acceptable agreement is observed when comparing the present and the previously reported \cite{seo2016} $\overline{Nu}$ values averaged over the surface of a hot cylinder\footnote{The $\overline{Nu}$ values corresponding to unsteady flows reported in \cite{seo2016} were averaged over time.} for the entire range of $R/L$ and $Ra$ values, as summarized in Table \ref{NuAvCyl}.
\begin{table}[]
 \fontsize{11}{11}
\selectfont
\centering
\caption{Comparison between the present and the previously published $\overline{Nu}$ values averaged over the surface of a hot cylinder placed within a cold cube.}
\label{NuAvCyl}
\begin{tabular}{ccccccc}
\hline
{}& \multicolumn{2}{c}{$Ra=10^4$} & \multicolumn{2}{c}{$Ra=10^5$}   & \multicolumn{2}{c}{$Ra=10^6$} \\
\cline{2-3}
\cline{4-5}
\cline{6-7}
\multicolumn{1}{c} { $R/L$} &    \multicolumn{1}{c} {Ref.\cite{seo2016}} & \multicolumn{1}{c}{Present}&    \multicolumn{1}{c} {Ref.\cite{seo2016}} & \multicolumn{1}{c}{Present}  &   \multicolumn{1}{c} {Ref.\cite{seo2016}} & \multicolumn{1}{c}{Present}\\ \hline
0.1      & 6.2493     & 6.4880 & 11.138   & 11.662   &18.326      &19.250  \\
0.2      & 5.1184     & 5.1500 & 7.2271   & 7.5800   &13.361      &13.937  \\
0.3      & 5.8084     & 5.7304 & 6.4790   & 6.5169   &11.272      &11.401  \\
0.4      & 8.7030     & 8.5544 & 8.7030   & 8.7643   &10.716      &10.832  \\ \hline

\end{tabular}
\end{table}

A spatial distribution of temperature isosurfaces corresponding to the steady state flows calculated for the entire range of $R/L$ and $Ra$ values is shown in Fig. \ref{fig:HotCylinderTempr}. For all the configurations and for the entire range of $Ra$ numbers the temperature distribution is symmetric relative to the $X-Z$ and $Y-Z$ planes. As expected, all the configurations are characterized by a linear temperature distribution with almost concentric isosurfaces for a small value of $Ra=10^3$. With an increase in the $Ra$ values, the flow convection becomes more dominant and the temperature isosurfaces acquire irregular shapes. For $R/L= 0.1$ and $R/L=0.2$ the shape of temperature isosurfaces elongates towards the top of the cube. The shape elongation is not uniform and is more pronounced close to the isothermal boundaries, which is a result of lower vertical velocities in these regions. This trend, however, is not observed for configurations with higher $R/L$ ratios. To understand the reason for this difference we visualize convection cells inherent to each configuration for the highest $Ra=10^6$ value by plotting the $\lambda_2$ criterion (see Fig. \ref{fig:HotCylinderLambda}). It can be seen that the configurations with $R/L=0.1$ and $R/L=0.2$ are characterized by two major convection cells which are symmetric relative to the $Y-Z$ plane and  occupy the whole upper part of the cube. This is in contrast to configurations with higher $R/L$ ratios (0.3 and 0.4), both having four major symmetric discrete convection cells located close to the top boundary of the cube.

\begin{figure}[H]
\centering

      \begin{subfigure}{\textwidth}
        \includegraphics[width=0.3\textwidth,clip=]{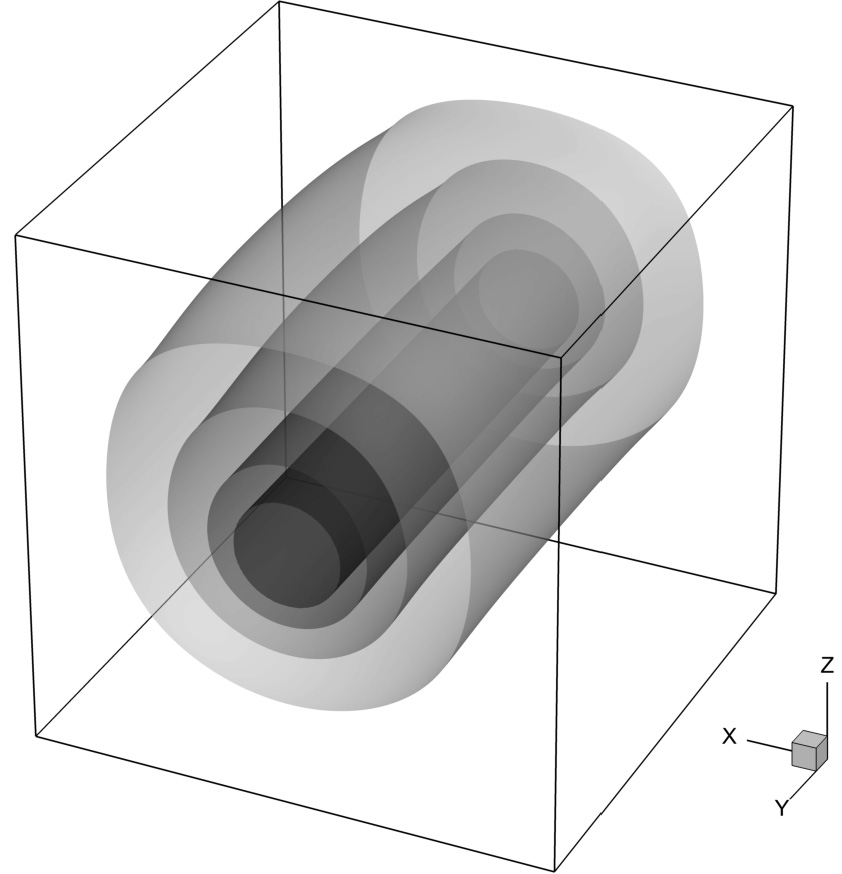}
        \includegraphics[width=0.3\textwidth,clip=]{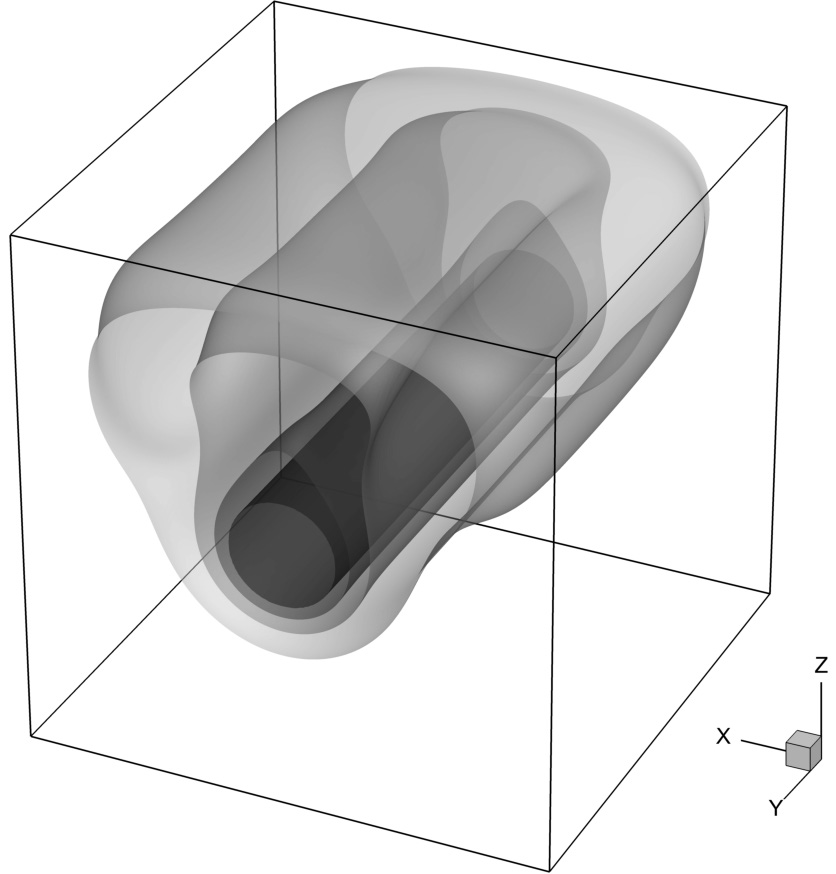}
        \includegraphics[width=0.3\textwidth,clip=]{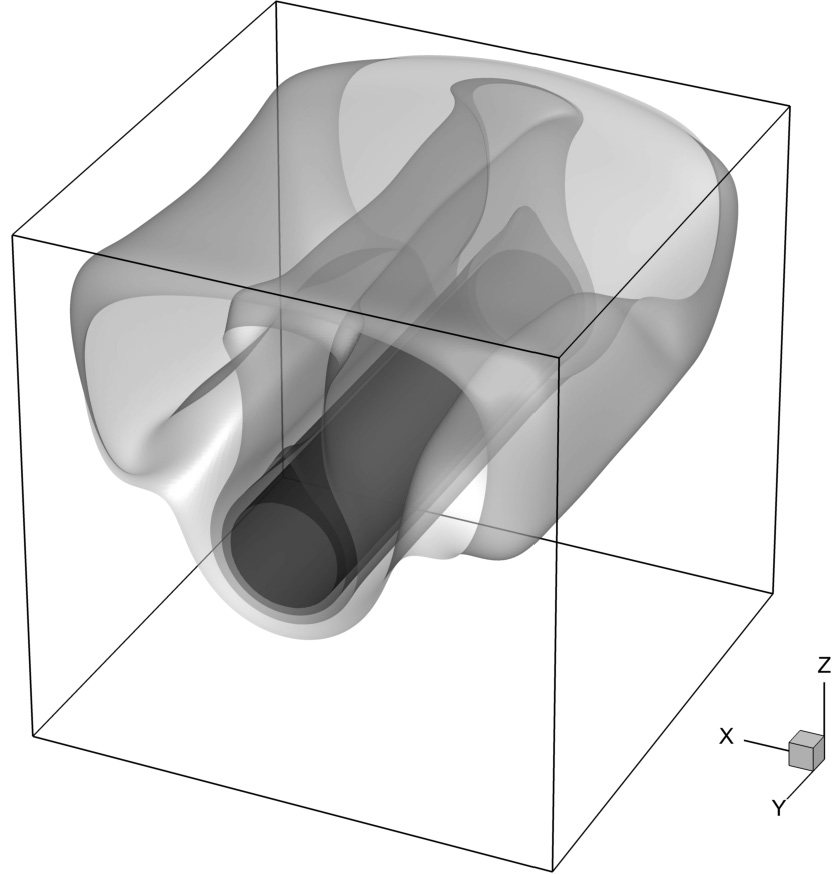}
         \caption{$R/L=0.1$, $Ra=10^4, 10^5, 10^6$ }
    \end{subfigure}
    \begin{subfigure}{\textwidth}
        \includegraphics[width=0.3\textwidth,clip=]{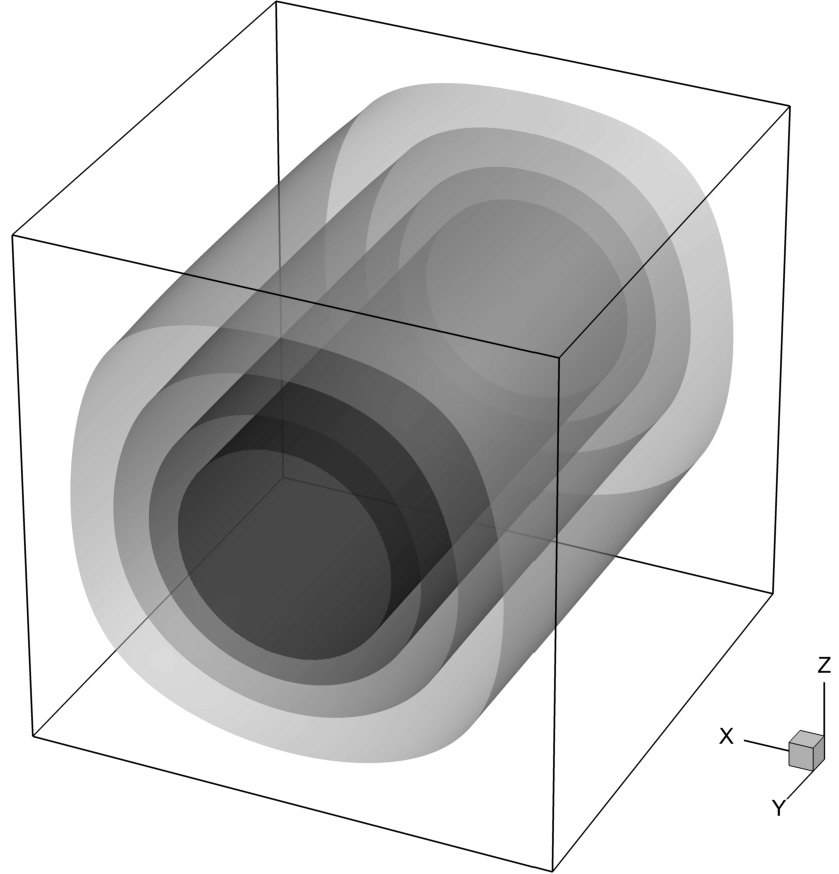}
        \includegraphics[width=0.3\textwidth,clip=]{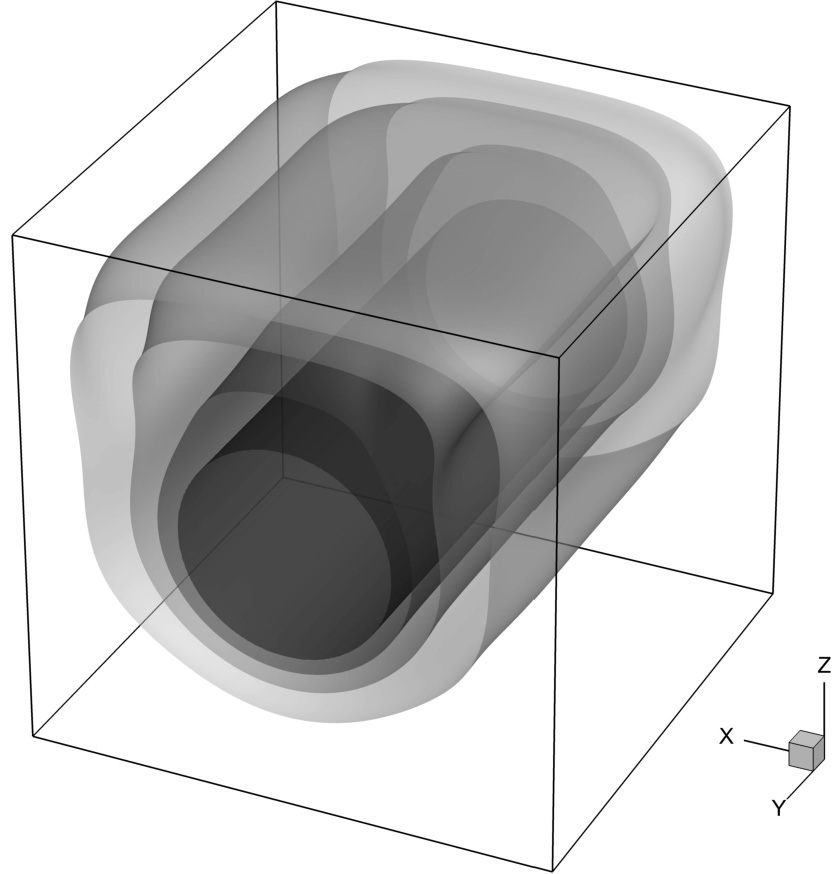}
        \includegraphics[width=0.3\textwidth,clip=]{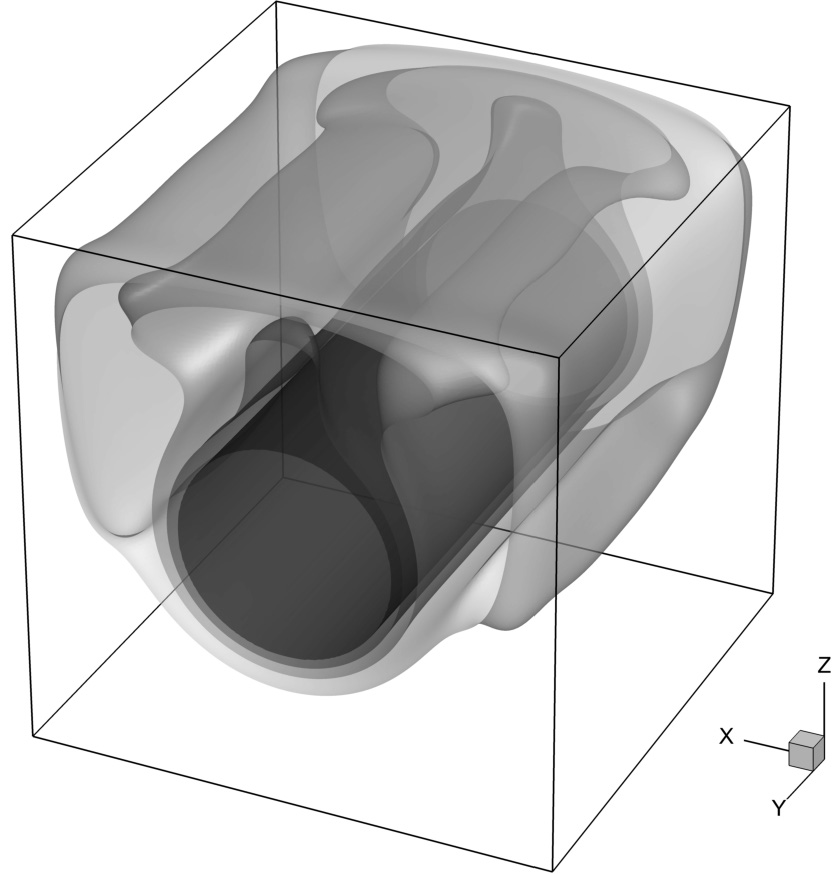}
        \caption{$R/L=0.2$, $Ra=10^4, 10^5, 10^6$ }
    \end{subfigure}
   \begin{subfigure}{\textwidth}
        \includegraphics[width=0.3\textwidth,clip=]{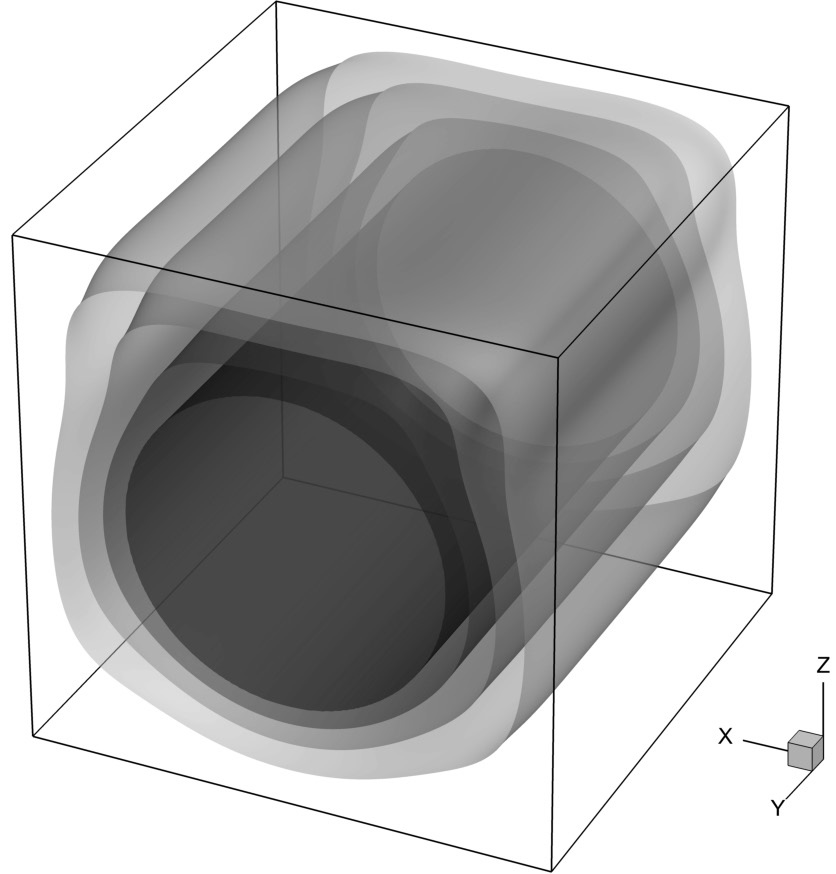}
        \includegraphics[width=0.3\textwidth,clip=]{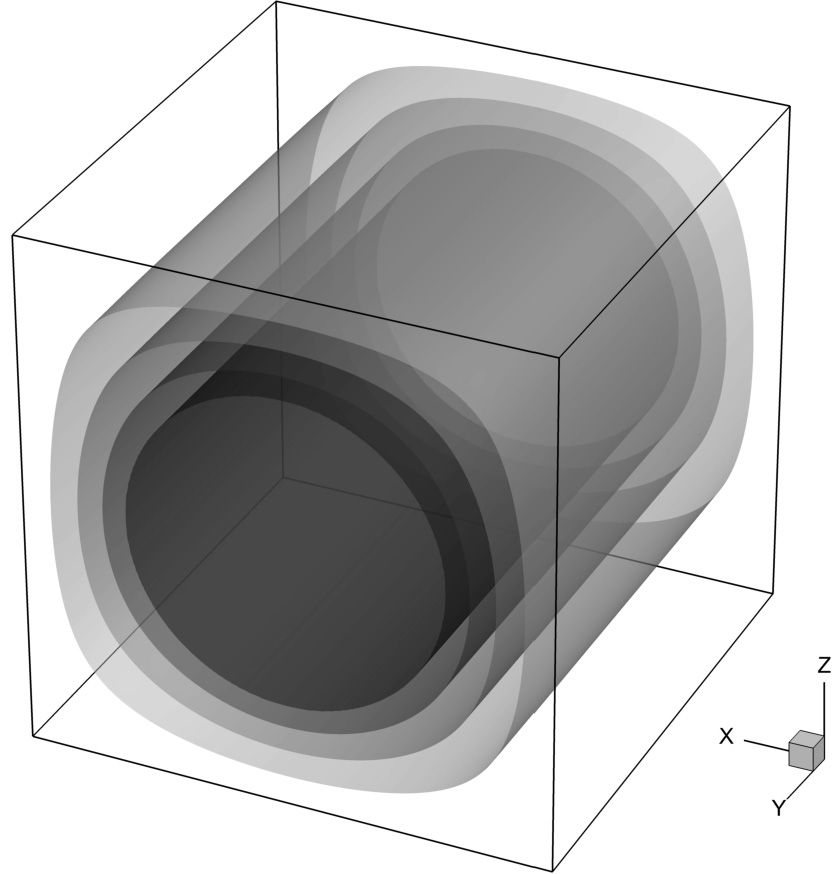}
        \includegraphics[width=0.3\textwidth,clip=]{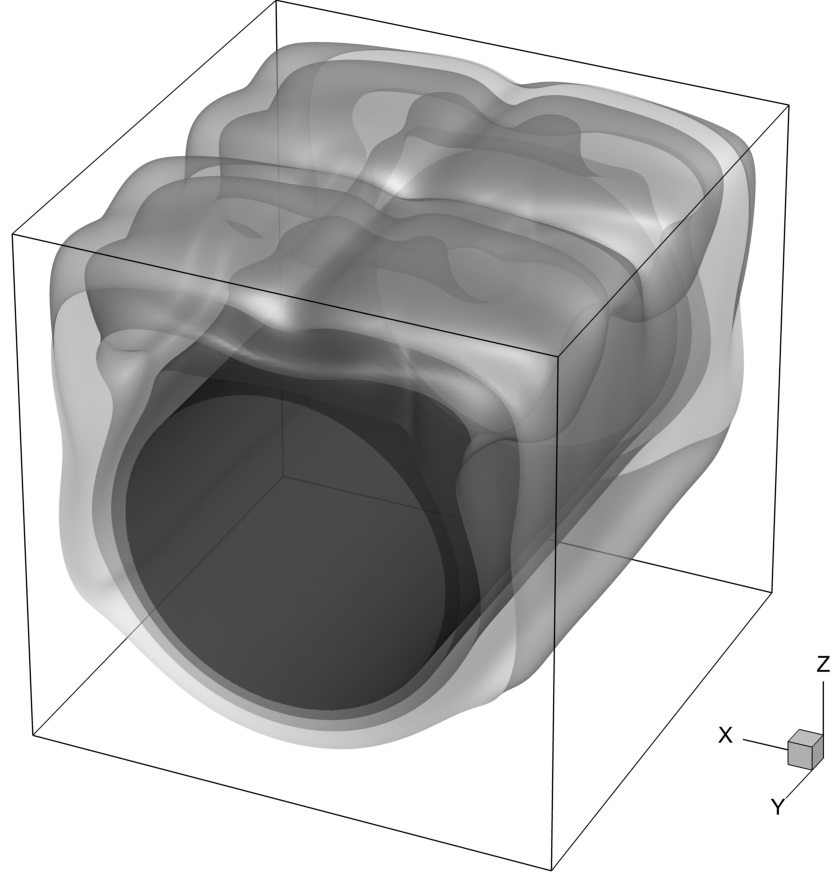}
         \caption{$R/L=0.3$, $Ra=10^4, 10^5, 10^6$ }
    \end{subfigure}
    \end{figure}
\begin{figure}[H]
    \ContinuedFloat 
    \begin{subfigure}{\textwidth}
        \includegraphics[width=0.3\textwidth,clip=]{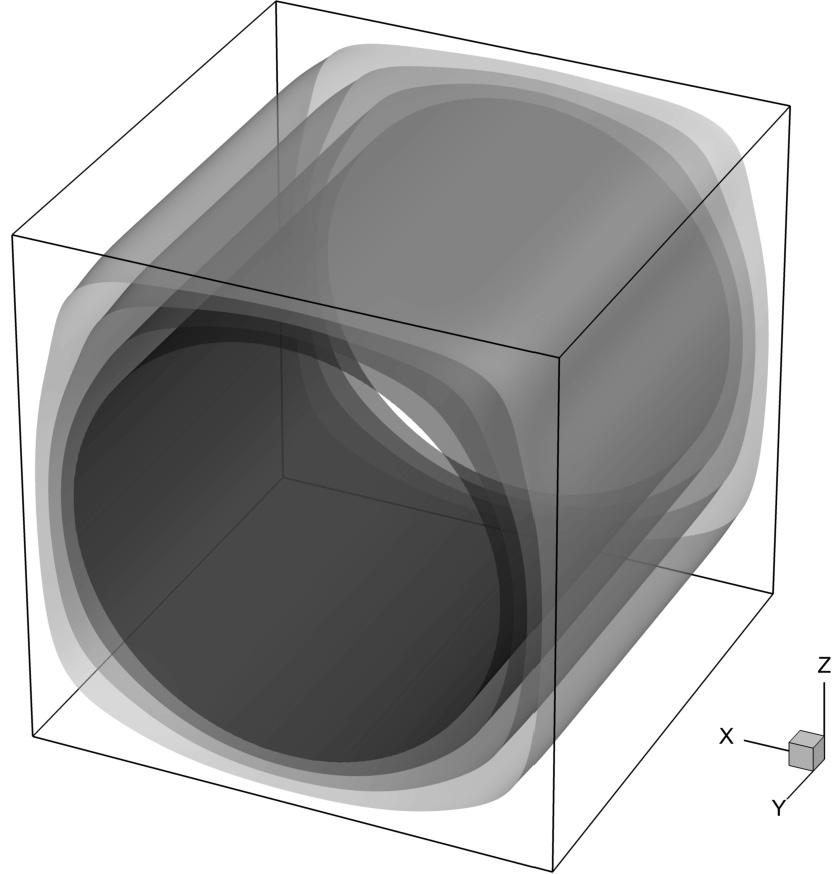}
        \includegraphics[width=0.3\textwidth,clip=]{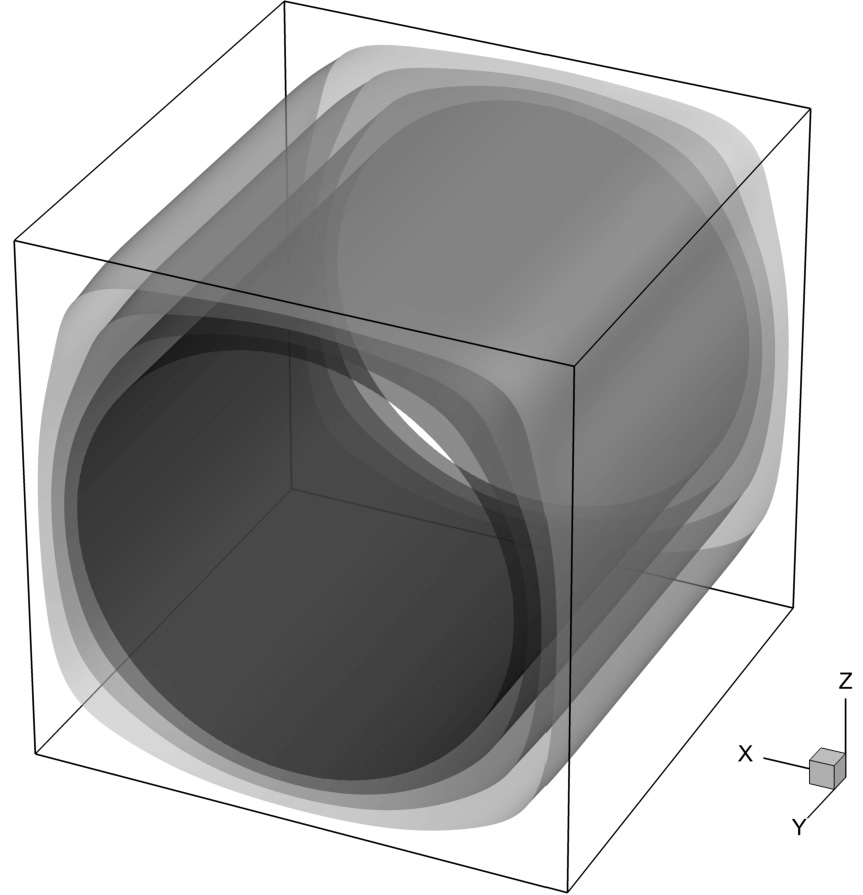}
        \includegraphics[width=0.3\textwidth,clip=]{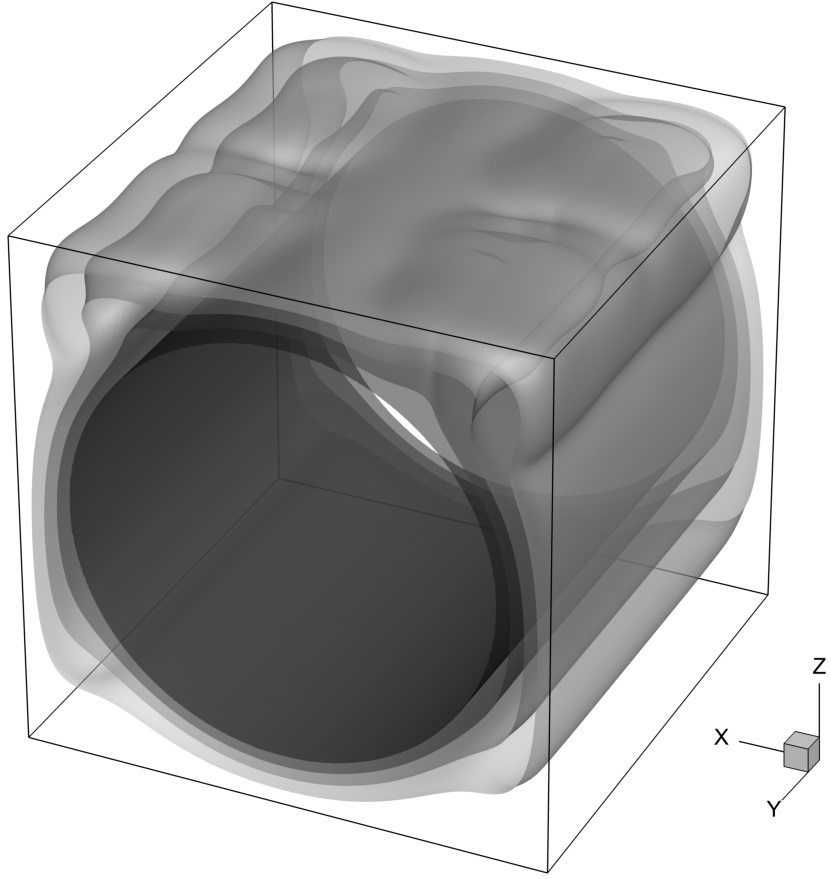}
        \caption{$R/L=0.4$, $Ra=10^4, 10^5, 10^6$ }
    \end{subfigure}

    \caption{Spatial distribution of temperature isosurfaces for the values of $\theta=0, 0.25, 0.5, 0.75$. }
    \label{fig:HotCylinderTempr}
\end{figure}

\begin{figure}[H]
\centering

      \begin{subfigure}{\textwidth}
        \includegraphics[width=0.4\textwidth,clip=]{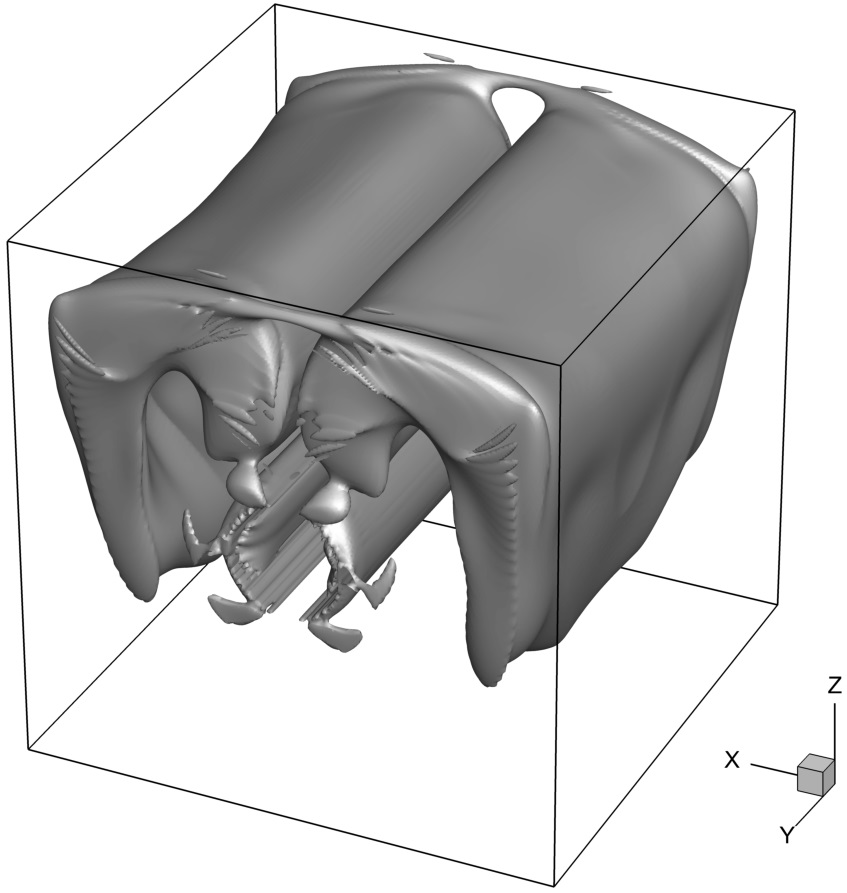}
        \includegraphics[width=0.45\textwidth,clip=]{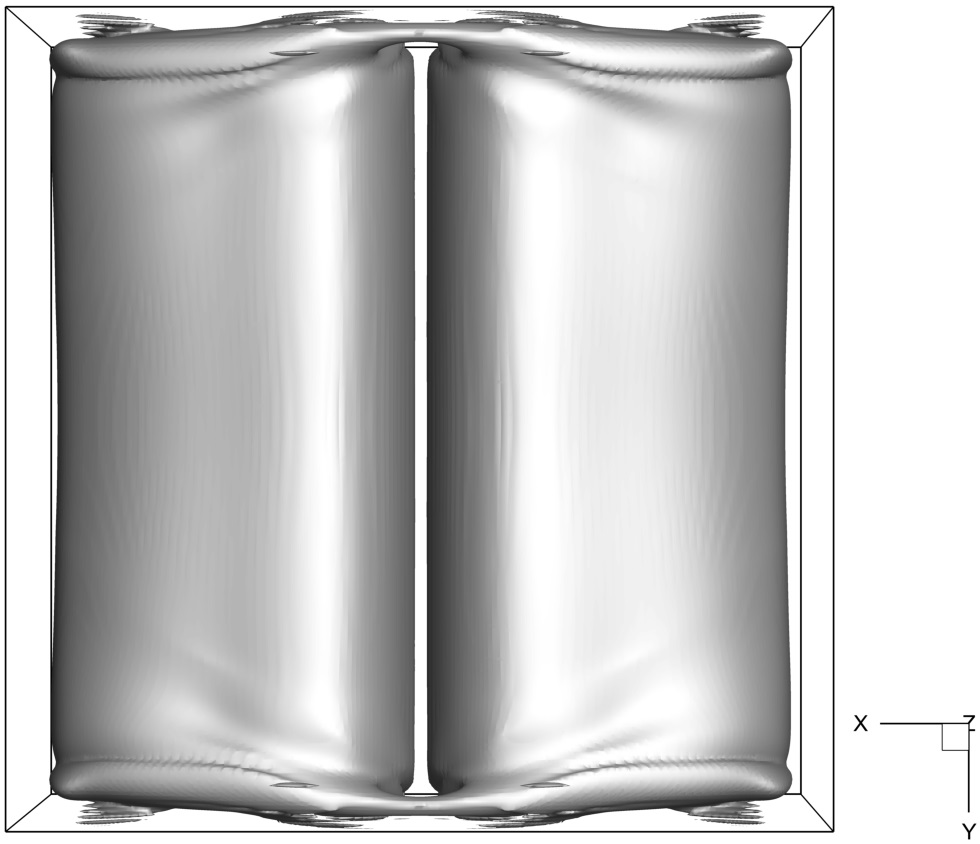}
         \caption{Isosurfaces of $\lambda_2=-0.1$, $R/L=0.1$ }
    \end{subfigure}
    \begin{subfigure}{\textwidth}
        \includegraphics[width=0.4\textwidth,clip=]{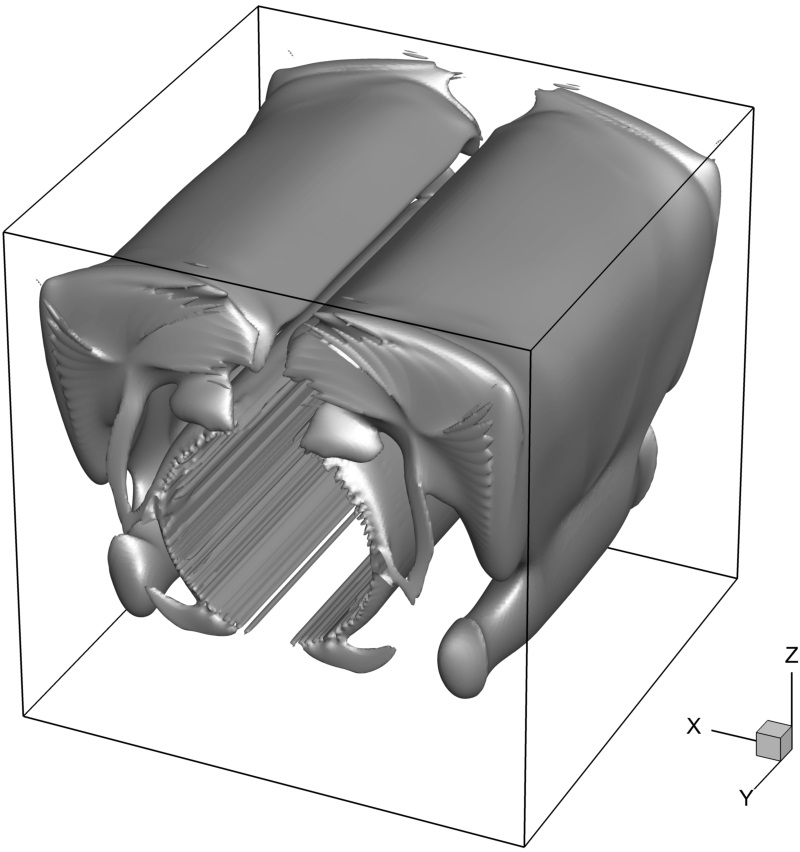}
        \includegraphics[width=0.45\textwidth,clip=]{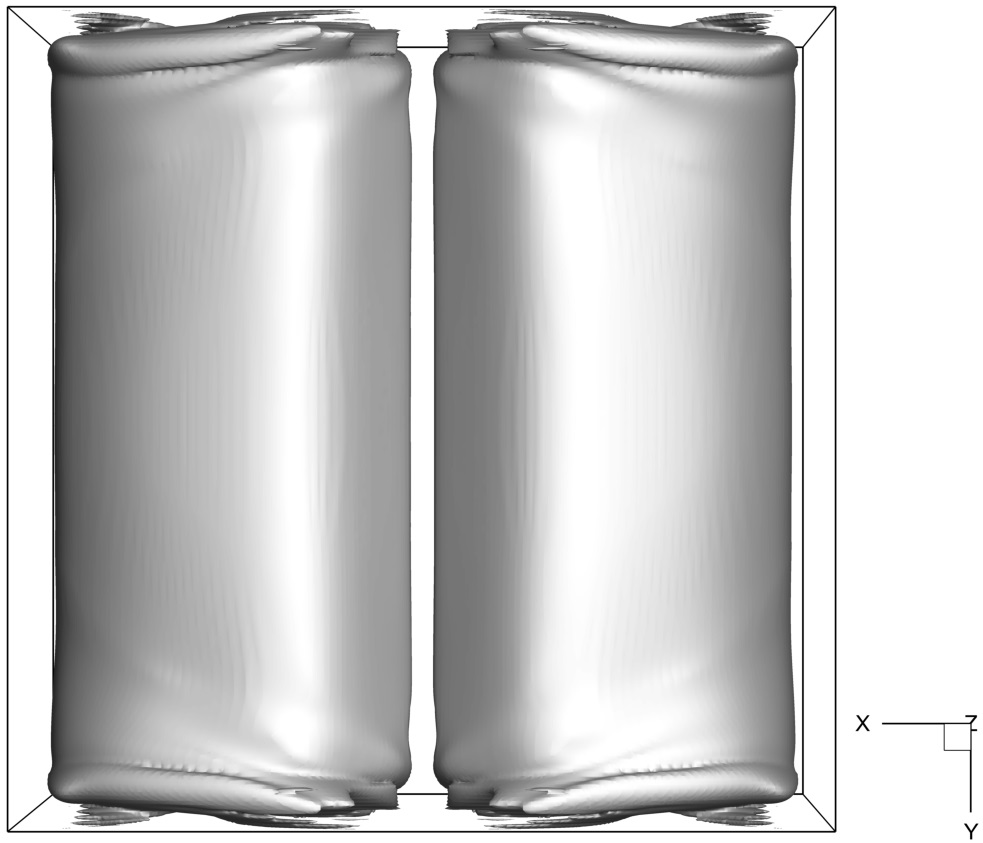}
        \caption{Isosurfaces of $\lambda_2=-0.1$,$R/L=0.2$  }
    \end{subfigure}
   \begin{subfigure}{\textwidth}
        \includegraphics[width=0.4\textwidth,clip=]{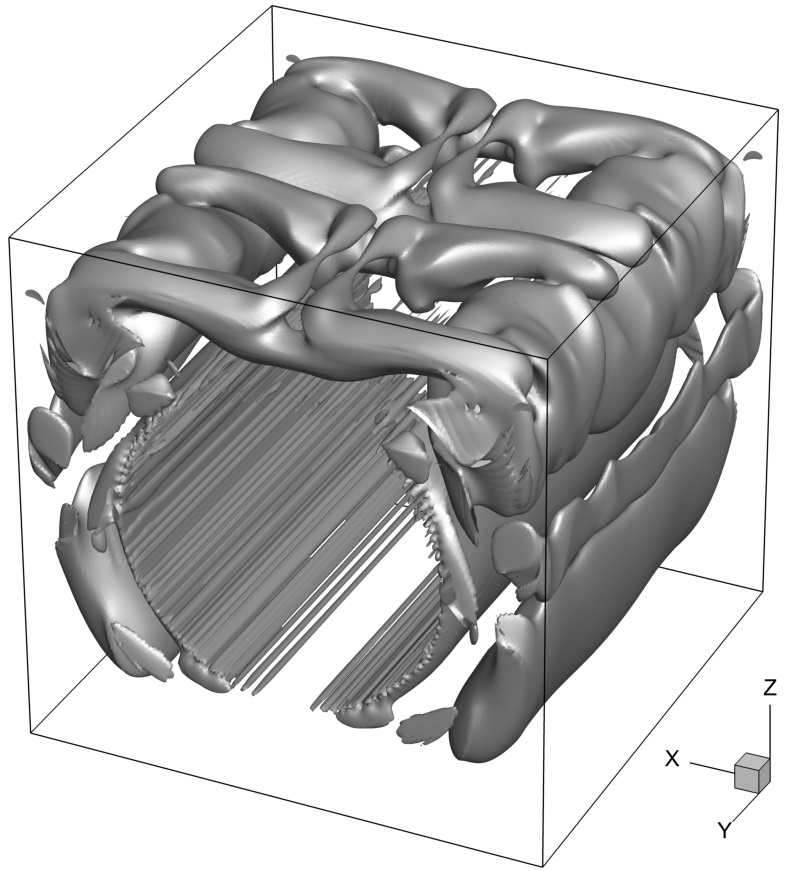}
        \includegraphics[width=0.45\textwidth,clip=]{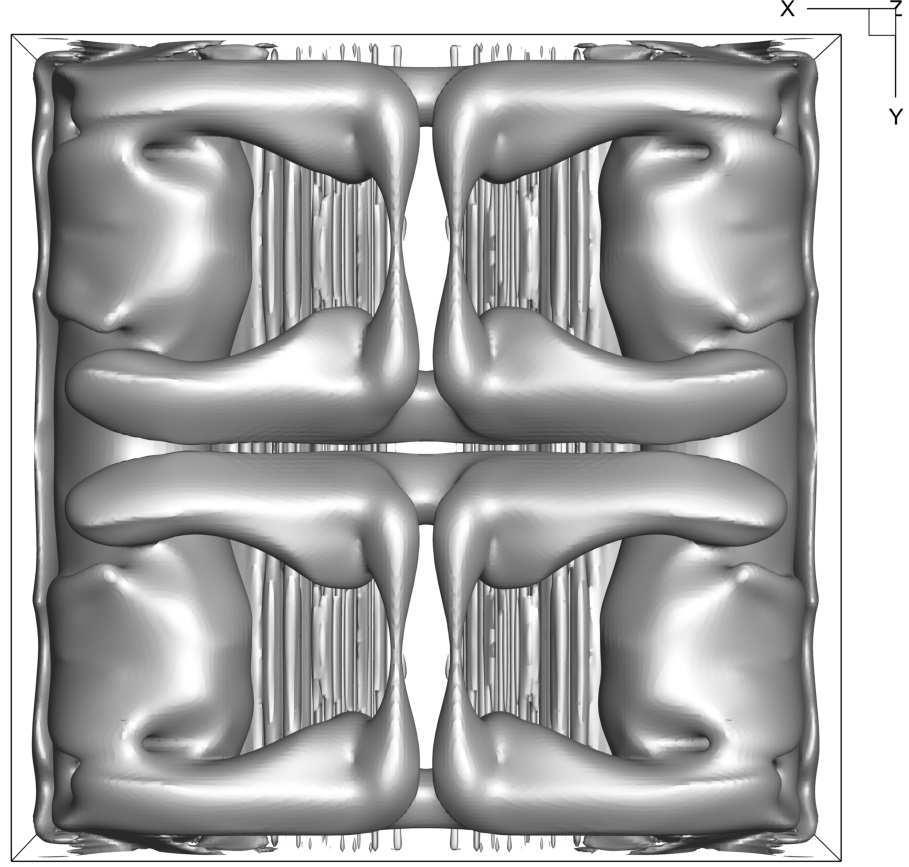}
         \caption{Isosurfaces of $\lambda_2=-0.1$,$R/L=0.3$  }
    \end{subfigure}
    \end{figure}
\begin{figure}[H]
    \ContinuedFloat 
    \begin{subfigure}{\textwidth}
        \includegraphics[width=0.4\textwidth,clip=]{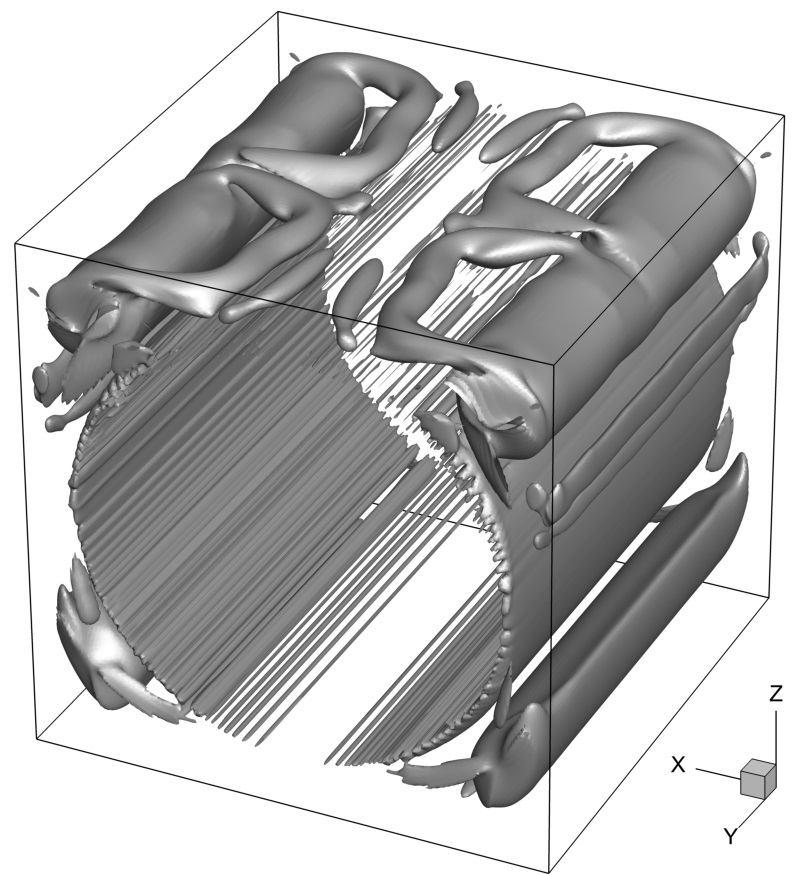}
        \includegraphics[width=0.45\textwidth,clip=]{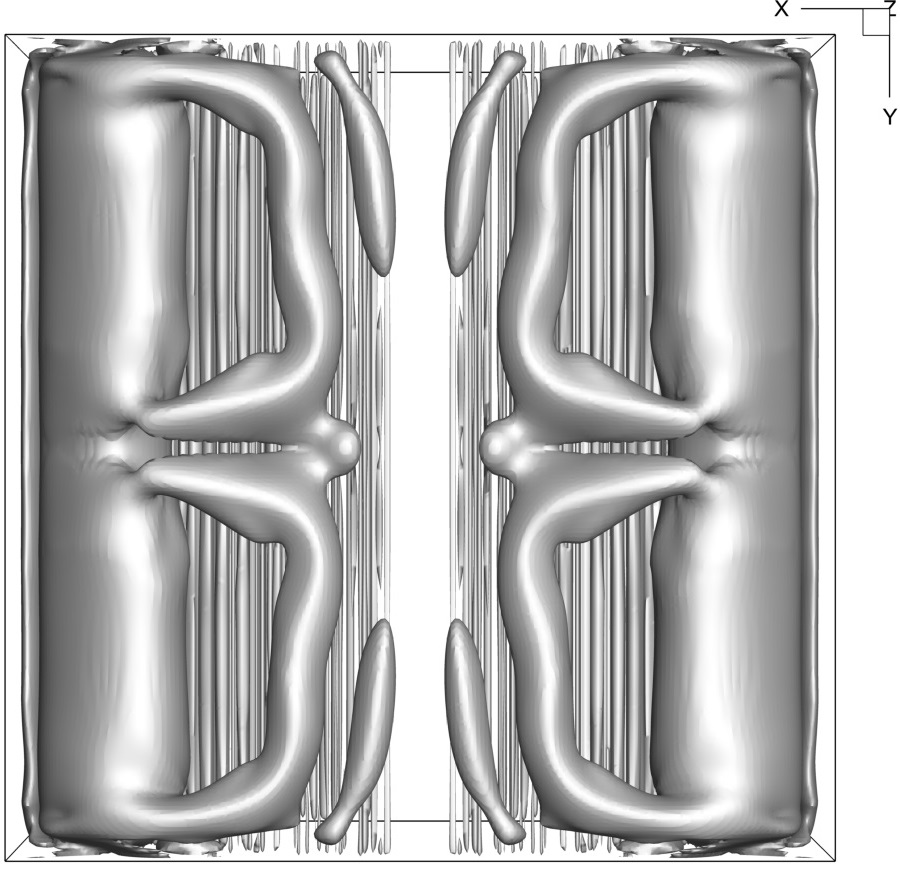}
        \caption{Isosurfaces of $\lambda_2=-0.1$,$R/L=0.4$  }
    \end{subfigure}

    \caption{Visualization of convection cells for the natural convection flow from a hot cylinder placed inside a cold cube.}
    \label{fig:HotCylinderLambda}
\end{figure}

We next focus on the differences between the present and the reported \cite{seo2016} time averaged spatial distributions of convection cells obtained for $Ra=10^6$ and $R/L=0.4$. Both  distributions are symmetric relative to the $X-Z$ and $Y-Z$ planes. The currently obtained distribution is characterized by four major discrete convection cells appearing close to the cube top boundary. In contrast, the distribution reported in \cite{seo2016} has only two major discrete convection cells occupying almost the entire region adjacent to the top boundary of the cube.  The above argument supports a possibility that the two flows belong to different branches, which can explain the observed  qualitative and qualitative differences between the corresponding flow characteristics. The qualitative discrepancies between the present and the reported \cite{seo2016} time averaged spatial distributions of convection cells obtained for $Ra=10^6$ and $R/L=0.3$ are not as significant as for $R/L=0.4$ and, apparently, can be attributed to the differences between the implemented numerical approaches. An accurately performed grid and time independence analysis further corroborates the currently obtained results.

Table \ref{SetupCyl} presents details regarding characteristics of the numerical solution as a function of the $Ra$ and $R/L$ values. As has been already observed for the configuration of a hot sphere inside a cold cube, the memory consumption varies inversely with the $Ra$ value. In the present case it scales as $Ra^{-0.16}$ ,$Ra^{-0.15}$,$Ra^{-0.12}$ and $Ra^{-0.09}$ for $R/L=0.1$,$R/L=0.2$,$R/L=0.3$, and $R/L=0.4$, respectively. At the same time, for the given $Ra$ value the memory consumption is linearly proportional to the number of Lagrangian points characterizing the surface of the immersed body. Regarding the computational times, no clear trend was observed as a function of the $Ra$ and $R/L$ values, while the average time required for performing a single time step is equal to 3.4 seconds.
\begin{table}[]
\fontsize{10}{10}
\selectfont
\centering
\caption{Setup and characteristics of the numerical solution obtained on a $200^3$ grid for the simulation of the natural convection from a hot cylinder placed inside a cold cube.}
\label{SetupCyl}
\begin{tabular}{llllllll}
\hline
\multicolumn{1}{>{}m{1cm}}{$Ra$} &\multicolumn{1}{>{}m{1cm}}{$R/L$}& \multicolumn{1}{>{}m{2cm}}{Time step duration, [s]} & \multicolumn{1}{>{}m{2cm}}  {RAM, [Gb]} & \multicolumn{1}{>{}m{1cm}} {$\Delta t$} &   \multicolumn{1}{>{}m{3cm}} {Method of solution of Eqs. \ref{SchurSplitted}-a} &   \multicolumn{1}{>{}m{2cm}} {Number of time steps} &   \multicolumn{1}{>{}m{2cm}} {Threshold for the filling of $[\textbf{\textit{I}}\textbf{\textit{H}}^{-1}\textbf{\textit{R}}]$}\\

\hline
\multicolumn{1}{c}{\multirow{4}{*}{$10^4$}}&$0.1$& 3.4 & 29.5&   \multicolumn{1}{c}{\multirow{4}{*}{$10^{-3}$}}&  \multicolumn{1}{c}{\multirow{4}{*}{Direct ($LU$)}}  &   $O(10^3)$ &  \multicolumn{1}{c}{\multirow{4}{*}{$10^{-25}$}}\\
\multicolumn{1}{c}{}& $0.2$& 3.2 & 50.3&   \multicolumn{1}{c}{}                          &  \multicolumn{1}{c}{}                                &   $O(10^3)$ & \multicolumn{1}{c}{}\\
\multicolumn{1}{c}{}& $0.3$& 7.5 & 68.0&   \multicolumn{1}{c}{}                          &  \multicolumn{1}{c}{}                                &   $O(10^3)$ & \multicolumn{1}{c}{}\\
\multicolumn{1}{c}{}& $0.4$& 5.2 & 92.1&   \multicolumn{1}{c}{}                          &  \multicolumn{1}{c}{}                                &   $O(10^3)$ & \multicolumn{1}{c}{}\\
\hline
\multicolumn{1}{c}{\multirow{4}{*}{$10^5$}}&$0.1$& 3.2 & 19.1&   \multicolumn{1}{c}{\multirow{4}{*}{$10^{-3}$}}&  \multicolumn{1}{c}{\multirow{4}{*}{Direct ($LU$)}}  &  $O(10^4)$ &  \multicolumn{1}{c}{\multirow{4}{*}{$10^{-25}$}}\\
\multicolumn{1}{c}{}& $0.2$& 2.9 & 32.5&   \multicolumn{1}{c}{}                          &  \multicolumn{1}{c}{}                                & $O(10^3)$ & \multicolumn{1}{c}{}\\
\multicolumn{1}{c}{}& $0.3$& 3.5 & 48.7&   \multicolumn{1}{c}{}                          &  \multicolumn{1}{c}{}                                & $O(10^3)$ & \multicolumn{1}{c}{}\\
\multicolumn{1}{c}{}& $0.4$& 4.6 & 68.6&   \multicolumn{1}{c}{}                          &  \multicolumn{1}{c}{}                                & $O(10^3)$ & \multicolumn{1}{c}{}\\
\hline
\multicolumn{1}{c}{\multirow{4}{*}{$10^6$}}&$0.1$& 2.4 & 14.1&   \multicolumn{1}{c}{\multirow{4}{*}{$10^{-3}$}}&  \multicolumn{1}{c}{\multirow{4}{*}{Direct ($LU$)}}  & $O(10^4)$ &  \multicolumn{1}{c}{\multirow{4}{*}{$10^{-25}$}}\\
\multicolumn{1}{c}{}& $0.2$& 2.8 & 25.5&   \multicolumn{1}{c}{}                          &  \multicolumn{1}{c}{}                                & $O(10^4)$ & \multicolumn{1}{c}{}\\
\multicolumn{1}{c}{}& $0.3$& 3.2 & 40.6&   \multicolumn{1}{c}{}                          &  \multicolumn{1}{c}{}                                & $O(10^4)$ & \multicolumn{1}{c}{}\\
\multicolumn{1}{c}{}& $0.4$& 4.3 & 62.1&   \multicolumn{1}{c}{}                          &  \multicolumn{1}{c}{}                                & $O(10^4)$ & \multicolumn{1}{c}{}\\
\hline
\end{tabular}
\end{table}

\subsection{Differentially heated spherical shell}
\begin{figure}
\centering
\caption{Physical model of the differentially heated concentric spherical shell located at the center of a cube.}
\includegraphics[width=0.7\textwidth,clip=]{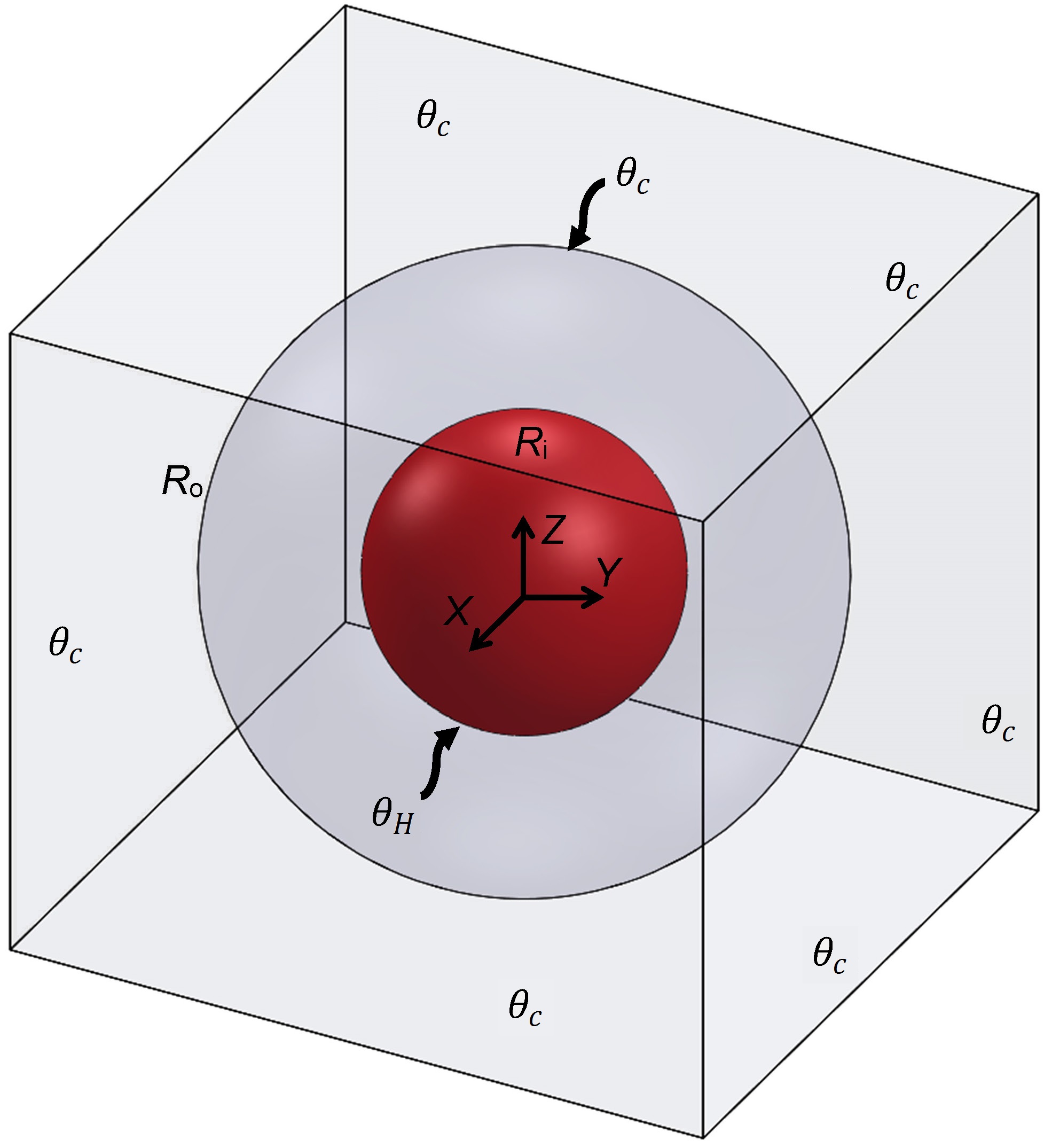}
\label{fig:PhysModelShell}
\end{figure}

We consider the natural convection flow in a spherical shell formed by two concentric spheres both located at the center of a cube, as shown in Fig. \ref{fig:PhysModelShell}. The internal sphere of radius $R_i$ is held at a constant hot temperature $\theta_H$, while the external sphere of radius $R_o$ is held at a constant cold temperature $\theta_C$. Non-slip boundary conditions for all the velocity components are imposed for all the cube walls, which are also held at a constant cold temperature $\theta_C$. The gravity force acts downwards. In all the simulations the minimal distance between the external sphere and the cube walls was at least 10 grid cells, in order to avoid high velocity and temperature gradients outside the external sphere. The difference between the temperatures of the internal and external spheres, $\theta_h - \theta_c$, and the difference between the external and internal radii, $L=R_o - R_i$, are used to scale the temperature and length fields. An additional parameter, $\phi=R_i/R_o$ is introduced to define the ratio between internal, $R_i$, and external, $R_o$, radii. Hence, the geometrical dimensions of the spherical gap are not constant and are determined by a non-dimensional distance between the external and internal radii (which is always equal to unity) and the value of parameter $\phi$. According to our previous research \cite{feldman2013}, a grid consisting of no less than $6 \times 10^6$ finite volumes is required to accurately capture both steady and unsteady flow phenomena for this kind of flow. For this reason all the simulations presented in this section were performed on a $300^3$ uniform grid.

\begin{figure}[H]
\centering
      \begin{subfigure}{\textwidth}
        \includegraphics[width=0.35\textwidth,clip=]{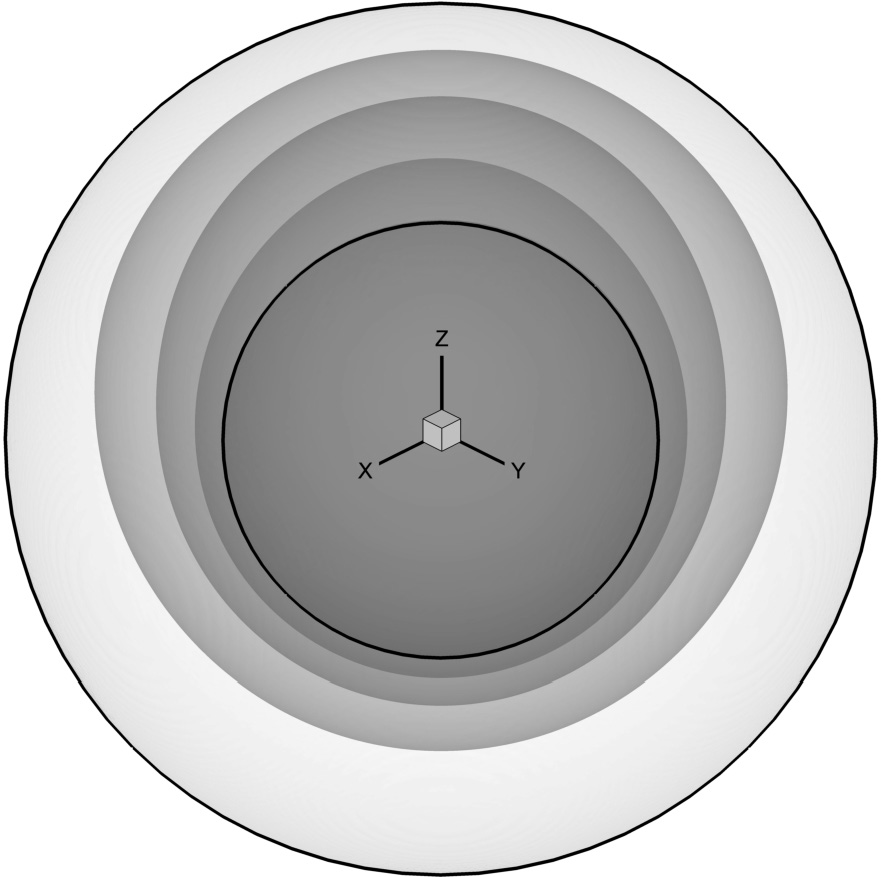}
        \includegraphics[width=0.44\textwidth,clip=]{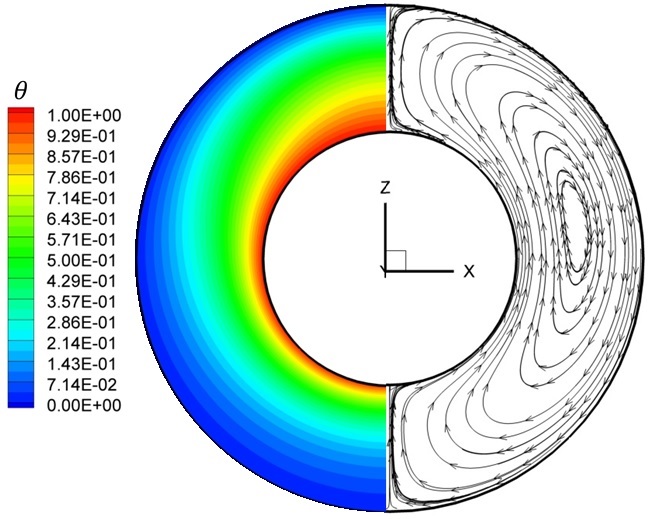}
         \caption{$Ra=10^3$}
    \end{subfigure}
    \begin{subfigure}{\textwidth}
        \includegraphics[width=0.35\textwidth,clip=]{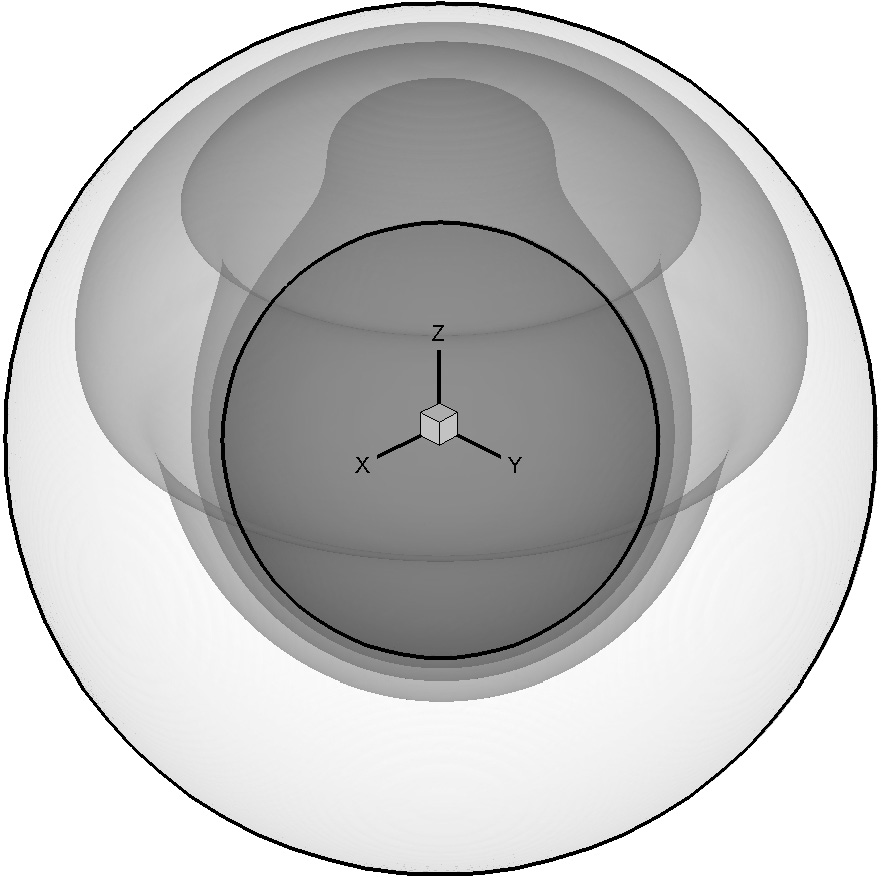}
        \includegraphics[width=0.44\textwidth,clip=]{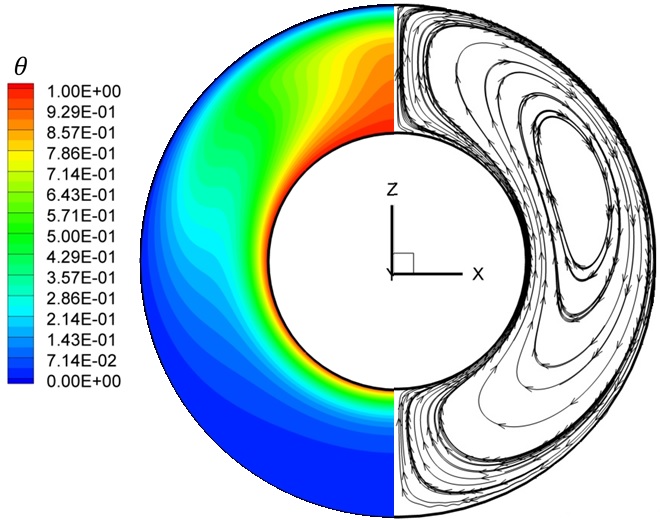}
        \caption{$Ra=10^4$}
    \end{subfigure}
      \begin{subfigure}{\textwidth}
        \includegraphics[width=0.35\textwidth,clip=]{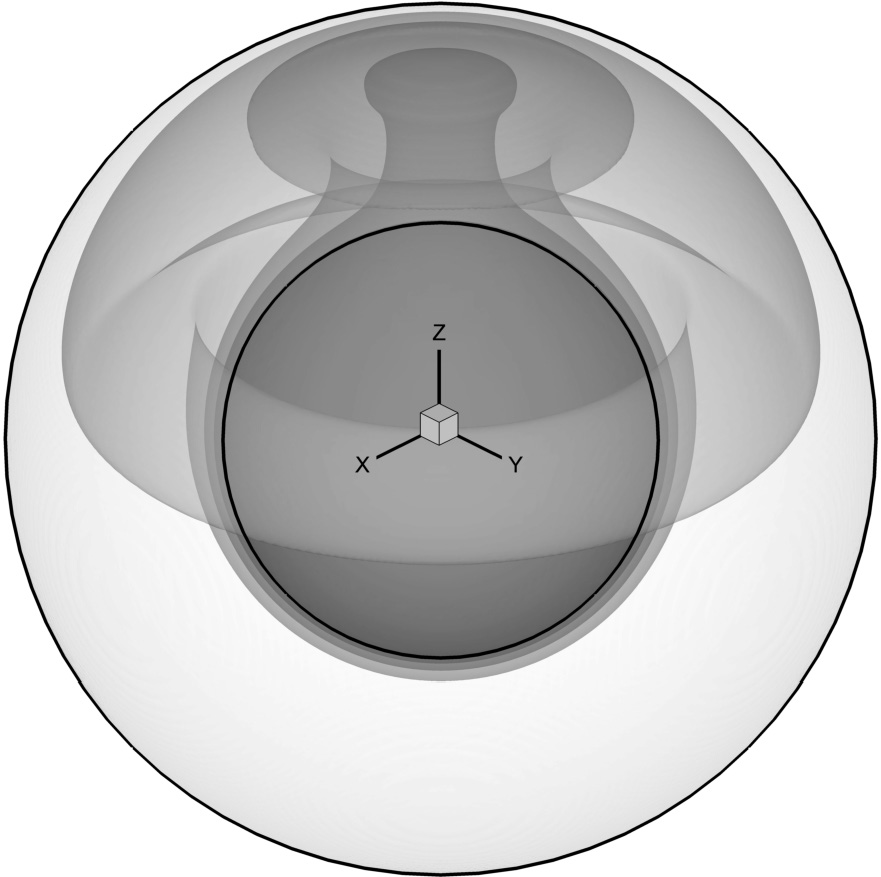}
        \includegraphics[width=0.44\textwidth,clip=]{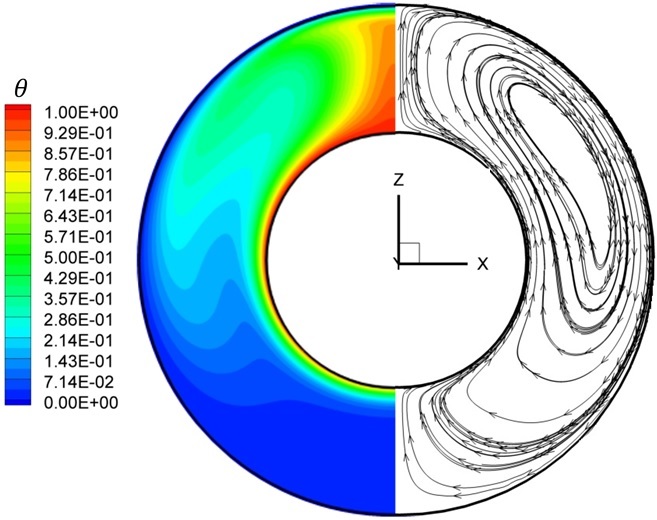}
        \caption{$Ra=10^5$}
    \end{subfigure}
    \caption{Flow characteristics of steady state natural convection flow in a spherical gap characterised by the value of $\phi=0.5$. Left column corresponds to 3D isosurfaces of temperature, $\theta = 1, 0.75, 0.5, 0.25, 0$; right column corresponds to the contours of temperature and path lines observed at $Z-X$ cross-section. }
    \label{fig:SteadySpherGap}
\end{figure}

We start our analysis with an investigation of the steady state flows developing inside the spherical shell characterized by the value of $\phi=0.5$ for the range of $10^4\leq Ra \leq 10^6$. Figure \ref{fig:SteadySpherGap} presents 3D  iso-surfaces of temperature distribution, along with the temperature contours and pathlines of the steady flow observed at the $X-Z$ cross section.

It can be clearly recognized that the obtained steady state flow is axi-symmetric for all $Ra$ values, in agreement with the previously published studies  \cite{gulb2015IJHMT,feldman2012aiaaa,chu1993heattransf,dehghan2010heattransfeng}. As expected the dominance of convective heat transfer increases with increasing $Ra$, which is reflected by a prevalence of thin thermal and velocity boundary layers near the surface of the hot sphere and higher temperature gradients inside the hot thermal plume located at the top region of the gap. It can also be observed that the center of the crescent convection cell typical of these kinds of flow is shifted upwards as the $Ra$ number increases. As can be seen from Table \ref{NuAvergSteady}, the values of the $\overline{Nu}$ number averaged over the surface of the outer sphere by
\begin{equation}
\overline{Nu}=\sqrt{PrRa}\Delta x\overline{Q}/\phi,
\label{HeatFlux}
\end{equation}
are in good agreement with the corresponding $\overline{Nu}$ values previously reported in the literature for the entire range of $Ra$ values. It should be noted here that there is a small (no more than 2\%) discrepancy between the average $\overline{Nu}$ values calculated  for internal and external spheres. This discrepancy is typical of numerical methods which do not directly resolve the thinnest boundary layers near the solid surfaces, such as IB and lattice Boltzman methods. For example, a discrepancy of about the same magnitude was reported by Gallegos and M\'{a}laga \cite{galleg2017EJMF}, who investigated the natural convection flow in eccentric spherical shells by the lattice Boltzman method. It is believed, however, that the observed discrepancy does not lead to incorrect conservation of the overall heat flux through the boundaries of the sphere (otherwise the steady state regime would not have been reached), rather, it only appears while postprocessing the obtained results and can be further minimized by increasing the number of Lagrangian points.
\begin{table}[]
\fontsize{10}{10}
\selectfont
\centering
\caption{The $\overline{Nu}$ number values averaged over the surface of an external sphere.}
\label{NuAvergSteady}
\begin{tabular}{lllllll}
\hline
\multicolumn{1}{>{}m{1cm}}{$\phi$} & \multicolumn{1}{>{}m{1cm}} {$Ra$} &   \multicolumn{1}{>{}m{2cm}} {Present} & \multicolumn{1}{>{}m{2cm}}{Ref. \cite{gulb2015IJHMT}} & \multicolumn{1}{>{}m{2cm}}  {Ref. \cite{feldman2012aiaaa}} &   \multicolumn{1}{>{}m{2cm}} {Ref. \cite{chu1993heattransf}} &   \multicolumn{1}{>{}m{2cm}} {Ref. \cite{dehghan2010heattransfeng}}\\
\hline
\multicolumn{1}{>{}m{1cm}}{} & \multicolumn{1}{>{}m{1cm}} {$10^3$} &   \multicolumn{1}{>{}m{2cm}} {1.1088} & \multicolumn{1}{>{}m{2cm}}{1.120} & \multicolumn{1}{>{}m{2cm}}  {1.104} &   \multicolumn{1}{>{}m{2cm}} {1.0990} &   \multicolumn{1}{>{}m{2cm}} {1.1310}\\
\multicolumn{1}{>{}m{1cm}}{0.5} & \multicolumn{1}{>{}m{1cm}} {$10^4$} &   \multicolumn{1}{>{}m{2cm}} {1.9648} & \multicolumn{1}{>{}m{2cm}}{1.987} & \multicolumn{1}{>{}m{2cm}}  {1.9665} &   \multicolumn{1}{>{}m{2cm}} {1.9730} &   \multicolumn{1}{>{}m{2cm}} {1.9495}\\
\multicolumn{1}{>{}m{1cm}}{} & \multicolumn{1}{>{}m{1cm}} {$10^5$} &   \multicolumn{1}{>{}m{2cm}} {3.4649} & \multicolumn{1}{>{}m{2cm}}{3.452} & \multicolumn{1}{>{}m{2cm}}  {3.4012} &   \multicolumn{1}{>{}m{2cm}} {3.4890} &   \multicolumn{1}{>{}m{2cm}} {3.4648}\\
\hline
\end{tabular}
\end{table}

\begin{table}[]
\fontsize{10}{10}
\selectfont
\centering
\caption{Setup and characteristics of the numerical solution obtained on a $200^3$ grid for the simulation of the natural convection inside a differentially heated spherical shell, characterized by the value of $\phi=0.5$.}
\label{SetupConcentrShell}
\begin{tabular}{lllllll}
\hline
\multicolumn{1}{>{}m{1cm}}{$Ra$} & \multicolumn{1}{>{}m{2cm}}{Time step duration, [s]} & \multicolumn{1}{>{}m{2cm}}  {RAM, [Gb]}& \multicolumn{1}{>{}m{1cm}} {$\Delta t$} &   \multicolumn{1}{>{}m{3cm}} {Method of solution of Eqs. \ref{SchurSplitted}-a} &   \multicolumn{1}{>{}m{2cm}} {Number of time steps} &   \multicolumn{1}{>{}m{2cm}} {Sparsing threshold for $[\textbf{\textit{I}}\textbf{\textit{H}}^{-1}\textbf{\textit{R}}]$} \\
\hline
$10^3$ & 2.5 & 20.70&     \multicolumn{1}{c}{\multirow{4}{*}{$10^{-3}$}}&  \multicolumn{1}{c}{\multirow{4}{*}{Direct ($LU$)}} & $O(10^4)$ & \multicolumn{1}{c}{\multirow{4}{*}{$10^{-16}$}}\\
$10^4$ & 2.4 & 15.04&    \multicolumn{1}{c}{}                          &  \multicolumn{1}{c}{}                               & $O(10^4)$ & \multicolumn{1}{c}{}\\
$10^5$ & 2.3 & 9.734&     \multicolumn{1}{c}{}&  \multicolumn{1}{c}{} & $O(10^5)$ & \multicolumn{1}{c}{}\\

\hline
\end{tabular}
\end{table}

\begin{figure}[H]
\centering
\begin{subfigure}{\textwidth}
        \includegraphics[width=0.45\textwidth,clip=]{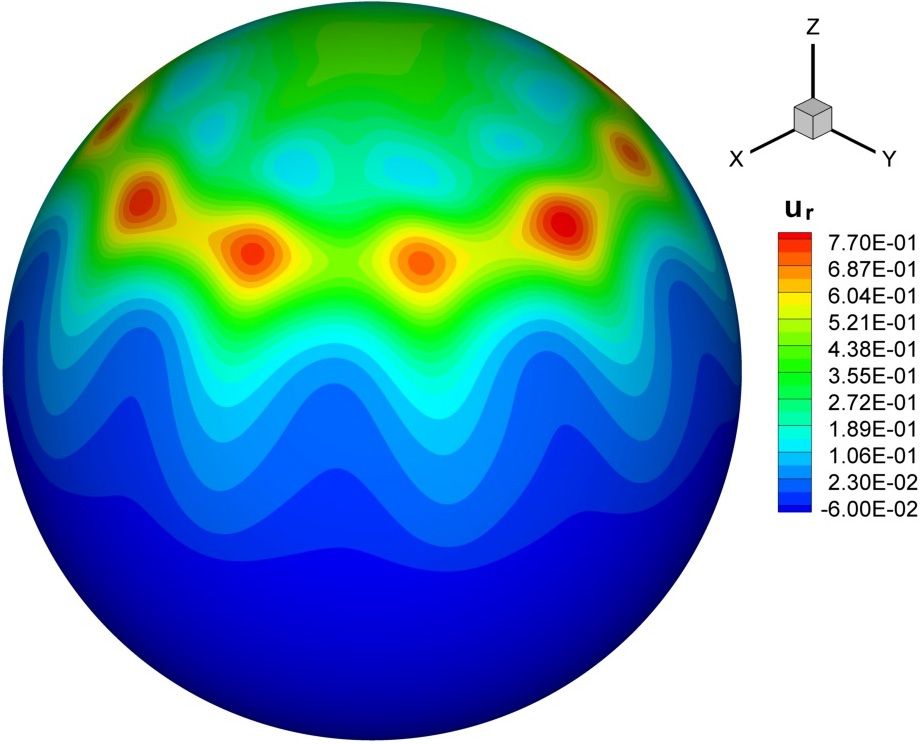}
        \includegraphics[width=0.455\textwidth,clip=]{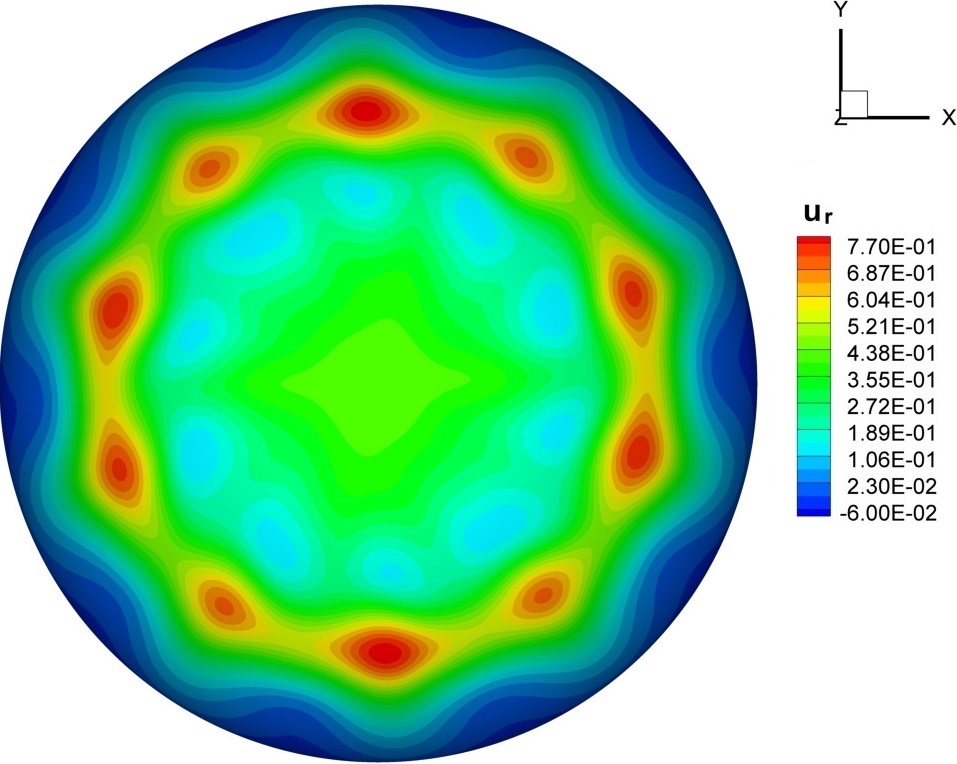}
        \caption{Snapshot of the radial velocity $v_r$  on the midrange $D=(D_i+D_o)/2$ spherical surface}
\end{subfigure}
\begin{subfigure}{\textwidth}
        \includegraphics[width=0.44\textwidth,clip=]{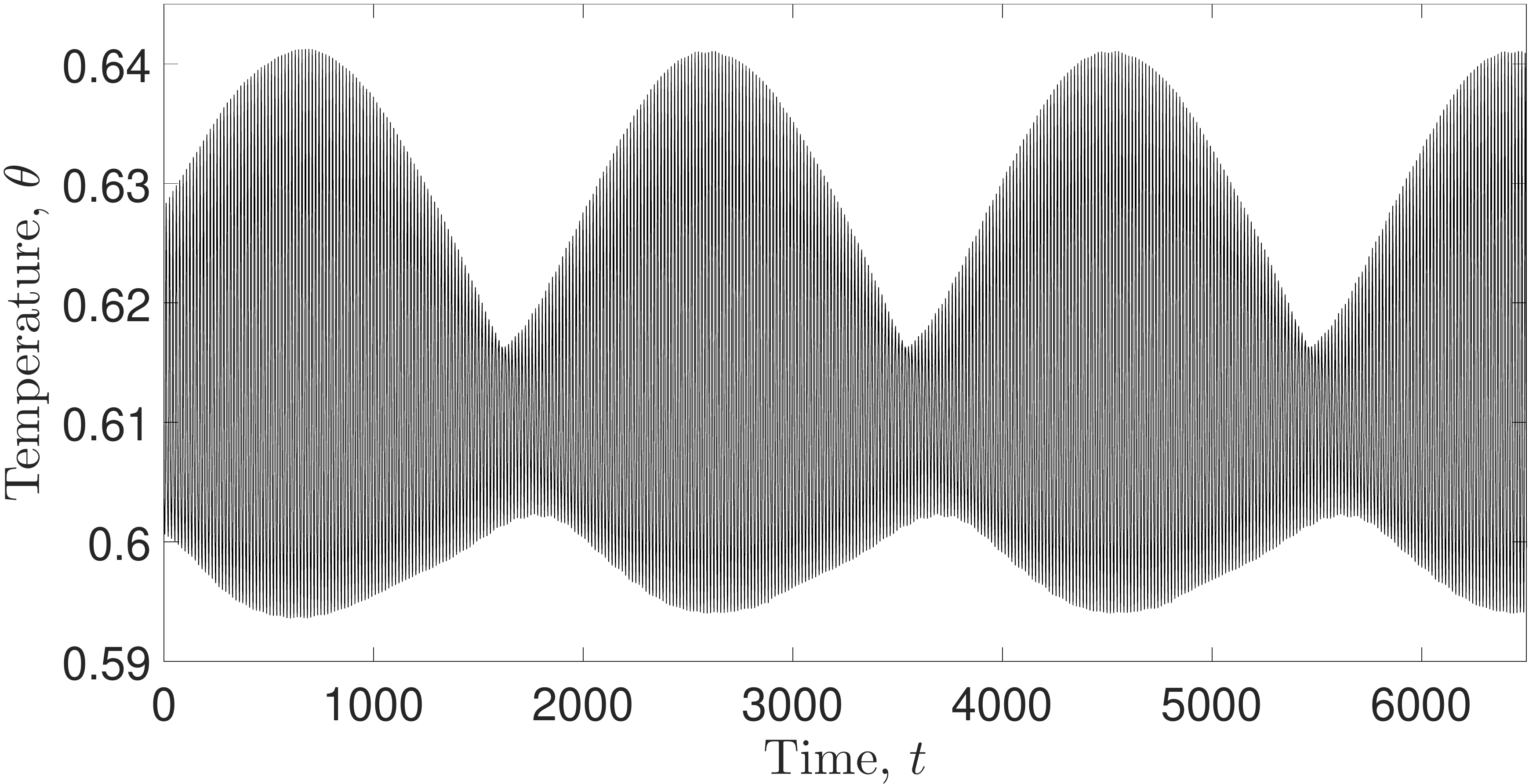}
        \includegraphics[width=0.45\textwidth,clip=]{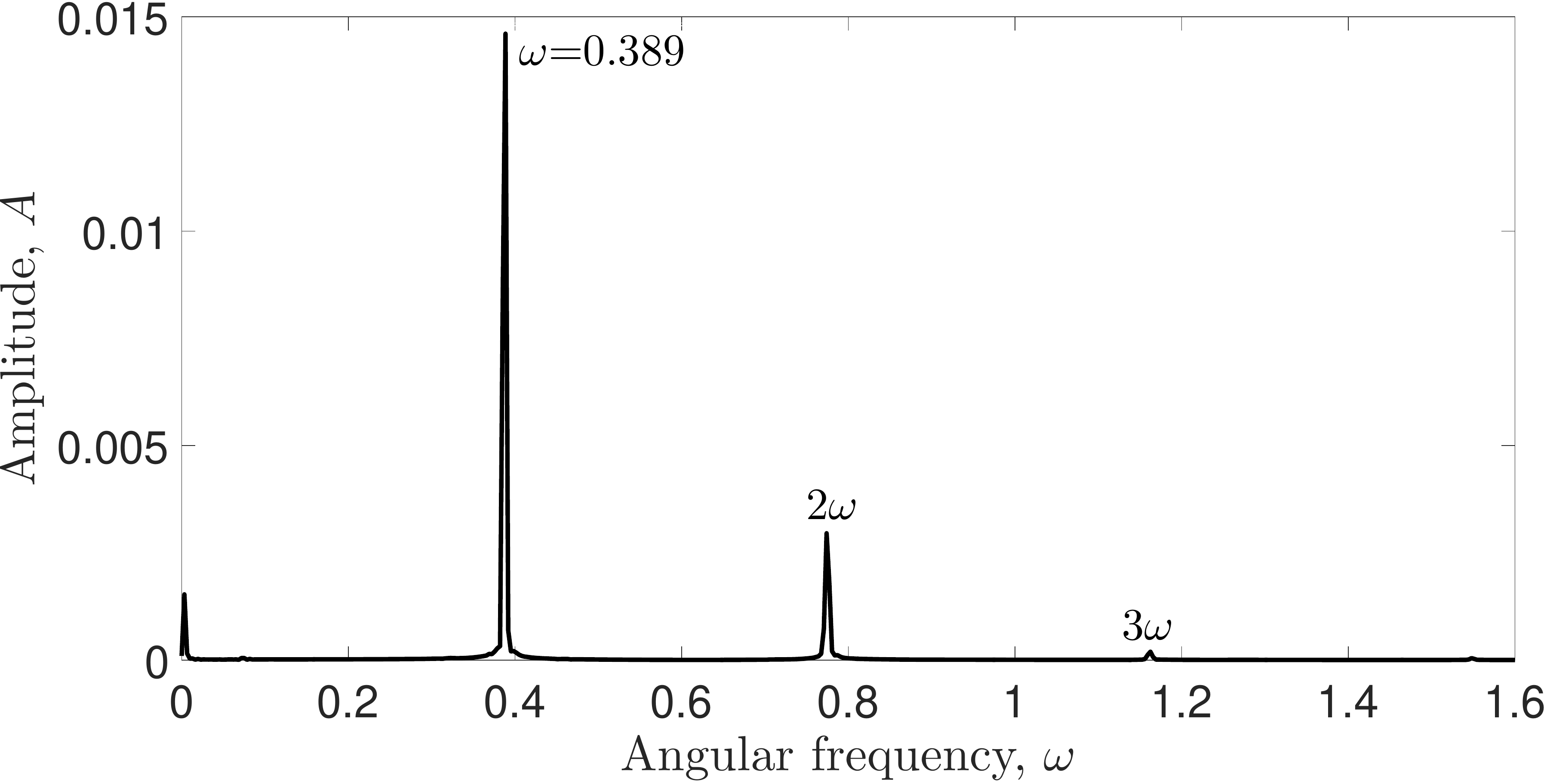}
        \caption{Time evolution and Fourier spectrum of the temperature acquired at point (0.01262, -2.1212, 2.1165)}
\end{subfigure}
\begin{subfigure}{\textwidth}
        \includegraphics[width=0.45\textwidth,clip=]{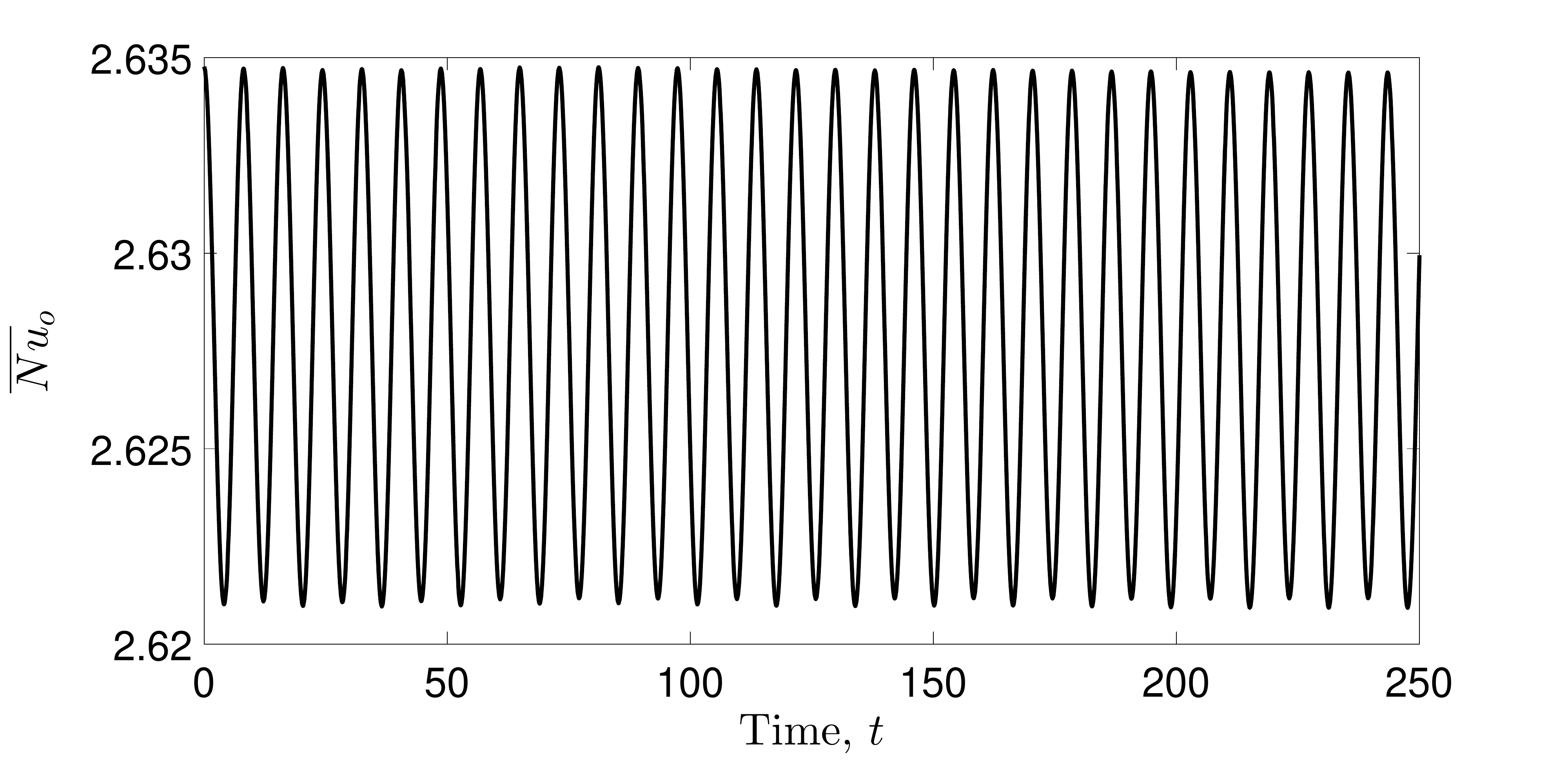}
        \includegraphics[width=0.45\textwidth,clip=]{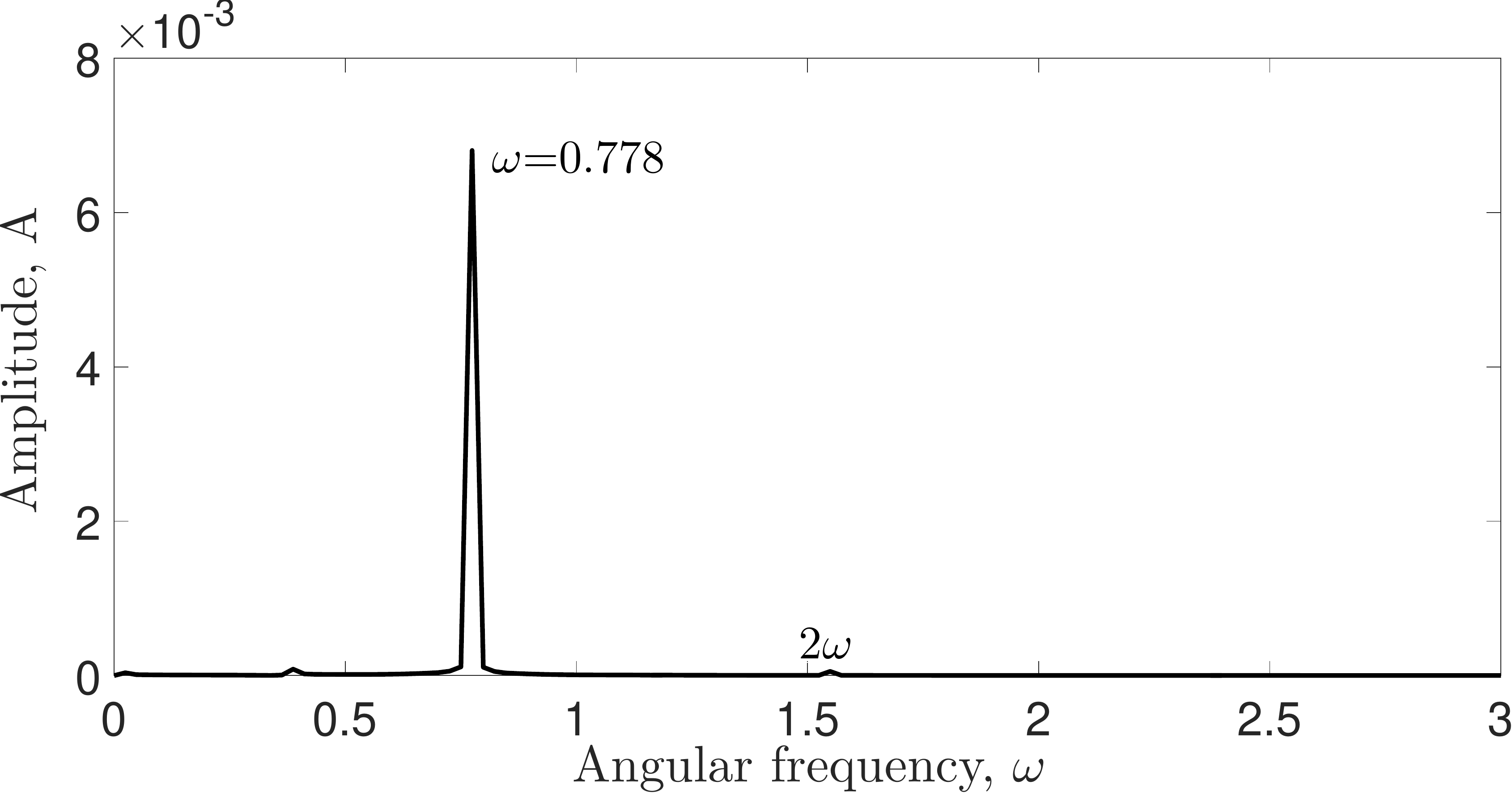}
        \caption{Time evaluation and Fourier spectrum of the Nusselt number averaged over the surface of the outer sphere, $\overline{Nu}_o$}
\end{subfigure}
    \caption{Flow characteristics of unsteady natural convection flow exhibiting travelling wave instability obtained for $Ra=4.6 \times 10^4$.}
    \label{fig:PeriodicSpherGap}
\end{figure}

To prove that the developed method can also accurately simulate unsteady fluid dynamics, the natural convection inside a concentric spherical shell characterized by the value of $\phi=0.714$ is simulated at $Ra=4.6\times 10^4$ and $Pr=0.7$. At these values, the natural convection flow developing inside the above configuration exhibits complex periodic multi cellular dynamics in the form of a travelling wave, as detailed in a number of previous studies \cite{scurtu2008jphys,feldman2013ijhmt,travnikov2015}. Figure \ref{fig:PeriodicSpherGap}a presents a snapshot of the distribution of radial velocity $v_r$ on the midrange $D=(D_i+D_o)/2$ spherical surface. The distribution is characterized by 10 clearly distinguished discrete convection cells in full agreement with the results obtained by the linear stability analysis of Travnikov et al. \cite{travnikov2015}, which  predict the highest growth rate for the wave number $m=10$ at $Ra_{cr}=3.9563\times 10^4$. Figure \ref{fig:PeriodicSpherGap}b presents the time evolution of temperature acquired at point (0.01262, -2.1212, 2.1165) located on the midrange spherical surface on the meridian circle along which the maximal $v_r$ component variations are observed. The Fourier transform of the temperature evolution yields the value of angular frequency of the bifurcated flow $\omega=0.389$. Note that this value  is in excellent agreement with the critical frequency value $\omega_{cr}$  reported in \cite{travnikov2015}\footnote{The $\omega_{cr}$ reported in \cite{travnikov2015} was renormalized by dividing it by $\sqrt{PrRa}$ to fit the scaling of the present study.}.It is remarkable that the average Nusselt number $\overline{Nu}$ is even more sensitive to the flow oscillations, and oscillates with an angular frequency twice that of the temperature (and the velocity ) fields (see Fig. \ref{fig:PeriodicSpherGap}c), in agreement with the recent study of Scurtu et al. \cite{scurtu2008jphys}.

The advantages of utilizing an iterative BICG algorithm for the solution of Eq. (\ref{SchurSplitted}a) for a large problem solved on a $300^3$ grid can be seen from the setup characteristics summarized in Table \ref{SetupGap}. In fact, despite the high resolution and the large amount  of Lagrangian points determining the surface of two immersed bodies (more than $3\times10^5$) the simulation only required 10Gb of RAM. This is comparable with the memory required when simulating convection flow from a single sphere in a cube on a $200^3$  grid (about $2\times 10^4$ Lagrangian points) and approximately 6 time less than maximal amount memory consumed when simulating convection flow  from a horizontal cylinder in a cube on a $200^3$ grid (about $1\times 10^5$ Lagrangian points)\footnote{Both simulations are based on direct $LU$ decomposition utilized for the solution of Eq. (\ref{SchurSplitted}a).}. The relatively long duration of a time step (15.1 s) is not surprising and is the result of both high grid resolution and a low value of convergence criterion for the the BICG algorithm, which was set to $\xi=10^{-7}$.

\begin{table}[]
\fontsize{10}{10}
\selectfont
\centering
\caption{Setup characteristics of the numerical solution obtained on a $300^3$ grid for the simulation of the natural convection inside a differentially heated spherical shell characterized by the value of $\phi=0.714$.}
\label{SetupGap}
\begin{tabular}{lllllll}
\hline
\multicolumn{1}{>{}m{1cm}}{$\phi$}& \multicolumn{1}{>{}m{2cm}}{Time step duration, [s]} & \multicolumn{1}{>{}m{2cm}}  {RAM, [Gb]} & \multicolumn{1}{>{}m{1cm}} {$\Delta t$} &   \multicolumn{1}{>{}m{3cm}} {Method of solution of Eqs. \ref{SchurSplitted}-a} &   \multicolumn{1}{>{}m{2cm}} {Number of time steps} &   \multicolumn{1}{>{}m{2cm}} {Threshold for the filling of $[\textbf{\textit{I}}\textbf{\textit{H}}^{-1}\textbf{\textit{R}}]$}\\
\hline
$0.714$ &15.1 & 10.1& $5\times10^{-3}$& BICG algorithm & $O(10^5)$ & $10^{-25}$\\
\hline
\end{tabular}
\end{table}

\section{Conclusions}
A novel formulation of the IB method based on a semi-implicit coupling between the Lagrangian force and heat flux densities and the Eulerian velocity and temperature fields has been developed and successfully verified for a number of confined 3D natural convection flows in the presence of immersed bodies of various geometries. The obtained results were favorably compared with available data for both steady and periodic flows. The existence of qualitative differences between the obtained and previously published study for the flow around a horizontal cylinder characterized by  high $R/L$ ratios and a value of $Ra=10^6$  was attributed to the flow non-linearity, which can exhibit different states at the same $Ra$ values.

The developed approach can be seen as a generic methodology for acquiring existing time marching solvers of NS equations based on a segregated pressure-velocity coupling (e.g. SIMPLE, fractional step,  projection methods and their derivatives) with the IB functionality. The implemented semi-implicit coupling between the Lagrangian force and heat flux densities and Eulerian velocity and temperature fields has proved to be a highly efficient and high fidelity methodology for simulating confined natural convection flows in the presence of immersed bodies at moderate $Ra$ values. The high efficiency of the methodology stems from the Schur complement decoupling of the governing equations, which allows for precomputing the most time consuming operations of the algorithm.

Performance of both  direct and iterative methods was investigated for the time marching stage of the algorithm. It was found that the time marching stage of the approach utilizing the direct method based on $LU$ decomposition is attractive for the problems solved on moderate (up tp $200^3$) grid resolutions. On the other hand, as a result of the poor scalability of the elimination and back substitution stages, the performance of the direct method based solvers deteriorates if high resolution grids are utilized. It was demonstrated that replacing a direct $LU$ factorization by an iterative BICG algorithm at the time marching stage of the algorithm provides a promising alternative if the solution is performed on high resolution grids.

Although the present study focused on a demonstration of the capabilities of the developed algorithm for stationary immersed bodies, it should be noted that the developed methodology can also be straightforwardly extended to configurations with periodically moving immersed bodies. Such an extension is of high relevance in the simulation of the undulatory motion of tiny sea creatures, as well as in various biomedical applications, and will be the focus of our future work. An emphasis  will be put on parallelizing the time marching stage of the algorithm by employing hybrid and distributed memory paradigms. Recalling that the time marching stage comprises a stand alone part completely detached from the original solver and physical model, it is expected that the above task can be accomplished with a reasonable programming effort.










\begin{thebibliography}{49}
\expandafter\ifx\csname natexlab\endcsname\relax\def\natexlab#1{#1}\fi
\providecommand{\bibinfo}[2]{#2}
\ifx\xfnm\relax \def\xfnm[#1]{\unskip,\space#1}\fi
\bibitem[{Peskin(1972)}]{peskin1972JCP}
\bibinfo{author}{C.~S. Peskin},
\newblock \bibinfo{title}{Flow patterns around heart valves: a numerical
  method},
\newblock \bibinfo{journal}{Journal of Computational Physics}
  \bibinfo{volume}{10} (\bibinfo{year}{1972}) \bibinfo{pages}{252--271}.
\bibitem[{Mohd-Yusof(1997)}]{Mohd-Yusof1997}
\bibinfo{author}{J.~Mohd-Yusof},
\newblock \bibinfo{title}{Combined immersed-boundary/b-spline methods for
  simulations of flow in complex geometries},
\newblock \bibinfo{journal}{Center for Turbulence Research, Annual Research
  Briefs}  (\bibinfo{year}{1997}) \bibinfo{pages}{317--327}.
\bibitem[{Faldun et~al.(2000)Faldun, Verzicco, Orlandi, and
  Mohd-Yusof}]{faldun2000JCP}
\bibinfo{author}{A.~Faldun, E.}, \bibinfo{author}{R.~Verzicco},
  \bibinfo{author}{P.~Orlandi}, \bibinfo{author}{J.~Mohd-Yusof},
\newblock \bibinfo{title}{Combined immersed-boundary finite-difference methods
  for three-dimensional complex flow simulations},
\newblock \bibinfo{journal}{Journal of Computational Physics}
  \bibinfo{volume}{161} (\bibinfo{year}{2000}) \bibinfo{pages}{35--60}.
\bibitem[{Le et~al.(2008)Le, Khoo, and Lim}]{le2007computmeth}
\bibinfo{author}{D.~V. Le}, \bibinfo{author}{B.~C. Khoo},
  \bibinfo{author}{K.~M. Lim},
\newblock \bibinfo{title}{{An implicit-forcing immersed boundary method for
  simulating viscous flows in irregular domains}},
\newblock \bibinfo{journal}{Comput. Meth. applied mechan. engineer.}
  \bibinfo{volume}{197} (\bibinfo{year}{2008}) \bibinfo{pages}{2119--2130}.
\bibitem[{Mittal and Iaccarino(2005)}]{mittal2005AnnRev}
\bibinfo{author}{R.~Mittal}, \bibinfo{author}{G.~Iaccarino},
\newblock \bibinfo{title}{The immersed boundary method},
\newblock \bibinfo{journal}{Ann. Rev. Fluid Mech.} \bibinfo{volume}{37}
  (\bibinfo{year}{2005}) \bibinfo{pages}{239--261}.
\bibitem[{Yoon et~al.(2010)Yoon, Yu, Ha, and Park}]{yoon2010heattransf}
\bibinfo{author}{H.~S. Yoon}, \bibinfo{author}{D.~H. Yu},
  \bibinfo{author}{M.~Y. Ha}, \bibinfo{author}{Y.~G. Park},
\newblock \bibinfo{title}{Three-dimensional natural convection in an enclosure
  with a sphere at different vertical locations},
\newblock \bibinfo{journal}{Int. J. Heat Mass Transfer} \bibinfo{volume}{53}
  (\bibinfo{year}{2010}) \bibinfo{pages}{3143--3155}.
\bibitem[{Ren et~al.(2012)Ren, Shu, and Yang}]{ren2012compufluid}
\bibinfo{author}{W.~W. Ren}, \bibinfo{author}{C.~Shu}, \bibinfo{author}{W.~M.
  Yang},
\newblock \bibinfo{title}{{Boundary condition-enforced immersed boundary method
  for thermal flow problems with Dirichlet temperature condition and its
  applications}},
\newblock \bibinfo{journal}{Comput. Fluid.} \bibinfo{volume}{57}
  (\bibinfo{year}{2012}) \bibinfo{pages}{40--51}.
\bibitem[{Ren et~al.(2013)Ren, Shu, and Y.}]{ren2013IJHMT}
\bibinfo{author}{W.~Ren}, \bibinfo{author}{C.~Shu}, \bibinfo{author}{W.~Y.},
\newblock \bibinfo{title}{An efficient immersed boundary method for thermal
  flow problems with heat flux boundary conditions},
\newblock \bibinfo{journal}{Int. J. Heat and Mass Transfer}
  \bibinfo{volume}{64} (\bibinfo{year}{2013}) \bibinfo{pages}{694--705}.
\bibitem[{Gulberg and Feldman(2015)}]{gulb2015IJHMT}
\bibinfo{author}{Y.~Gulberg}, \bibinfo{author}{Y.~Feldman},
\newblock \bibinfo{title}{On laminar natural convection inside multi-layered
  spherical shells},
\newblock \bibinfo{journal}{Int. J. Heat and Mass Transfer}
  \bibinfo{volume}{91} (\bibinfo{year}{2015}) \bibinfo{pages}{908--921}.
\bibitem[{Su et~al.(2007)Su, Lai, and Lin}]{su2007compufluid}
\bibinfo{author}{S.~W. Su}, \bibinfo{author}{M.~C. Lai}, \bibinfo{author}{C.~A.
  Lin},
\newblock \bibinfo{title}{An implicit-forcing immersed boundary method for
  simulating viscous flows in irregular domains},
\newblock \bibinfo{journal}{Comput. Fluid.} \bibinfo{volume}{36}
  (\bibinfo{year}{2007}) \bibinfo{pages}{313--324}.
\bibitem[{Wang et~al.(2008)Wang, Fan, and Luo}]{wang2008MTPHFLOW}
\bibinfo{author}{Z.~Wang}, \bibinfo{author}{J.~Fan}, \bibinfo{author}{K.~Luo},
\newblock \bibinfo{title}{{Combined multi-direct forcing and immersed boundary
  method for simulating flows with moving particles}},
\newblock \bibinfo{journal}{Int. J. Multiphase Flow} \bibinfo{volume}{34}
  (\bibinfo{year}{2008}) \bibinfo{pages}{283--302}.
\bibitem[{Kempe and Fr{\"{o}hlich, J.}(2012)}]{kempe2012JCP}
\bibinfo{author}{T.~Kempe}, \bibinfo{author}{Fr{\"{o}hlich, J.}},
\newblock \bibinfo{title}{An improved immersed boundary method with direct
  forcing for the simulation of particle laden flows},
\newblock \bibinfo{journal}{Journal of Computational Physics}
  \bibinfo{volume}{231} (\bibinfo{year}{2012}) \bibinfo{pages}{3663--3684}.
\bibitem[{Kempe et~al.(2015)Kempe, Lennartz, Schwarz, and Fr{\"{o}hlich,
  J.}}]{kempe2015JCP}
\bibinfo{author}{T.~Kempe}, \bibinfo{author}{M.~Lennartz},
  \bibinfo{author}{S.~Schwarz}, \bibinfo{author}{Fr{\"{o}hlich, J.}},
\newblock \bibinfo{title}{Imposing the free-slip condition with a continuous
  forcing immersed boundary method},
\newblock \bibinfo{journal}{Journal of Computational Physics}
  \bibinfo{volume}{282} (\bibinfo{year}{2015}) \bibinfo{pages}{183--209}.
\bibitem[{Glowinski et~al.(1998)Glowinski, Pan, and
  P\'eriaux}]{glowinski1998ComputMethApplMechEngrg}
\bibinfo{author}{R.~Glowinski}, \bibinfo{author}{T.~Pan},
  \bibinfo{author}{J.~P\'eriaux},
\newblock \bibinfo{title}{{Distributed Lagrange multiplier methods for
  incompressible viscous flow around moving rigid bodies}},
\newblock \bibinfo{journal}{Comput. Methods Appl. Mech. Engrg.}
  \bibinfo{volume}{151} (\bibinfo{year}{1998}) \bibinfo{pages}{181--194}.
\bibitem[{Glowinski et~al.(2001)Glowinski, Pan, Hesla, Joseph, and
  Periaux}]{Glowinski2001JCP}
\bibinfo{author}{R.~Glowinski}, \bibinfo{author}{T.~Pan},
  \bibinfo{author}{T.~Hesla}, \bibinfo{author}{D.~Joseph},
  \bibinfo{author}{A.~Periaux},
\newblock \bibinfo{title}{A fictitious domain approach to the direct numerical
  simulation of incompressible viscous flow past moving rigid bodies:
  application to particulate flow},
\newblock \bibinfo{journal}{J. Comput. Phys.} \bibinfo{volume}{169}
  (\bibinfo{year}{2001}) \bibinfo{pages}{363--426}.
\bibitem[{Yu et~al.(2002)Yu, Phan-Thien, Fan, and Tanner}]{Yu2002JNNFM}
\bibinfo{author}{Z.~Yu}, \bibinfo{author}{N.~Phan-Thien},
  \bibinfo{author}{Y.~Fan}, \bibinfo{author}{R.~Tanner},
\newblock \bibinfo{title}{Viscoelastic mobility problem of a system of
  particles},
\newblock \bibinfo{journal}{J. Non-Newtonian Fluid Mech.} \bibinfo{volume}{104}
  (\bibinfo{year}{2002}) \bibinfo{pages}{87--124}.
\bibitem[{Yu et~al.(2004)Yu, Phan-Thien, and Tanner}]{Yu2004JFM}
\bibinfo{author}{Z.~Yu}, \bibinfo{author}{N.~Phan-Thien},
  \bibinfo{author}{R.~Tanner},
\newblock \bibinfo{title}{Dynamical simulation of sphere motion in a vertical
  tube},
\newblock \bibinfo{journal}{J. Fluid Mech.} \bibinfo{volume}{518}
  (\bibinfo{year}{2004}) \bibinfo{pages}{61--93}.
\bibitem[{Yu(2005)}]{Yu2005JCP}
\bibinfo{author}{Z.~Yu},
\newblock \bibinfo{title}{{A DLM/FD method for fluid/flexible-body
  interactions}},
\newblock \bibinfo{journal}{J. Comput. Phys.} \bibinfo{volume}{207}
  (\bibinfo{year}{2005}) \bibinfo{pages}{1--27}.
\bibitem[{Taira and Colonius(2007)}]{taira2007JCP}
\bibinfo{author}{K.~Taira}, \bibinfo{author}{T.~Colonius},
\newblock \bibinfo{title}{The immersed boundary method: A projection approach},
\newblock \bibinfo{journal}{Journal of Computational Physics}
  \bibinfo{volume}{225} (\bibinfo{year}{2007}) \bibinfo{pages}{3121--3133}.
\bibitem[{Choi et~al.(2015)Choi, Colonius, and Williams}]{choi2015jfm}
\bibinfo{author}{J.~Choi}, \bibinfo{author}{T.~Colonius},
  \bibinfo{author}{D.~R. Williams},
\newblock \bibinfo{title}{{Surging and plunging oscillations of an airfoil at
  low Reynolds number}},
\newblock \bibinfo{journal}{Journal of Fluid Mechanics} \bibinfo{volume}{763}
  (\bibinfo{year}{2015}) \bibinfo{pages}{237--253}.
\bibitem[{Samanta et~al.(2010)Samanta, Appelo, Colonius, Nott, and
  Hall}]{samanta2010aiaa}
\bibinfo{author}{A.~Samanta}, \bibinfo{author}{D.~Appelo},
  \bibinfo{author}{T.~Colonius}, \bibinfo{author}{J.~Nott},
  \bibinfo{author}{J.~Hall},
\newblock \bibinfo{title}{Computational modeling and experiments of natural
  convection for a titan mongolfiere},
\newblock \bibinfo{journal}{AIAA Journal} \bibinfo{volume}{48}
  (\bibinfo{year}{2010}) \bibinfo{pages}{1007--1016}.
\bibitem[{Wang and Eldredge(2915)}]{wang2015JCP}
\bibinfo{author}{C.~Wang}, \bibinfo{author}{J.~D. Eldredge},
\newblock \bibinfo{title}{Strongly coupled dynamics of fluids and rigid-body
  systems with the immersed boundary projection method},
\newblock \bibinfo{journal}{Journal of Computational Physics}
  \bibinfo{volume}{295} (\bibinfo{year}{2915}) \bibinfo{pages}{87--113}.
\bibitem[{Kallemov et~al.(2016)Kallemov, Bhalla, Griffith, and
  Donev}]{kallemov2016CAMcOS}
\bibinfo{author}{B.~Kallemov}, \bibinfo{author}{A.~Bhalla},
  \bibinfo{author}{B.~Griffith}, \bibinfo{author}{A.~Donev},
\newblock \bibinfo{title}{{An immersed boundary method for rigid bodies}},
\newblock \bibinfo{journal}{Comm. App. Math. Comp. Sci.} \bibinfo{volume}{11}
  (\bibinfo{year}{2016}) \bibinfo{pages}{79--141}.
\bibitem[{Stein et~al.(2016)Stein, Guy, and Thomases}]{Stein2016JCP}
\bibinfo{author}{D.~Stein}, \bibinfo{author}{R.~Guy},
  \bibinfo{author}{B.~Thomases},
\newblock \bibinfo{title}{{Immersed boundary smooth extension: A high-order
  method for solving PDE on arbitrary smooth domains using Fourier spectral
  methods}},
\newblock \bibinfo{journal}{J. Comput. Phys.} \bibinfo{volume}{304}
  (\bibinfo{year}{2016}) \bibinfo{pages}{252--274}.
\bibitem[{Feldman and Gulbeg(2016)}]{feldman2016JCP}
\bibinfo{author}{Y.~Feldman}, \bibinfo{author}{Y.~Gulbeg},
\newblock \bibinfo{title}{{An extension of the immersed boundary method based
  on the distributed Lagrange multiplier approach}},
\newblock \bibinfo{journal}{Journal of Computational Physics}
  \bibinfo{volume}{322} (\bibinfo{year}{2016}) \bibinfo{pages}{248--266}.
\bibitem[{Liska and Colonius(2016)}]{Liska2016JCP}
\bibinfo{author}{S.~Liska}, \bibinfo{author}{T.~Colonius},
\newblock \bibinfo{title}{{A fast immersed boundary method for external
  incompressible viscous flows using lattice Green's functions}},
\newblock \bibinfo{journal}{J. Comput. Phys.} \bibinfo{volume}{331}
  (\bibinfo{year}{2016}) \bibinfo{pages}{257--279}.
\bibitem[{Bao et~al.(2017)Bao, Donev, Griffith, McQueen, and
  Peskin}]{Bao2017JCP}
\bibinfo{author}{Y.~Bao}, \bibinfo{author}{A.~Donev},
  \bibinfo{author}{B.~Griffith}, \bibinfo{author}{D.~McQueen},
  \bibinfo{author}{C.~Peskin},
\newblock \bibinfo{title}{{An Immersed Boundary method with divergence-free
  velocity interpolation and force spreading}},
\newblock \bibinfo{journal}{J. Comput. Phys.} \bibinfo{volume}{347}
  (\bibinfo{year}{2017}) \bibinfo{pages}{183--206}.
\bibitem[{Stein et~al.(2017)Stein, Guy, and Thomases}]{Stein2017JCP}
\bibinfo{author}{D.~Stein}, \bibinfo{author}{R.~Guy},
  \bibinfo{author}{B.~Thomases},
\newblock \bibinfo{title}{{Immersed Boundary Smooth Extension (IBSE): A
  high-order method for solving incompressible flows in arbitrary smooth
  domains}},
\newblock \bibinfo{journal}{J. Comput. Phys.} \bibinfo{volume}{335}
  (\bibinfo{year}{2017}) \bibinfo{pages}{155--178}.
\bibitem[{Park et~al.(2016)Park, Pan, Lee, and Choi}]{Park2016JCP}
\bibinfo{author}{H.~Park}, \bibinfo{author}{X.~Pan}, \bibinfo{author}{C.~Lee},
  \bibinfo{author}{J.-I. Choi},
\newblock \bibinfo{title}{A pre-conditioned implicit direct forcing based
  immersed boundary method for incompressible viscous flows},
\newblock \bibinfo{journal}{J. Comput. Phys.} \bibinfo{volume}{314}
  (\bibinfo{year}{2016}) \bibinfo{pages}{774--799}.
\bibitem[{Le et~al.(2008)Le, Khoo, and Lim}]{lee2008ComputMethApplMechEngrg}
\bibinfo{author}{D.~Le}, \bibinfo{author}{B.~Khoo}, \bibinfo{author}{K.~Lim},
\newblock \bibinfo{title}{{An implicit-forcing immersed boundary method for
  simulating viscous flows in irregular domains}},
\newblock \bibinfo{journal}{Comput. Methods Appl. Mech. Engrg.}
  \bibinfo{volume}{197} (\bibinfo{year}{2008}) \bibinfo{pages}{2119--2130}.
\bibitem[{Adams et~al.(1999)Adams, Swarztrauber, and Sweet}]{fishpack}
\bibinfo{author}{J.~Adams}, \bibinfo{author}{R.~Swarztrauber},
  \bibinfo{author}{R.~Sweet},
\newblock \bibinfo{title}{{FISHPACK: Efficient FORTRAN subprograms for the
  solution of separable eliptic partial differential equations }},
\newblock \bibinfo{journal}{http://www.scd.ucar.edu/css/software/fishpack/}
  (\bibinfo{year}{1999}).
\bibitem[{Vitoshkin and Gelfgat(2013)}]{vitoshkin2013}
\bibinfo{author}{H.~Vitoshkin}, \bibinfo{author}{A.~Y. Gelfgat},
\newblock \bibinfo{title}{{On direct inverse of Stokes, Helmholtz and Laplacian
  operators in view of time-stepper-based Newton and Arnoldi solvers in
  incompressible CFD}},
\newblock \bibinfo{journal}{Communications in Computational Physics}
  \bibinfo{volume}{14} (\bibinfo{year}{2013}) \bibinfo{pages}{1103--1119}.
\bibitem[{Patankar and Spalding(1972)}]{patankar197IJHMT}
\bibinfo{author}{S.~V. Patankar}, \bibinfo{author}{D.~B. Spalding},
\newblock \bibinfo{title}{{A calculation procedure for heat, mass and momentum
  in three-dimensional parabolic fows}},
\newblock \bibinfo{journal}{Int. J. Heat Mass Transf.} \bibinfo{volume}{15}
  (\bibinfo{year}{1972}) \bibinfo{pages}{1787--1806}.
\bibitem[{Roma et~al.(1999)Roma, Peskin, and Berger}]{roma1999JCP}
\bibinfo{author}{A.~Roma}, \bibinfo{author}{C.~S. Peskin},
  \bibinfo{author}{M.~J. Berger},
\newblock \bibinfo{title}{An adaptive version of the immersed boundary method},
\newblock \bibinfo{journal}{Journal of Computational Physics}
  \bibinfo{volume}{153} (\bibinfo{year}{1999}) \bibinfo{pages}{509--534}.
\bibitem[{Uhlmann(2005)}]{uhlmann2005JCP}
\bibinfo{author}{M.~Uhlmann},
\newblock \bibinfo{title}{An immersed boundary method with direct forcing for
  the simulation of particulate flows},
\newblock \bibinfo{journal}{Journal of Computational Physics}
  \bibinfo{volume}{209} (\bibinfo{year}{2005}) \bibinfo{pages}{448--476}.
\bibitem[{Patankar(1980)}]{patankar1980}
\bibinfo{author}{S.~V. Patankar}, \bibinfo{title}{{Numerical heat transfer and
  fluid flow}}, \bibinfo{publisher}{McGraw-Hill}, \bibinfo{year}{1980}.
\bibitem[{Amestoy et~al.(1998)Amestoy, Duff, L\'Excellent, and
  Koster}]{amestoy1998ComputMethApplMechEngrg}
\bibinfo{author}{P.~Amestoy}, \bibinfo{author}{I.~Duff},
  \bibinfo{author}{J.~L\'Excellent}, \bibinfo{author}{J.~Koster},
\newblock \bibinfo{title}{{Multifrontal parallel distributed symmetric and
  unsymmetric solvers}},
\newblock \bibinfo{journal}{Comput. Methods Appl. Mech. Engrg.}
  \bibinfo{volume}{184} (\bibinfo{year}{1998}) \bibinfo{pages}{501--520}.
\bibitem[{Amestoy et~al.(2001)Amestoy, Duff, L\'Excellent, and
  Koster}]{amestoy2001SIAMJMatr}
\bibinfo{author}{P.~Amestoy}, \bibinfo{author}{I.~Duff},
  \bibinfo{author}{J.~L\'Excellent}, \bibinfo{author}{J.~Koster},
\newblock \bibinfo{title}{{A fully asynchronous multifrontal solver using
  distributed dynamic scheduling}},
\newblock \bibinfo{journal}{{SIAM J. Matrix Anal. Appl.}} \bibinfo{volume}{23}
  (\bibinfo{year}{2001}) \bibinfo{pages}{15--41}.
\bibitem[{Leopardi(2006)}]{Leopardi2006TRANS}
\bibinfo{author}{P.~Leopardi},
\newblock \bibinfo{title}{A partition of the unit sphere into regions of equal
  area and small diameter},
\newblock \bibinfo{journal}{Electron. Trans. Numer. Anal.} \bibinfo{volume}{25}
  (\bibinfo{year}{2006}) \bibinfo{pages}{309--327}.
\bibitem[{Jeong and Hussain(1995)}]{Jeong1995JFM}
\bibinfo{author}{J.~Jeong}, \bibinfo{author}{F.~Hussain},
\newblock \bibinfo{title}{On the identification of a vortex},
\newblock \bibinfo{journal}{J. Fluid Mech.} \bibinfo{volume}{285}
  (\bibinfo{year}{1995}) \bibinfo{pages}{69--94}.
\bibitem[{Seo et~al.(2016)Seo, Doo, and Ha}]{seo2016}
\bibinfo{author}{Y.~Seo}, \bibinfo{author}{J.~Doo}, \bibinfo{author}{M.~Ha},
\newblock \bibinfo{title}{Three-dimensional flow instability of natural
  convection induced by variation in radius of inner circular cylinder inside
  cubic enclosure},
\newblock \bibinfo{journal}{Int. J. Heat and Mass Transfer}
  \bibinfo{volume}{95} (\bibinfo{year}{2016}) \bibinfo{pages}{566--578}.
\bibitem[{Feldman and Colonius(2013)}]{feldman2013}
\bibinfo{author}{Y.~Feldman}, \bibinfo{author}{T.~Colonius},
\newblock \bibinfo{title}{On a transitional and turbulent natural convection in
  spherical shells},
\newblock \bibinfo{journal}{Int. J. Heat and Mass Transfer}
  \bibinfo{volume}{64} (\bibinfo{year}{2013}) \bibinfo{pages}{514--525}.
\bibitem[{Feldman et~al.(2012)Feldman, Colonius, Pauken, Hall, and
  Jones}]{feldman2012aiaaa}
\bibinfo{author}{Y.~Feldman}, \bibinfo{author}{T.~Colonius},
  \bibinfo{author}{M.~Pauken}, \bibinfo{author}{J.~Hall},
  \bibinfo{author}{J.~Jones},
\newblock \bibinfo{title}{Simulation and cryogenic experiments of natural
  convection for the {Titan} {Montgolfiere}},
\newblock \bibinfo{journal}{accepted for publication at AIAA Journal}
  (\bibinfo{year}{2012}).
\bibitem[{Chu and Lee(1993)}]{chu1993heattransf}
\bibinfo{author}{H.~S. Chu}, \bibinfo{author}{T.~S. Lee},
\newblock \bibinfo{title}{Transient natural convection heat transfer between
  concentric spheres},
\newblock \bibinfo{journal}{Int. J. Heat Mass Transfer} \bibinfo{volume}{36}
  (\bibinfo{year}{1993}) \bibinfo{pages}{3159--3170}.
\bibitem[{Dehghan and Masih(2010)}]{dehghan2010heattransfeng}
\bibinfo{author}{A.~A. Dehghan}, \bibinfo{author}{K.~Masih},
\newblock \bibinfo{title}{Numerical simulation of buoyancy-induced turbulent
  flow between two concentric isothermal spheres},
\newblock \bibinfo{journal}{Heat Transfer Eng.} \bibinfo{volume}{31}
  (\bibinfo{year}{2010}) \bibinfo{pages}{33--44}.
\bibitem[{Gallegos and Malaga(2017)}]{galleg2017EJMF}
\bibinfo{author}{A.~D. Gallegos}, \bibinfo{author}{C.~Malaga},
\newblock \bibinfo{title}{Natural convection in eccentric spherical annuli},
\newblock \bibinfo{journal}{Europ. J. Mechics - B/Fluids} \bibinfo{volume}{65}
  (\bibinfo{year}{2017}) \bibinfo{pages}{464--471}.
\bibitem[{Scurtu et~al.(2008)Scurtu, Futterer, and Egbers}]{scurtu2008jphys}
\bibinfo{author}{N.~Scurtu}, \bibinfo{author}{B.~Futterer},
  \bibinfo{author}{C.~Egbers},
\newblock \bibinfo{title}{Three-dimensional natural convection in spherical
  annuli},
\newblock \bibinfo{journal}{Journal of Physics: Conference Series}
  \bibinfo{volume}{137,012017} (\bibinfo{year}{2008}) \bibinfo{pages}{1--9}.
\bibitem[{Feldman and Colonius(2013)}]{feldman2013ijhmt}
\bibinfo{author}{Y.~Feldman}, \bibinfo{author}{T.~Colonius},
\newblock \bibinfo{title}{On a transitional and turbulent natural convection in
  spherical shells},
\newblock \bibinfo{journal}{Int. J. Heat Mass Transf.} \bibinfo{volume}{64}
  (\bibinfo{year}{2013}) \bibinfo{pages}{514--525}.
\bibitem[{Travnikov et~al.(2015)Travnikov, Eckert, and
  Odenbach}]{travnikov2015}
\bibinfo{author}{V.~Travnikov}, \bibinfo{author}{K.~Eckert},
  \bibinfo{author}{S.~Odenbach},
\newblock \bibinfo{title}{Linear stability analysis of the convective flow in a
  spherical gap with $\eta$ = 0.714},
\newblock \bibinfo{journal}{Int. J. Heat and Mass Transfer}
  \bibinfo{volume}{80} (\bibinfo{year}{2015}) \bibinfo{pages}{266--273}.

\end{thebibliography}
\end{document}